\documentclass[twocolumn, twocolappendix]{aastex7}

\usepackage{wrapfig}
\usepackage{float}

\usepackage{mathptmx}
\usepackage[T1]{fontenc}
\usepackage{ae,aecompl}
\usepackage{graphicx}	
\usepackage{amsmath}	
\usepackage{amssymb}	
\usepackage{subfigmat}
\usepackage{enumerate}
\usepackage{natbib}
\usepackage{verbatim}
\usepackage{tabularx, booktabs}
\usepackage{rotating}
\usepackage{longtable}
\usepackage{ltablex}
\usepackage{threeparttablex}

\usepackage{listings}
\usepackage{siunitx}

\newcommand{\degree}{$^{\circ}$}
\newcommand{\rom}[1]{\uppercase\expandafter{\romannumeral#1}}
\newcommand{\msun}{\mbox{M$_\odot$}}

\begin{document}

\title{EPISODE III: The Nested Jet/Outflow Morphology of EC 53 Revealed by JWST and ALMA}

\author[0000-0001-6324-8482]{Seonjae Lee}
\email{sunjae627@snu.ac.kr}
\affil{Department of Physics and Astronomy, Seoul National University, 1 Gwanak-ro, Gwanak-gu, Seoul 08826, Republic of Korea}

\author[0000-0003-3119-2087]{Jeong-Eun Lee}
\email{lee.jeongeun@snu.ac.kr}
\affil{Department of Physics and Astronomy, Seoul National University, 1 Gwanak-ro, Gwanak-gu, Seoul 08826, Republic of Korea}
\affil{SNU Astronomy Research Center, Seoul National University, 1 Gwanak-ro, Gwanak-gu, Seoul 08826, Republic of Korea}


\author[0000-0002-2523-3762]{Chul-Hwan Kim}
\email{chkim9407@snu.ac.kr}
\affil{Department of Physics and Astronomy, Seoul National University, 1 Gwanak-ro, Gwanak-gu, Seoul 08826, Republic of Korea}

\author[0000-0002-0226-9295]{Seokho Lee}
\email{seokholee@kasi.re.kr}
\affil{Korea Astronomy and Space Science Institute, 776 Daedeok-daero, Yuseong-gu, Daejeon 34055, Republic of Korea}

\author[0000-0002-6773-459X]{Doug Johnstone}
\email{doug.johnstone@nrc-cnrc.gc.ca}
\affil{NRC Herzberg Astronomy and Astrophysics, 5071 West Saanich Rd, Victoria, BC V9E 2E7, Canada}
\affil{Department of Physics and Astronomy, University of Victoria, 3800 Finnerty Rd, Elliot Building, Victoria, BC V8P 5C2, Canada}

\author[0000-0002-7154-6065]{Gregory J. Herczeg}
\email{gherczeg1@gmail.com}
\affil{Kavli Institute for Astronomy and Astrophysics, Peking University, Yiheyuan Lu 5, Haidian Qu, 100871 Beijing, People’s Republic of China}
\affil{Department of Astronomy, Peking University, Yiheyuan 5, Haidian Qu, 100871 Beijing, People’s Republic of China}

\author[0000-0003-1665-5709]{Joel Green}
\email{jgreen@stsci.edu}
\affil{STScI, 3700 San Martin Drive, Baltimore, MD 21218, USA}

\author[0000-0001-8822-6327]{Logan Francis}
\email{francis@strw.leidenuniv.nl}
\affil{Leiden Observatory, Leiden University, PO Box 9513, 2300 RA Leiden, The Netherlands}

\author[0000-0001-8227-2816]{Yao-Lun Yang}
\email{yao-lun.yang@riken.jp}
\affil{Star and Planet Formation Laboratory, Pioneering Research Institute, RIKEN, 2-1 Hirosawa, Wako, Saitama, 351-0198, Japan}

\author[0009-0007-5845-6974]{Hyundong Lee}
\email{dennishd123@snu.ac.kr}
\affil{Department of Physics and Astronomy, Seoul National University, 1 Gwanak-ro, Gwanak-gu, Seoul 08826, Republic of Korea}

\author[0000-0003-0998-5064]{Nagayoshi Ohashi}
\email{ohashi@asiaa.sinica.edu.tw}
\affil{Academia Sinica Institute of Astronomy \& Astrophysics, 11F of Astronomy-Mathematics Building, AS/NTU, No. 1, Sec. 4, Roosevelt Rd., Taipei 10617, Taiwan,
R.O.C.}

\correspondingauthor{Jeong-Eun Lee}
\email{lee.jeongeun@snu.ac.kr}





\begin{abstract}

We present an extensive study of the structure and kinematics of the jet and outflow of EC 53, a Class I protostar with a quasi-periodic variability, using combined James Webb Space Telescope (JWST) and Atacama Large Millimeter/submillimeter Array (ALMA) observations. ALMA continuum observations resolve a compact disk with a radius of $\sim$0.14\arcsec\ (60\,au). Scattered light from the outflow cavity is prominent in the short-wavelength NIRCam and NIRSpec observations, revealing only the southeast nearside lobe. We detected 27 H$_2$ emission lines tracing a narrow, cone-shaped structure within the outflow cavity. A high-velocity ionized jet is detected in several forbidden atomic lines, characterized by a position angle of 142\degree, an opening angle of 1.4\degree, and an estimated geometric launching radius of at most $\sim$40\,au. 
Mid-infrared CO ro-vibrational emission lines, stronger in the P-branch, show a similar distribution to the H$_2$ emission and are likely to originate from hot gas within the outflow cavity. CO and C$_2$H emission lines detected by ALMA trace slower, colder outflow components and cavity walls. The spatial and kinematic stratification between the hot atomic and molecular components and the colder molecular gas is consistent with predictions from MHD disk wind models,  although envelope material entrained by a wide-angle wind or jet may also contribute. Our analysis highlights the powerful synergy between JWST and ALMA in advancing the understanding of protostellar jets and outflows across multiple spatial and physical scales.
\end{abstract}

\keywords{}

\section{Introduction} \label{sec:intro}

The formation of stars is initiated by the gravitational collapse of dense cores within molecular clouds \citep{Shu1987, Terebey1984}. As these cores collapse, conservation of angular momentum increases the rotational velocity, which, if unregulated, can bring the nascent star close to the so-called break-up limit, where centrifugal forces exceed the gravitational pull \citep{Bodenheimer1995}. This scenario underscores the classical angular momentum problem in star formation; without efficient angular momentum removal, protostars cannot form \citep{McKee2007}. This issue is mitigated through the combined roles of two key components: accretion disks and bipolar ejections. The disk acts as a reservoir of angular momentum, allowing mass transport toward the protostar, while jets and outflows serve as channels for expelling excess angular momentum from the system \citep{Frank2014, Pudritz2007}.

Bipolar ejection phenomena are typically divided into high-velocity, well-collimated jets and lower-velocity, wide-angle outflows \citep{Arce2007, Frank2014}. Jets are often associated with shock features such as knots and bow shocks and are believed to originate in the inner regions of the star-disk system, near the dust sublimation radius. These jets are traced by atomic or ionic emission lines, including [Fe II], [Ni II], and [Ne II] \citep[see review by][]{Bally2016, vanDishoeck2025}. Molecular jets, which are more common in deeply embedded Class 0 sources, can be traced with molecular lines such as CO, SiO, SO, and ro-vibrationally excited H$_2$ \citep{Tafalla2000, Hirano2010, cfLee2017, CarattioGaratti2024}. On the other hand, outflows are typically traced with cooler molecular species, such as purely rotational H$_2$, CO, H$_2$CO, and CH$_3$OH \citep{Bachiller1996, Jorgensen2007, Tychoniec2021}.

Multiple theoretical models have been proposed to explain the launching and collimation of jets and outflows \citep{Pudritz2019, Ray2021}. The stellar wind model proposes that winds are launched directly from the stellar surface, carrying away angular momentum and regulating stellar spin \citep{Matt2005, Matt2008}. The X-wind model suggests that wide-angle high-velocity winds power outflows and originate near the corotation radius, where the stellar magnetosphere truncates the disk \citep{shu1994}. These models emphasize angular momentum removal near the inner disk, allowing mass to reach the protostar without a catastrophic increase in angular momentum, but leave unexplained the process by which accretion through the disk occurs. The magnetohydrodynamic (MHD) disk wind model offers a promising solution by enabling angular momentum extraction over a broad radial extent of the disk and creating an inward flow of material through the disk \citep{Blandford1982, Ferreira2006, Bai2013}.

MRI (magnetorotational instability) has also been identified as a key mechanism for driving turbulence and accretion in ionized regions of the disk \citep{Balbus1991}. MRI is efficient in hot, ionized regions but becomes ineffective in the colder, neutral midplane of the outer disk—known as the `dead zone' \citep{Lesur2021, Pascucci2023}. In the dead zone, MHD disk winds are suggested as being dominant. Other outflow-driving mechanisms, such as photoevaporative (PE) disk winds \citep{Ercolano2017}, wind-driven entrainment \citep{cfLee2000, Liang2020}, and jet-driven entrainment \citep{Rabenanahary2022}, may also operate simultaneously, but are unlikely to play a major role in enabling disk accretion.

Observationally, the synergy between ALMA and JWST provides a comprehensive, multi-wavelength view of protostellar jet and outflow systems. ALMA excels at tracing low-temperature molecular lines with high spatial and spectral resolution, revealing velocity gradients across outflows that are often interpreted as signatures of jet rotation or magnetically driven disk winds \citep{cfLee2017, cfLee2018, deValon2020, Lopez-Vazquez2024}. Complementing this, JWST’s NIRSpec/IFU and MIRI/MRS instruments probe hot gas and ionic emission, uncovering the fine structure of jets and outflows in unprecedented detail \citep{Ray2023, Harsono2023, Narang2024, Nisini2024, Assani2024, CarattioGaratti2024, Narang2025, Navarro2025}. With both facilities offering sub-arcsecond spatial resolution, their combined capabilities enable a deeper understanding of the physical conditions, dynamics, and morphology of the protostellar outflow environment \citep{Delabrosse2024, Tychoniec2024, Okoda2025, vanDishoeck2025}.

To leverage this synergy, we conducted both ALMA and JWST observations of EC 53 (V371 Ser), a quasi-periodically variable Class I protostar \citep{Hodapp2012, YLee2020} located in the Serpens Main star-forming region \citep[d=436\,pc;][]{Ortiz-Leon2017}. EC 53 is known to drive a well-developed bipolar outflow oriented northwest (redshifted) to southeast (blueshifted), and harbors a compact disk with a radius of approximately 100\,au. The southeastern side of the disk is oriented toward the observer and is seen through the blueshifted outflow cavity, along which scattered near-IR light has been observed \citep{Hodapp2012}. Previous ALMA observations suggest a minimum central mass of $\sim$0.3$\pm$0.1 \msun\ and a disk inclination of 34\degree.8$\pm$2\degree.1 \citep[$0^\circ$ is face-on;][]{shlee2020}.

In this paper, we present the results of our coordinated ALMA and JWST observations aimed at characterizing the jet and outflow structures in EC 53 through a range of diagnostic tracers. This paper is the third in a series of papers on the JWST program, {\it EPISODE: EC 53, the only known Periodically variable Infant Star to chase the Outburst in the next Dynamical Event}. Paper I \citep{Lee2026} highlights silicate crystallization in the inner disk. Paper II (Seokho Lee et al., submitted) analyzes the variability of gas absorption lines. Paper IV (Kim et al., submitted) focuses on the ice absorption in the envelope. \par

Section~\ref{sec:observation} details the observational setup and data reduction procedures. In Section~\ref{sec:results}, we describe the morphological components of the jet and outflow system. Section~\ref{sec:analysis} presents the derived physical properties, and Section~\ref{sec:discussion} explores their implications for the origin and launching mechanisms of the observed outflows. We conclude with a summary of our main findings in Section~\ref{sec:summary}.

\begin{deluxetable*}{ccccc}
    \tabletypesize{\scriptsize}
    \tablecaption{Description of the ALMA data used in this work. \label{tb:ALMA}}
    \tablehead{
    \colhead{Type} & \colhead{Spectral Resolution} &  \colhead{Beam size}& \colhead{$\sigma$ [mJy\,beam$^{-1}$]} & \colhead{Program} }
    \startdata
 	CO J=2--1 & 61\,kHz & 0.90\arcsec $\times$ 0.69\arcsec & 18.9 & 2019.1.01792.S \\
    C$_2$H N=4--3 & 244\,kHz & 0.31\arcsec $\times$ 0.30\arcsec & 2.87 & 2016.1.01304.T\\
    Continuum & - & 0.018\arcsec $\times$ 0.015\arcsec & 0.043 & 2022.1.00800.S\\
    CO J=2--1 & 122\,kHz & 0.12\arcsec $\times$ 0.10\arcsec & 1.68 & 2022.1.00800.S\\
    CO J=2--1\tablenotemark{a} & 384\,kHz & 0.31\arcsec $\times$ 0.31\arcsec & 1.97 & 2022.1.00800.S
    \enddata
    \tablenotetext{a}{UV Tapered.}
\end{deluxetable*}

\section{Observation \& Data Reduction} \label{sec:observation}

\subsection{ALMA}
To trace the outflow and its cavity walls, we used molecular line observations from ALMA. The CO 2--1 line was used to probe both large- and small-scale outflows, combining data from Cycle 7 with a beam size of $\sim$1\arcsec\ \citep[Project ID: 2019.1.01792.S, PI: Diego Mardones;][]{Hsieh2024} and Cycle 9 with a beam size of $\sim$0.1\arcsec\ (Project ID: 2022.1.00800.S, PI: Seokho Lee). To trace the outflow cavity walls, we used Cycle 4 data with a beam size of 0.3\arcsec\ \citep[Project ID: 2016.1.01304.T, PI: Jeong-Eun Lee;][]{shlee2020}. For further details on the Cycle 4 and Cycle 7 observations, we refer the reader to \citet{shlee2020} and \citet{Hsieh2024}, respectively. Table~\ref{tb:ALMA} summarizes the ALMA observations used in this study.

The Cycle 9 line and continuum observations were conducted on 2023 May 24 and 2023 September 1, respectively. The observations covered baselines ranging from 27.5 m to 3.6 km with 43 12-m antennas for the former, and from 83.1 m to 15.2 km with 41 12-m antennas for the latter. The on-source integration times were 43.1 minutes and 46.1 minutes, respectively. The data were initially calibrated using the CASA 6.1.1.15 pipeline \citep{McMullin2007}. The quasar J1924–2914 was used as the bandpass and amplitude calibrator, and the nearby quasar J1851+0035 served as the phase calibrator.
To improve image quality of the molecular line data, the continuum image was generated using line-free channels, followed by three rounds of phase self-calibration with solution intervals of 300\,s, 150\,s, and 56.5\,s (`int'), respectively. A final round of amplitude self-calibration was also performed. The CO 2--1 image was cleaned using natural weighting, with and without a UV tapering of 0.2\arcsec. The continuum image was cleaned using Briggs weighting with a robust parameter of 0.5, yielding a beam size of 0.02\arcsec.

\subsection{JWST}

The NIRSpec IFU and MIRI MRS observations of EC 53 during the burst phase were obtained as part of the General Observer (GO) program 3477 (PI: Jeong-Eun Lee) on May 10, 2024. The NIRSpec IFU observations cover a wavelength range from 1.7 to 5.3\,$\mu$m with a resolving power of 2700, while the MIRI MRS observations cover a wavelength range from 4.9 to 27.9\,$\mu$m (R=1500--3500). These observations enable the investigation of the jet/outflow using various atomic and molecular lines. A detailed description of the observations and data reduction procedures is provided in \citet{Lee2026} for MIRI and Seokho Lee et al. (submitted; Paper II) for NIRSpec. \par
Supplementary JWST NIRCam images covering EC 53 were obtained from the Cycle 1 GO program 1611 (PI: Klaus M. Pontoppidan). \cite{Green2024} presents four images taken with different medium-band filters: F140M, F210M, F360M, and F480M. Each filter covers various emission lines known to trace protostellar outflows. The F480M filter includes H2 0--0 S(9) at 4.6947\,$\mu$m, [Fe II] at 4.8891\,$\mu$m, and the CO fundamental line forest. The F360M filter contains high energy transitions of H2 0--0 lines, while the F210M filter covers the H2 1--0 S(1) 2.1218\,$\mu$m line. The F140M filter is primarily dominated by scattered light. For detailed observation and preprocessing strategy of the archival NIRCam data, we refer to \cite{Green2024}.

\begin{figure*}
    \centering
    \includegraphics[width=\linewidth]{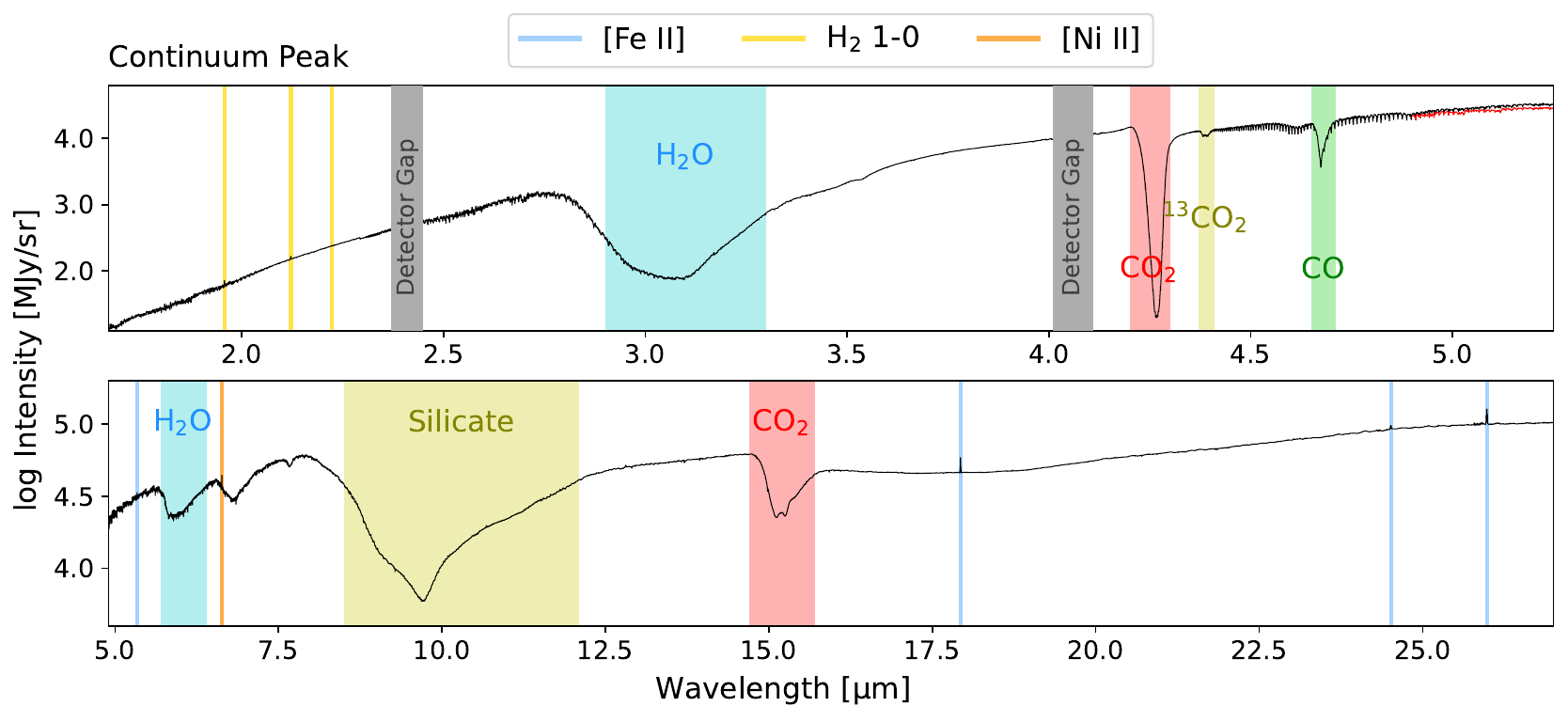}
    \includegraphics[width=\linewidth]{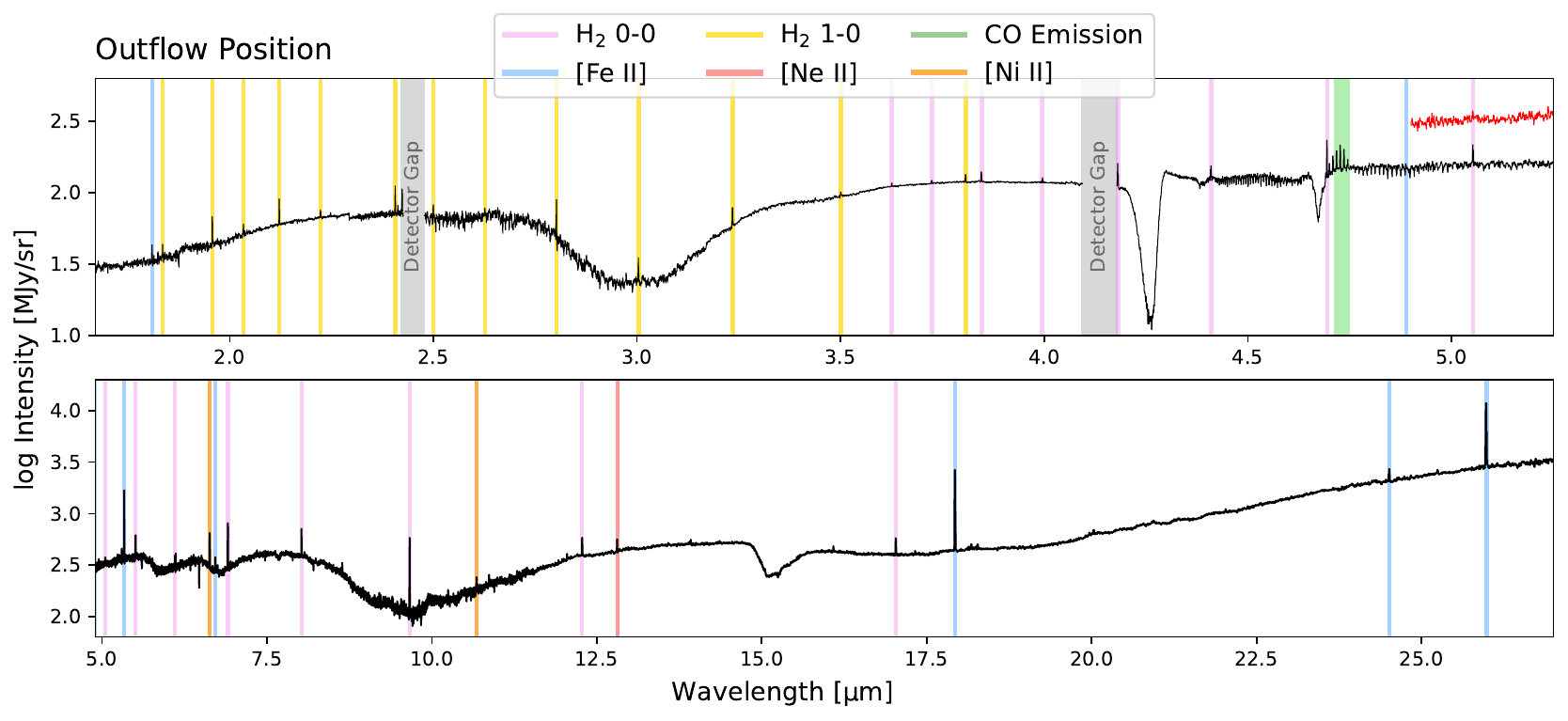}
    \caption{Near through mid-IR spectra toward EC 53 extracted from two different positions using a 0.4\arcsec\ aperture. The top two panels show the spectrum obtained at the ALMA continuum peak (18:29:51.175, +01:16:40.324), with the NIRSpec spectrum on top and the MIRI spectrum below. The bottom two panels show the same for an outflow position (18:29:51.26, +01:16:38.80), which corresponds to the mid-IR CO emission peak. Detector gaps, major ice absorption features, and emission lines are highlighted in various colors. In both NIRSpec panels, the overlapping portion of the MIRI spectrum is overplotted in red for comparison.}
    \label{fig:spectrum_center_outflow}
\end{figure*}

\subsubsection{Spectral Extraction and Flux Calibration} \label{sec:fluxcal}

While the basic data reduction and astrometric correction methods are described in Papers I and II, here we outline the additional steps used to extract spectra from individual spaxels. First, we applied a custom hot/cold pixel reduction method to remove residual hot and cold pixels that remained after processing with the JWST Calibration Pipeline. Further details on this method are provided in Appendix~\ref{sec:hotpixel}.

For comparison, we extracted representative 1D spectra at the position of the central source and at a selected outflow position coincident with the mid-IR CO emission peak, using an aperture radius of 0.4\arcsec. The resulting spectra are shown in Figure~\ref{fig:spectrum_center_outflow}; the top two panels correspond to the central source, and the bottom two panels to the outflow position. There are small discrepancies in flux levels between different spectral bands --- this is a known issue in JWST data, previously reported in the literature \citep[e.g.,][]{Yang2022, Gelder2024, Nisini2024}. To correct for this, we scaled each band to match the flux level of MIRI channel 1. However, the mismatch between the shortest wavelength in MIRI and the longest in NIRSpec is much larger than in other subbands, and the scale factor varies across positions, ranging from approximately 0.8 to 6. We do not apply scale factors between NIRSpec and MIRI, but perform analysis separately for each instrument. Additional details about the flux discrepancies are provided in Appendix~\ref{sec:app_fluxcal}. \par

All spectra used in this study were further corrected for residual fringe patterns using the \lstinline{fit_residual_fringes_1d} routine, a post-pipeline 1D fringe correction method. Near the central source, absorption features from simple ices and silicates are clearly visible. In the outflow region, absorption features persist, but numerous emission lines from various atomic and molecular species are also present.

\begin{figure*}
    \centering
    \includegraphics[width=\linewidth]{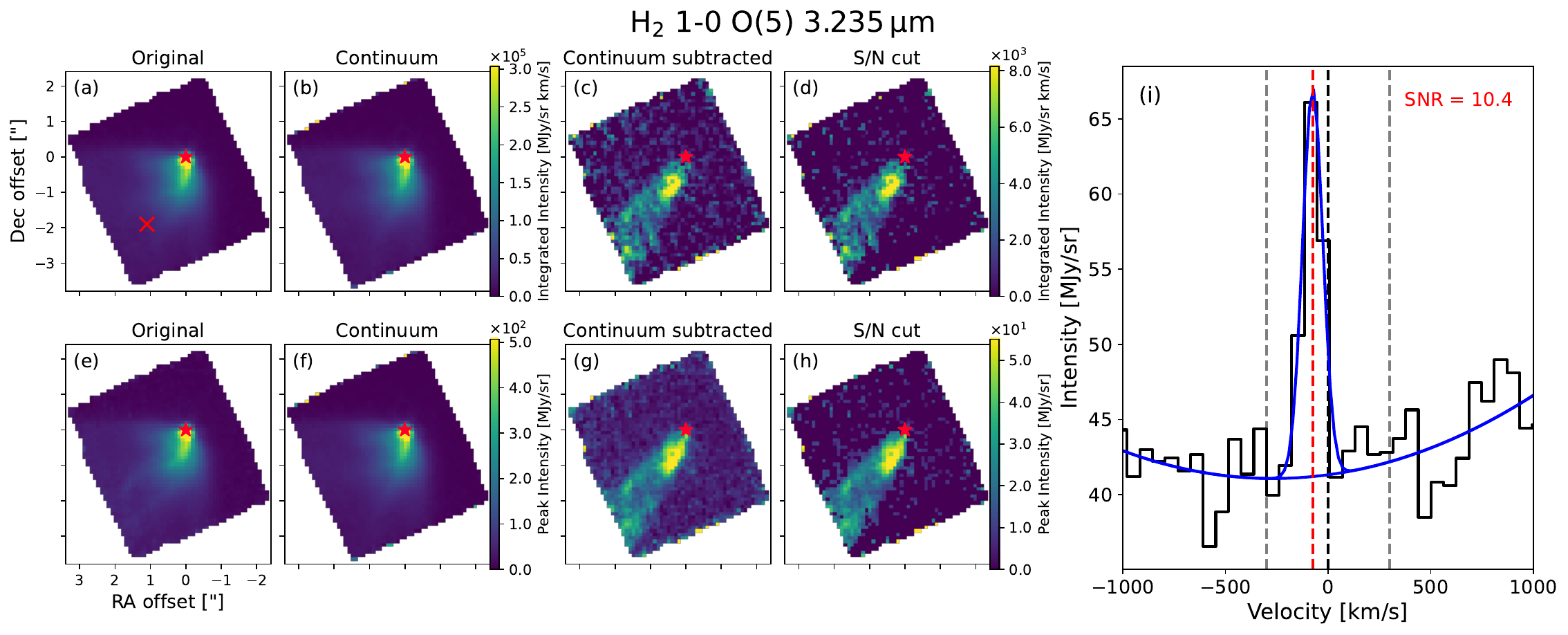}
    \caption{Reduction process of the H$_2$ 3.235\,$\mu$m line emission. (a–d) Integrated intensity maps at each stage of the reduction: (a) after hot pixel removal and flux calibration; (b) quadratic baseline fitting result; (c) after continuum subtraction; and (d) after applying a signal-to-noise (S/N) cut of 3. The image center is shifted southeastward to better display the outflow structure. (e–h) Corresponding peak intensity maps for each stage, respectively. The ALMA continuum center is marked with a red star. (i) Spectrum at the location marked with a red cross in panel (a). Gray vertical dashed lines indicate velocities of $-300$\,km\,s$^{-1}$ and $+300$\,km\,s$^{-1}$, used to define the emission channels. The blue line shows the fitted continuum and Gaussian profile. The red vertical dashed line indicates the Gaussian centroid, which is blueshifted relative to the barycentric velocity of the source ($-8.9$\,km\,s$^{-1}$, thick black dashed line).}
    \label{fig:mmap_simp}
\end{figure*}

\subsubsection{Reduction of Emission Lines}

We adopted a method similar to that of \cite{Assani2024} to recover the line emission distribution. For each spaxel, we defined the channels with a velocity between $-300$\,km\,s$^{-1}$ and $300$\,km\,s$^{-1}$ as the `emission channels', and channels between $-1000$\,km\,s$^{-1}$ and $1000$\,km \,s$^{-1}$, excluding the emission channels, as the `continuum channels'. A quadratic function was fitted to the continuum channels to estimate the continuum level at each spaxel. We used the median value as the constant continuum level for spaxels for which a valid quadratic fit could not be obtained. \par
Using the continuum fit at each spaxel, we generated a continuum cube and a continuum-subtracted line cube for each emission line. From the line cube, we produced integrated-intensity and peak-intensity maps. We masked out spaxels with a signal-to-noise ratio (SNR) below 3 and manually excluded spaxels with unphysically high intensities. The noise level for each spaxel was estimated with the root-mean-square (RMS) of the residuals in the continuum channels, and SNR was then calculated as the ratio of the peak intensity to the noise. Figure \ref{fig:mmap_simp} shows the full process for generating the H$_2$ 3.235\,$\mu$m emission map.

\section{Results} \label{sec:results}

\begin{figure}
    \centering
    \includegraphics[width=\linewidth]{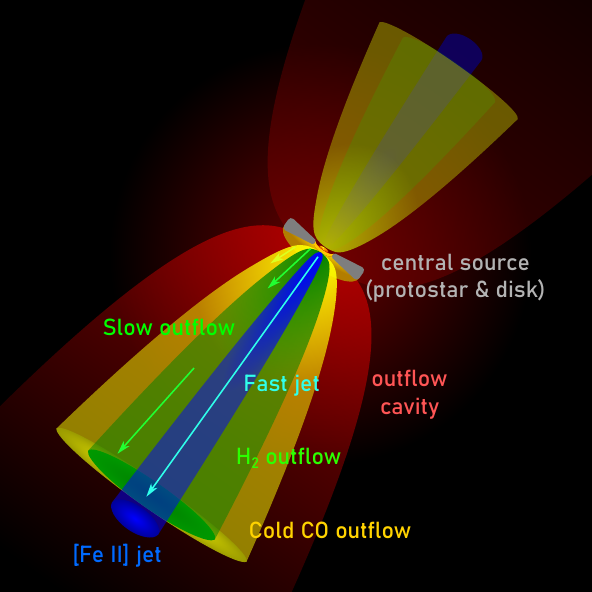}
    \caption{Schematic view of EC 53, highlighting each component analyzed in this work.}
    \label{fig:EC53_schematic}
\end{figure}

\begin{figure*}[ht]
    \centering
    \includegraphics[width=\linewidth]{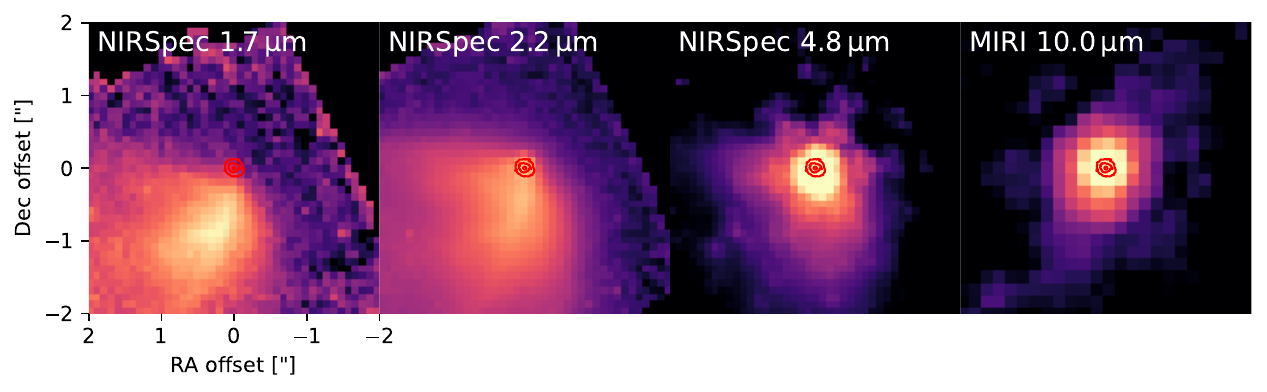}
    \includegraphics[width=\linewidth]{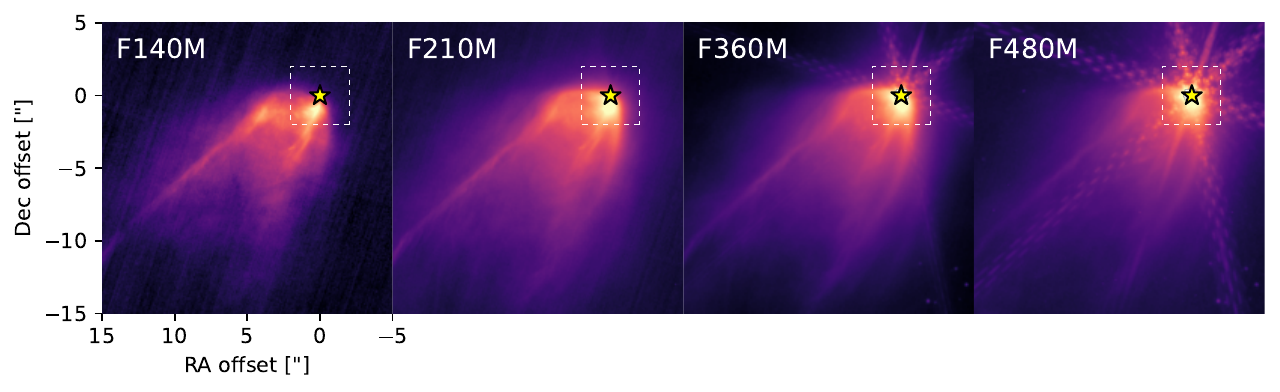}
    \caption{(Top) Continuum of EC 53 observed with JWST NIRSpec and MIRI at different wavelengths. The ALMA continuum is overlaid as red contours. (Bottom) Scattered light around EC 53 imaged by JWST NIRCam in different filters. A white dashed square indicates the area shown in the top panels. The position of the central source is marked with a yellow star.  The ALMA continuum peak coincides with the peak positions in all images at wavelengths longer than 2.1\,$\mu$m. The colorscale is logarithmic across all panels to accentuate faint details.}
    \label{fig:cont_spec}
\end{figure*}

In this section, we describe five distinct emission structures observed in EC 53: the envelope cavity and central source detected in the continuum, wide-angle H$_2$ outflow, collimated ionic jet, hot mid-IR CO emission, and ALMA outflow tracers. Figure \ref{fig:EC53_schematic} shows a schematic view of the jet and outflow system for EC 53.

\subsection{Envelope Cavity and Central Source detected in Continuum} \label{sec:continuum}
The scattered light from the central source appears as continuum emission within the outflow cavity at short wavelengths. However, since the dust scattering cross-section decreases with increasing wavelength, the continuum morphology varies significantly across wavelengths. \par

The top panels of Figure \ref{fig:cont_spec} show the continuum emission at different wavelengths as observed with NIRSpec and MIRI. In the image of the shortest wavelength (1.7\,$\mu$m), the most prominent feature is the wide outflow cavity. Only the southeast cavity, which corresponds to the blue-shifted lobe facing the observer, is visible. No strong emission is detected near the central source. At 2.2\,$\mu$m, the outflow cavity remains visible, but the brightest point of emission shifts northward toward the central source. In contrast, at the longer wavelengths of 4.8\,$\mu$m and 10\,$\mu$m, outflow structures are no longer visible, as scattered light becomes negligible. Additionally, the centrally peaked emission is unresolved at these wavelengths because the PSF full width at half maximum (FWHM) exceeds the source size. \par

This wavelength-dependent trend is also seen in the photometric images. The bottom panels of Figure \ref{fig:cont_spec} show JWST NIRCam images, which cover a larger field of view than a single NIRSpec cube slice (outlined in white dashed lines). The outflow cavity extends farther from the central source in these images. Comparing the NIRCam images, it is clear that the outflow and outflow cavity features dominate at shorter wavelengths, whereas the emission from the central source is prominent at longer wavelengths. In the F360M and F480M images, diffraction patterns from the central source are visible. Although scattered light from the cavity remains, it is significantly dominated by emission from the central source. \par

High angular resolution ALMA observations resolve the central source with a beam size of $\sim$0.02\arcsec. The ALMA continuum shown in  Figure \ref{fig:EC53_ALMA_continuum} and the top panels of Figure \ref{fig:cont_spec} trace the central protostar and its protoplanetary disk. We fitted the disk with a two-dimensional Gaussian using the \lstinline{imfit} task in CASA. The central position of the continuum is at $\alpha = 18^{\rm h}29^{\rm m}51.17537^{\rm s}$, $\delta = 1^{\rm d}16^{\rm m}40.32810^{\rm s}$ (J2000). After deconvolving the beam, the fitted disk has a semimajor axis of 142.5$\pm$2.5\,mas (62.1$\pm$1.1\,au), a semiminor axis of 121.0$\pm$2.1\,mas (52.8$\pm$0.9\,au), and a position angle of 60.6$\pm$4.4$^\circ$. Assuming a circular disk, the resulting inclination derived from the axis ratio is 31.9$\pm$3.2$^\circ$. These values are consistent with previous estimates from lower-resolution ALMA data \citep{shlee2020}. The MIRI emission peak coincides with the ALMA continuum. \par

\subsection{Wide Angle Outflow traced by H$_2$}
A total of 27 H$_2$ emission lines are detected with S/N>3 from our observations. Table \ref{tb:h2lines} lists all the lines detected. The H$_2$ line information is compiled from the HITRAN database \citep{Gordon2026}.   
Both vibrational transitions (v=0--0 and 1--0) are detected from the O, Q, and S branches, with upper energy levels reaching as high as 20,000\,K. No v=2--1 lines or HD lines are detected with S/N>3. \par 
The H$_2$ emission shows a cone-like morphology, narrower than the outflow cavity traced by scattered light, as shown in Figure \ref{fig:mmap_simp}. 
Figures \ref{fig:H2_nirspec_total_maps} and \ref{fig:H2_miri_total_maps} present the integrated intensity maps of each H$_2$ line obtained with NIRSpec and MIRI, respectively. 
Further discussion on the morphology and physical parameters of the H$_2$ outflow is discussed in Sections \ref{sec:H2_morph} and \ref{sec:H2_diagram}. \par 

\begin{figure*}
    \centering
    \includegraphics[width=\linewidth]{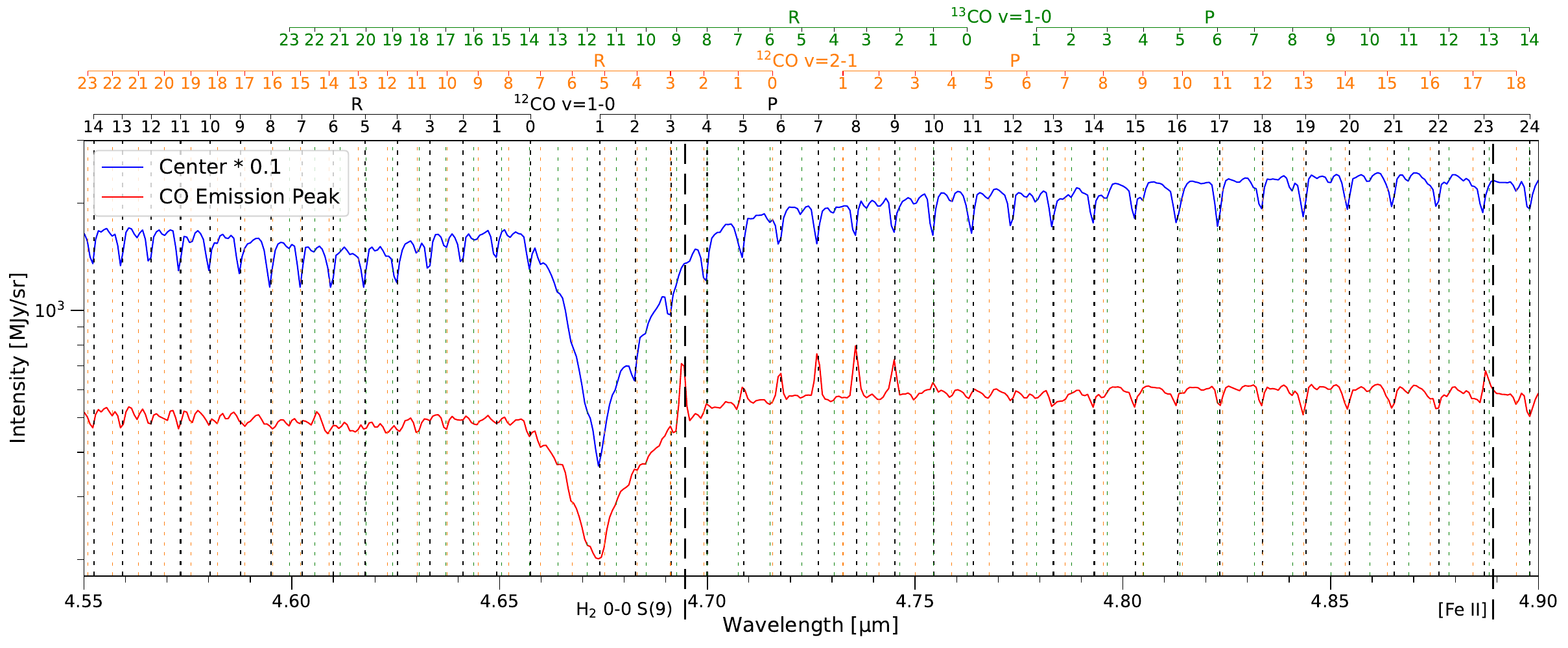}
    \caption{Spectra between 4.55\,$\mu$m and 4.90\,$\mu$m for the central source (blue) and CO emission peak (red). Both spectra are extracted using a 0.3\arcsec\ aperture. The wavelengths of the $^{12}$CO v=1--0, v=2--1, and $^{13}$CO v=1--0 transitions are marked with black, green, and orange dotted lines, respectively. Two transitions from other species, H$_2$ 0--0 S(9) and [Fe II], are indicated with black dashed lines. The line positions are shown in the rest frame without applying velocity shift corrections.}
    \label{fig:CO_spec}
\end{figure*}

\begin{figure}
    \centering
    \includegraphics[width=\linewidth]{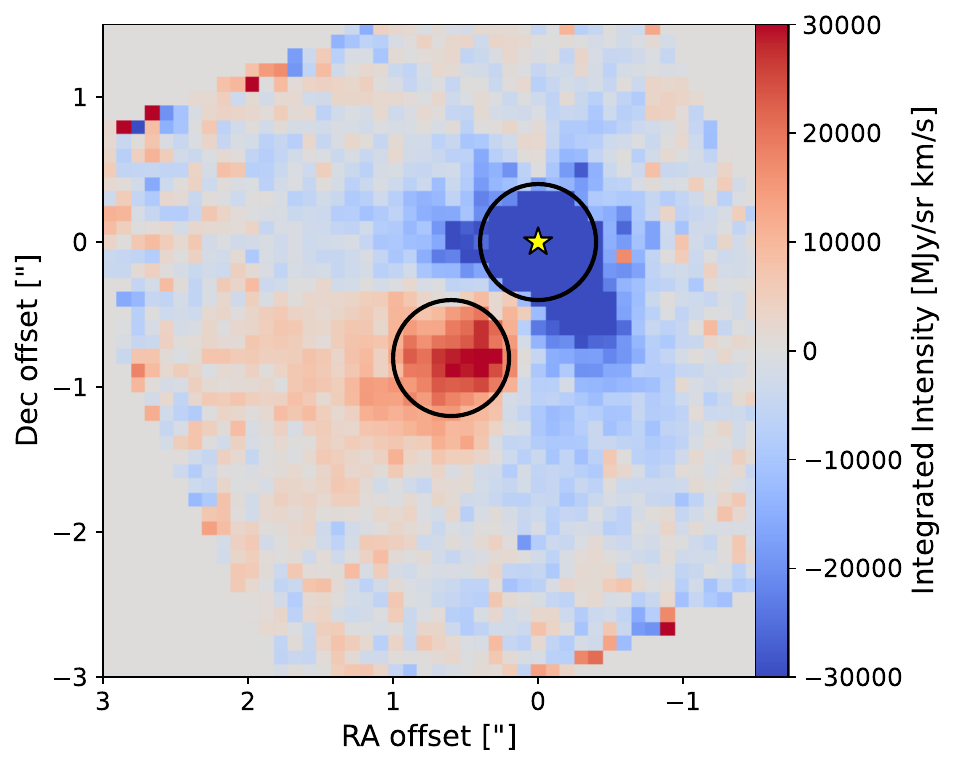}
    \caption{Integrated intensity map of the $^{12}$CO v=1--0 P(8) line. The central region appears in absorption while the surrounding regions exhibit emission. The apertures used for extracting the spectra of Figures \ref{fig:spectrum_center_outflow} and \ref{fig:CO_spec} are marked with black circles. The ALMA center position is marked with a yellow star.}
    \label{fig:CO_mmap}
\end{figure}

\subsection{Ionic Jet}

Several forbidden atomic lines, including [Fe II], [Ne II], and [Ni II], are detected. No hydrogen recombination, helium, or argon lines are detected. Table \ref{tb:atomlines} lists all the atomic lines detected in our observations. Line information is from the \href{https://physics.nist.gov/PhysRefData/ASD/lines_form.html}{NIST Atomic Spectra Database} \citep{NIST2023}. \par
Figure \ref{fig:Atomic_total_maps} shows the continuum-subtracted integrated intensity maps of the detected atomic emission lines. All atomic lines exhibit a similar collimated structure aligned along the blueshifted side of the outflow, regardless of the species. Two [Fe II] lines (17.94 and 25.99\,$\mu$m) also show redshifted emission on the northeast side. 

\subsection{Hot CO Emission and Absorption}
The spectral resolution of NIRSpec and MIRI is sufficient to resolve individual lines within the CO forest around 4.7\,$\mu$m (see Figure \ref{fig:CO_spec}). Numerous lines from various vibrational bands ($^{12}$CO v=1--0, v=2--1, $^{13}$CO v=1--0) are detected, alongside CO ice absorption and H$_2$ and [Fe II] emission lines. The spectra extracted from the central source position show that all CO lines are in absorption. A detailed analysis of these absorption features will be presented in a separate study (Seokho Lee et al., submitted). On the contrary, in the spectra from the outflow, the $^{12}$CO v=1--0 P-branch transitions from J=4 to J=11 appear in emission, while the R-branch lines and other CO bands remain in absorption. Figure \ref{fig:CO_mmap} shows the integrated intensity map of the $^{12}$CO v=1--0 P(8) line at 4.7359\,$\mu$m. Near the central source, the CO line shows in absorption. However, a distinct area southeast of the central source exhibits emission.  The origin of this emission and P-R asymmetry is discussed further in Section \ref{sec:CO_origin}. \par

\begin{figure*}
    \centering
    \includegraphics[width=\linewidth]{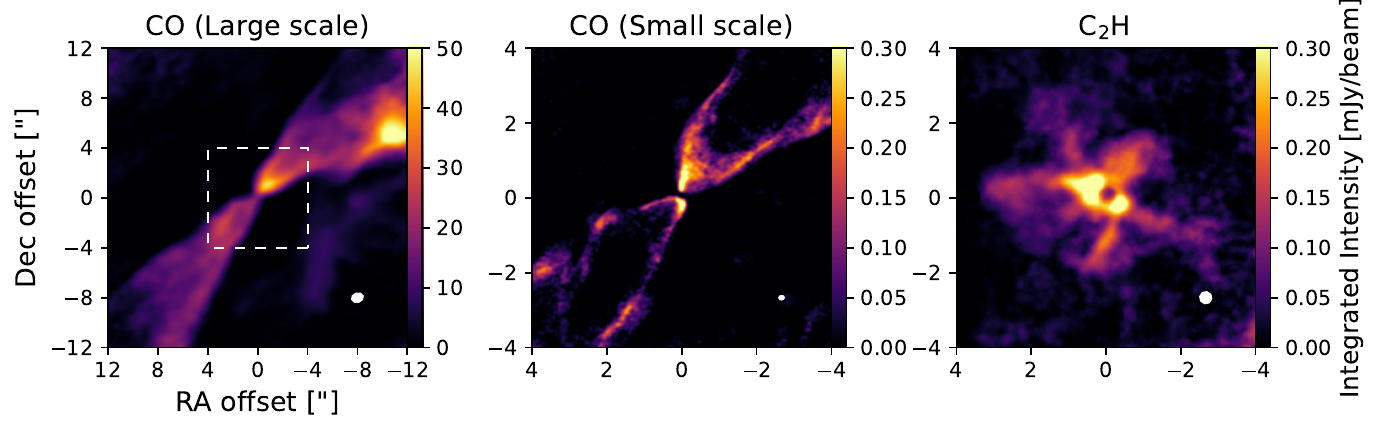}
    \caption{Integrated intensity maps of CO and C$_2$H from ALMA. A white ellipse indicates the beam size for each map. The leftmost panel covers a larger field of view (24\arcsec$\times$24\arcsec), while the other panels show zoomed-in regions (8\arcsec$\times$8\arcsec). The white dashed box in the leftmost panel outlines the area shown in the smaller panels.}
    \label{fig:ALMA_mom0}
\end{figure*}

\subsection{ALMA Outflow Tracers}
Figure \ref{fig:ALMA_mom0} presents the integrated intensity maps of CO and C$_2$H in EC 53. The CO outflow appears as two lobes extending widely (>10\arcsec) on both sides of the central protostar. An additional extended emission is present in the same region as the redshifted lobe. This emission is also detected in the NIRCam images and has been identified as associated with S68Nb2 \citep{Green2024}. Its blueshifted velocity further supports the interpretation that it is unrelated to the EC 53 outflow. \par
Due to the high spatial resolution of the CO observations (beam size $\sim$0.1\arcsec), large-scale structures (>1\arcsec) are partially resolved out. Nevertheless, the overall morphology of the outflow is consistent with that seen in lower-resolution data. As reported in Figure 3 of \citet{shlee2020}, C$_2$H traces the outflow cavity walls, which are broader than the CO outflow. C$_2$H can be enhanced by UV photons \citep{Fuente1993}, while it can be destroyed by gas phase reactions with H$_2$ in the warm envelope \citep{Aikawa2012}. This explains why C$_2$H is typically observed along cavity walls \citep{Tychoniec2021}. The spatial correlation between the C$_2$H emission and the near-IR scattered light supports the interpretation that the scattered light traces the outflow cavity structure.

\section{Analysis} \label{sec:analysis}

\subsection{H$_2$ Emission Morphology} \label{sec:H2_morph}

\begin{figure}
    \centering
    \includegraphics[width=0.95\linewidth]{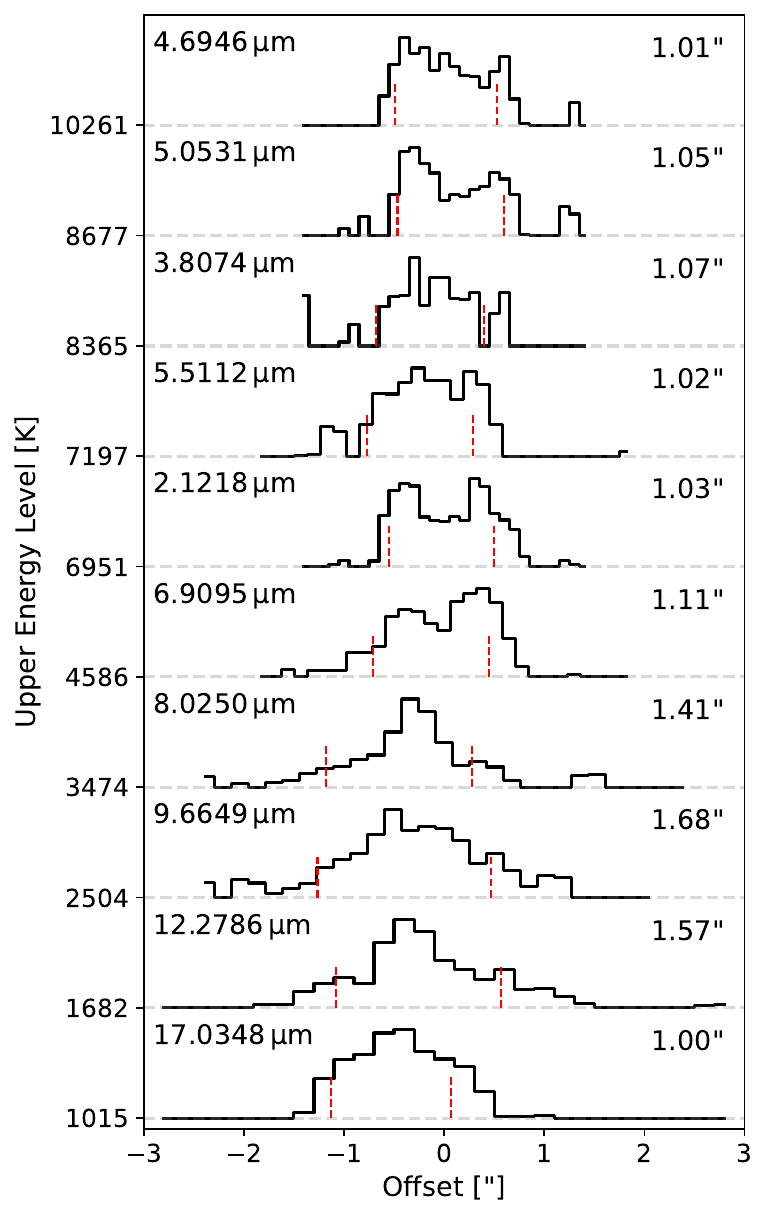}
    \caption{Normalized transverse cut profiles across the blueshifted outflow of EC 53 at 2\arcsec\ distance from the central source for each of the H$_2$ transitions. For each transition, the observed FWHM is marked with red dashed lines, and the PSF deconvolved FWHM is written in the upper right corner.}
    \label{fig:H2_width}
\end{figure}

Figure \ref{fig:H2_width} shows transverse cut profiles for various H$_2$ transitions. Since the H$_2$ transitions are scattered over a large range of wavelengths from ro-vibrational lines at $\sim$2\,$\mu$m to v=0--0 S(1) at 17.035\,$\mu$m, it is critical to account for the wavelength dependent PSF before comparing the line widths. The PSF FWHM as a function of wavelength for MIRI is calculated from \cite{Law2023}:
\begin{equation}
    \theta_\lambda = 0.033 (\lambda / {\rm micron}) + 0.106\arcsec. \label{eq:FWHM}
\end{equation}
The PSF FWHM of NIRSpec IFU is not well constrained and differs by dither patterns. In this work, we use a conservative constant value of 0.15\arcsec \citep{dEugenio2024}. We deconvolve the observed line width using the following equation:
\begin{equation}
    \theta_{\rm Real} = \sqrt{\theta_{\rm Apparent}^2 - \theta_{\rm PSF}^2}. \label{eq:deconv}
\end{equation}
After deconvolving the PSF from the observed line profile, there is a trend between excitation energy and spatial extent for the majority of the lines. Excluding the v=0--0 S(1) line with the lowest energy, there is a decreasing trend in line width with increasing upper energy. The lower energy lines ($E_u<5000$\,K) show a decrease in width from $\sim 1.7$\arcsec\ to $\sim 1.1$\arcsec, while the higher energy lines have a similar width of $\sim1.1$\arcsec. This suggests a nested structure, where more highly excited gas is more centrally concentrated.

\subsection{H$_2$ Emission Excitation Diagram Analysis} \label{sec:H2_diagram}

\begin{figure*}[t]
    \centering

    \includegraphics[width=0.49\linewidth]{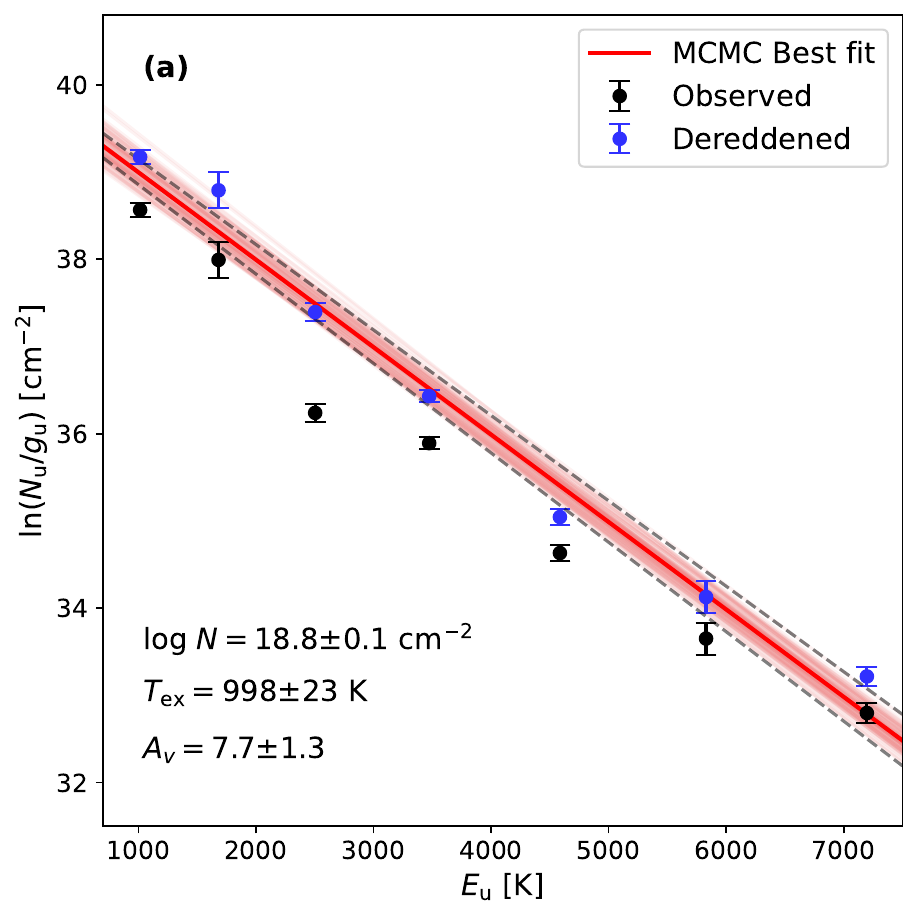}
    \includegraphics[width=0.49\linewidth]{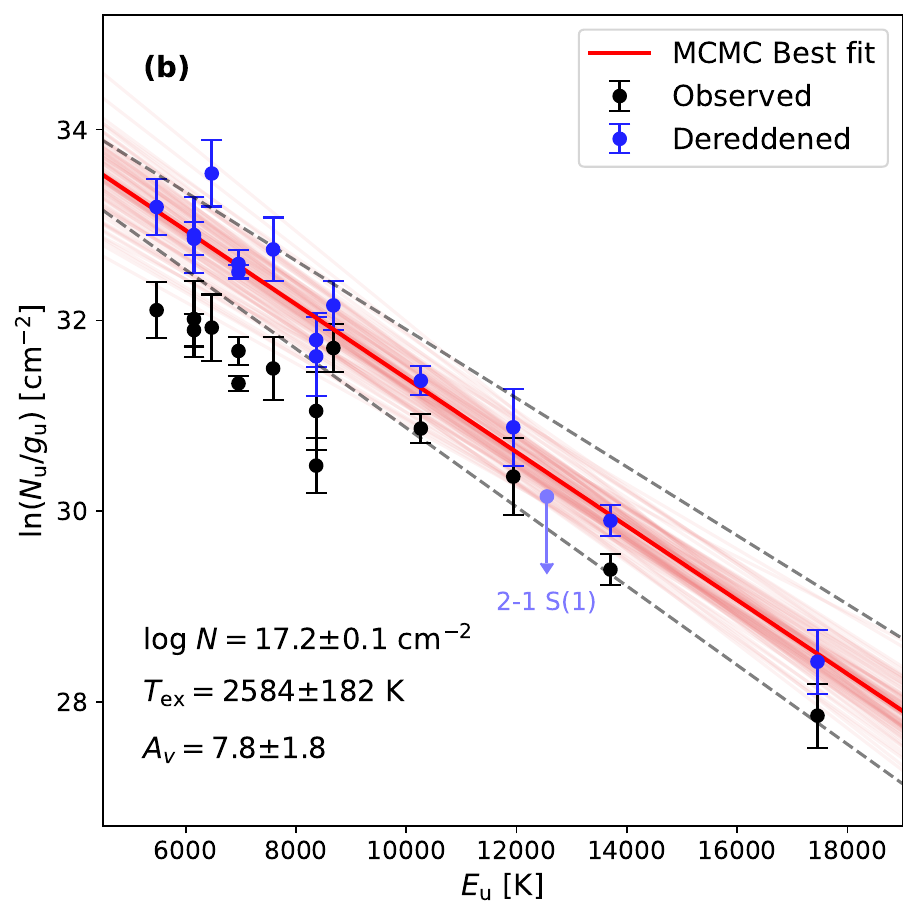}
    \includegraphics[width=\linewidth]{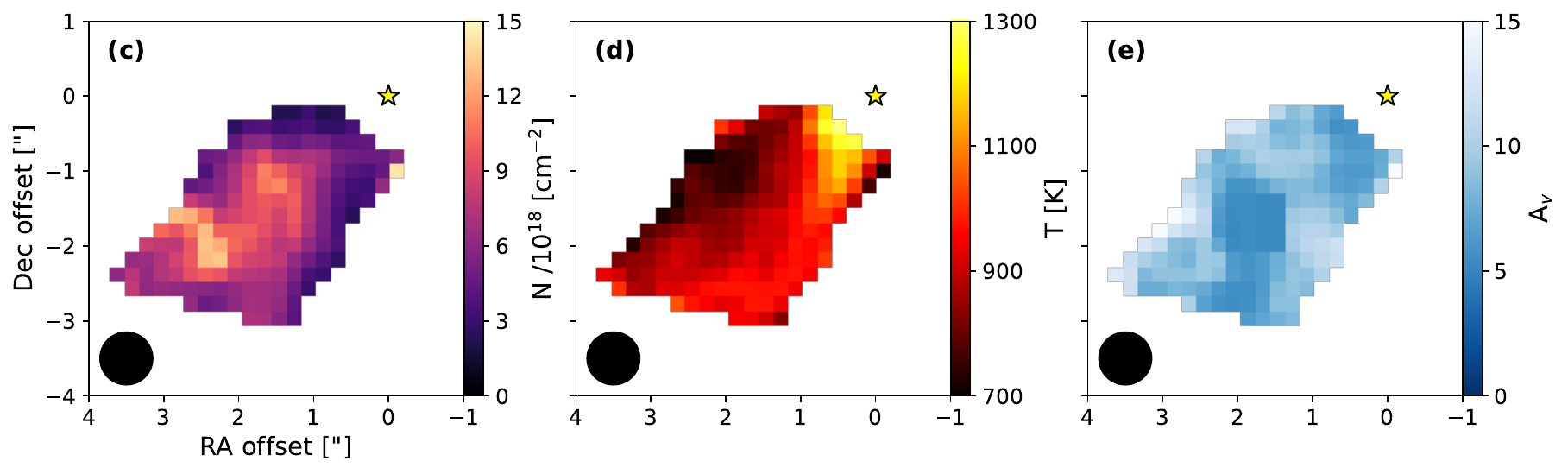}
    \includegraphics[width=\linewidth]{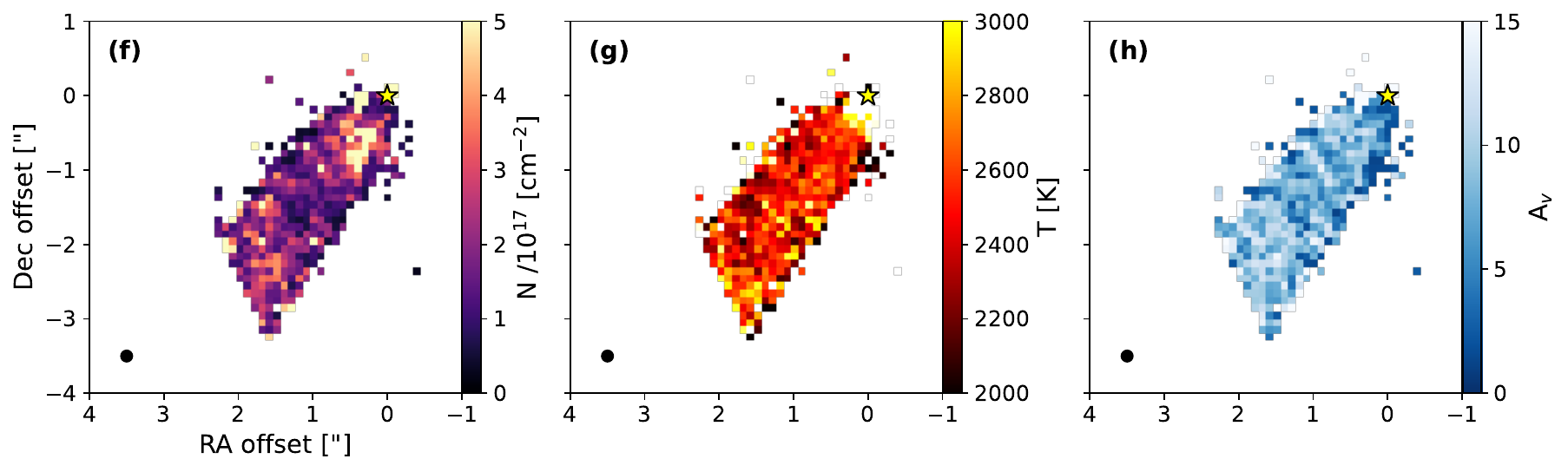}
    \caption{(a) MIRI H$_2$ excitation diagram result from a representative position of the outflow. The observed transitions are marked with black points, and blue points represent the measurements after extinction correction. The red line is the best-fit model, and the fainter lines are the posterior samples. Gray dashed lines indicate the models corresponding to the 16th and 84th percentiles. (b) Same with (a), but for NIRSpec transitions. The 3$\sigma$ limit for the undetected 2--1 S(1) transition is marked in lighter blue. (c, d, e) 2D distributions of H$_2$ column density, temperature, and extinction from the MIRI excitation diagram analysis. The position of the central source is marked with a yellow star. The black circle indicates the size of the convolved PSF. (f, g, h) Same as the row above, but for NIRSpec.}
    \label{fig:H2_rotation}
\end{figure*}

Since multiple emission lines from the same species are detected, we perform an excitation diagram analysis to derive the physical properties at each position of the outflow. Assuming that H$_2$ in the outflow is in local thermodynamical equilibrium (LTE) and that all lines are optically thin, the column density, excitation temperature, and extinction can be obtained using the following equations:
\begin{gather}
    N_i = \frac{4\pi}{hcA_{ij}}\int I_\nu dV,\\
    \ln\left(\frac{N_i}{g_i}\right)=\ln\left(\frac{N}{Q}\right)-\frac{E_i}{kT_{\rm ex}}.
\end{gather}
Here, $N_i$, $g_i$, and $E_i$ are the column density, degeneracy, and the energy of the upper level, respectively. $A_{ij}$ is the Einstein coefficient of the transition. $Q$ is the partition function, which depends on the excitation temperature $T_{\rm ex}$.

The observed emission is attenuated by foreground material, necessitating additional care before the analysis. While the SED fitting result of \cite{Dunham2015} estimates a visual extinction of $A_V$=9.6 mag, we treat $A_V$ as a free parameter in the excitation diagram fitting. We adopt the KP5 extinction curve, which is representative of icy dust in dense molecular clouds \citep{Pontoppidan2024}. For the KP5 extinction curve, the optical depth at the wavelength of H$_2$ S(3) line is calculated as $\tau_{\rm 9.7} = 0.1509A_V$. The impact of using different extinction curves is discussed in Appendix \ref{sec:H2_extinction}. The relation between the intrinsic intensity and the observed intensity is given by:
\begin{equation}
    I_{\rm obs} = 10^{-0.4A_\lambda}I_\nu = 10^{-0.4A_V (A_\lambda/A_V)}I_\nu.
\end{equation}

Since the angular resolution and pixel scale vary with wavelength, we convolve all the line maps to match the size of the PSF of the transition with the longest wavelength. Although the PSF is not strictly Gaussian, we assume the observed images as convolutions of the true emission with a Gaussian kernel with FWHM from Eq. \ref{eq:FWHM}. To match the PSF of the image from the shorter wavelength $\lambda$ to the PSF of representative wavelength $\lambda_0$, we apply a Gaussian convolution to the shorter wavelength map with a FWHM calculated as follows:
\begin{equation}
    \rm{FWHM} = \sqrt{\theta_{\lambda_0}^2 - \theta_\lambda^2}.
\end{equation}
Finally, all line maps are re-gridded to match the spatial grid of the transition with the longest wavelength. This procedure enables pixel-to-pixel analysis of excitation diagrams and yields spatially resolved maps of excitation temperature, column density, and extinction. The transitions within MIRI wavelengths and NIRSpec wavelengths are analyzed separately, since the spatial resolution of NIRSpec is much better, and there is a large flux discrepancy between the two instruments. The PSF FWHM and pixel size are 0.7\arcsec\ and 0.3\arcsec\ for MIRI wavelengths respectively, and 0.15\arcsec\ and 0.1\arcsec\ for NIRSpec, respectively.\par

For each pixel, we used the Markov-Chain Monte Carlo (MCMC) method to simultaneously fit the excitation temperature, column density, and extinction, along with their uncertainties. For detailed information about the MCMC fitting, refer to Appendix \ref{sec:H2_MCMC}. \par
The top panels of Figure \ref{fig:H2_rotation} show examples for the excitation diagram. The black points from the original datacubes do not align well along a straight line. However, after correcting for extinction, each excitation diagram is well explained with a single line. \par
Panels (c) - (e) show the results of the excitation diagram using MIRI wavelengths. The analysis was performed only on pixels where more than 4 H$_2$ lines were detected with S/N>3. The median column density of H$_2$, excitation temperature, and extinction are $6.4\times 10^{18}$\,cm$^{-2}$, 900\,K, and 8.3, respectively. The H$_2$ outflow has a roughly constant temperature between 700--1000\,K, except near the outflow base, where the temperature is significantly higher, exceeding 1200\,K.\par

Panels (f) - (h) show similar maps, but for NIRSpec wavelengths. The emission region appears narrower than in the MIRI excitation maps, but this is partially due to the PSF being much smaller (0.15\arcsec) than that of MIRI (0.7\arcsec). The median column density of H$_2$, excitation temperature, and extinction are $1.8\times 10^{17}$\,cm$^{-2}$, 2500\,K, and 8.1, respectively. Assuming the ro-vibrational transitions are in LTE and that this excitation diagram correctly reflects the gas kinetic temperature, the NIRSpec lines trace a hotter component than the MIRI lines. The column density of the hotter component is $\sim35$ times less than the cooler component, so the cooler component represents the bulk of the outflowing material. The extinctions derived from both analyses agree well. \par

The 2--1 S(1) transition at 2.2477\,$\mu$m is not detected above the 3$\sigma$ level in any pixel. This line is often used as a diagnostic of UV excitation or strong J-type shocks, since these mechanisms are expected to increase the 2--1/1--0 intensity ratio \citep{Kaplan2021, Kristensen2023}. Figure \ref{fig:H2_rotation}(b) shows the 3$\sigma$ upper limit of the observed 2--1 S(1) line. This upper limit is consistent with the excitation diagram prediction, supporting a scenario in which the H$_2$ emission is predominantly thermally excited. If significant UV fluorescence or strong J-type shocks were present, the 2–-1 S(1) flux would be expected to exceed the excitation-diagram prediction. No such excess is observed.

\subsection{[Fe II] Jet Morphology}

\begin{figure*}[t]
    \centering
    \includegraphics[width=\linewidth]{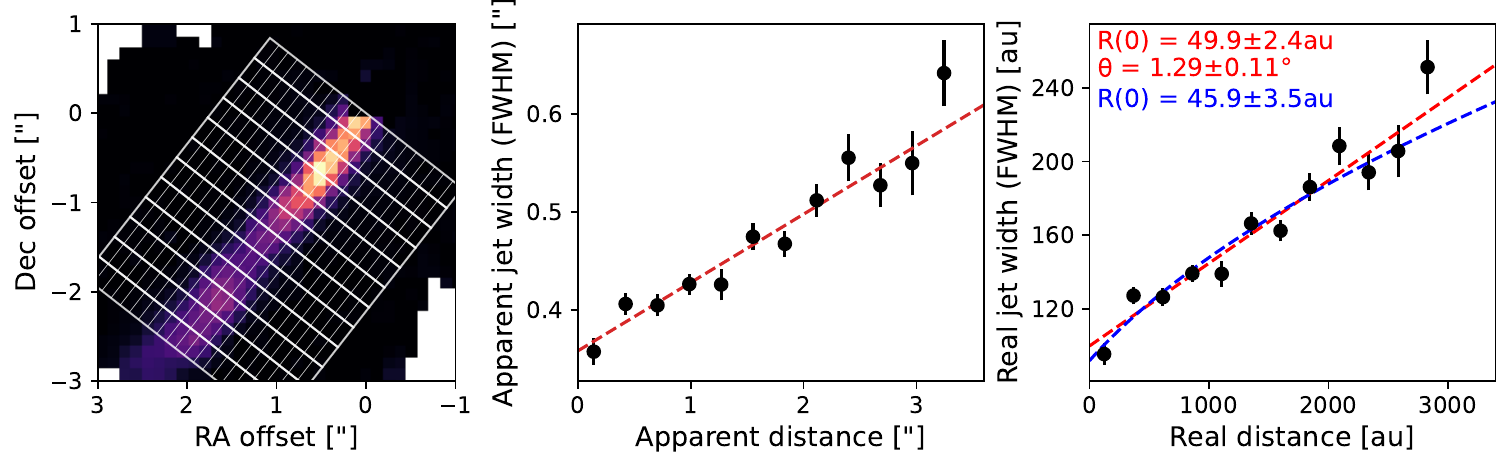}
    \caption{(Left) Apertures used for jet morphology analysis (white rectangles) overlaid on the [Fe II] 5.34\,$\mu$m peak intensity map. (Middle) Apparent FWHM of the jet as a function of distance from the central source, shown in angular units. A red dashed line indicates the best-fit linear function. (Right) Deconvolved, real FWHM of the jet as a function of distance from the central source, shown in distance units. The red- and blue-dashed lines indicate the best-fit linear function and the best-fit parabolic function, respectively. Fit parameters are written in the corresponding colors.}
    \label{fig:jet_morph}
\end{figure*}
We analyzed the [Fe II] 5.34$\,\mu$m emission line to determine the position angle (PA), opening angle, and launching radius of the atomic jet. For each horizontal row of pixels in the integrated intensity map, we measured the peak emission position and its width. A linear fit to the ALMA continuum center was used to determine the jet axis. The estimated PA is $142.2\pm0.4^\circ$. This is consistent with the PA of the Serpens filament ($139^\circ$) and the typical PA of the outflows found in Serpens Main \citep[$136\pm25^\circ$;][]{Green2024}. We estimate the jet's opening angle and launching radius, following the method described in \cite{Narang2024}. We first extracted transverse slices perpendicular to the jet axis, with a width along the jet equal to the empirical FWHM of the MIRI-MRS PSF at this wavelength ($\sim$0.28\arcsec). Each slice was further divided into 21 narrower pixel-width smaller apertures (0.13\arcsec). We extracted the spectrum from each aperture and measured the integrated line flux. The jet width in each slice was then determined by fitting a Gaussian profile to the transverse flux distribution. A linear fit to the measured jet widths as a function of distance from the source yielded the jet opening angle ($\theta$) and the base width ($R(0)$). \par

The middle panel of Figure \ref{fig:jet_morph} shows the apparent jet width increasing with distance from the central source. However, the observed jet width is broadened by the instrumental PSF, so we deconvolve the measured widths using Equation \ref{eq:deconv}. 
Assuming a distance of 436\,pc to EC 53 \citep{Ortiz-Leon2017} and a disk inclination angle of 32$^\circ$ derived at Section \ref{sec:continuum}, we calculated the intrinsic jet width and deprojected distance from the source for each slice. Finally, we derived the jet width at the base (half of the y-intercept) and the opening angle (using the slope) by using a linear fit; our linear fit to the deconvolved widths resulted in a base jet width (half of the y-intercept) of $49.9 \pm 2.4$\,au and an opening angle of $1.29 \pm 0.11^\circ$. Alternatively, a parabolic fit yields a base width of $45.9 \pm 3.5$\,au, which we adopt as an upper limit on the jet-launching radius. \par

The launching radius derived from our analysis (45.9\,au) is comparable to values reported for other low-mass protostars, such as IRAS 16253-2429 (23\,au; \citealt{Narang2024}), IRAS 15398-3359 (20.2\,au), or B335 (37.1\,au; \citealt{Federman2024}). However, these estimated jet launching radii are greater than the ones typically inferred from kinematic studies using SiO or optical tracers, which often yield values below 10\,au \citep{cfLee2020}. 
In the case of EC 53, the ALMA data do not reveal a collimated, jet-like high-velocity CO component, which limits our ability to directly constrain the jet size and its launching radius from CO kinematics alone. Moreover, it is important to note that the geometrically derived launching radius generally exceeds the MHD wind launching radius obtained through angular momentum measurements \citep{Pascucci2023} and should, therefore, be considered only as an upper limit.

\begin{figure*}
    \centering
    \includegraphics[width=0.67\linewidth]{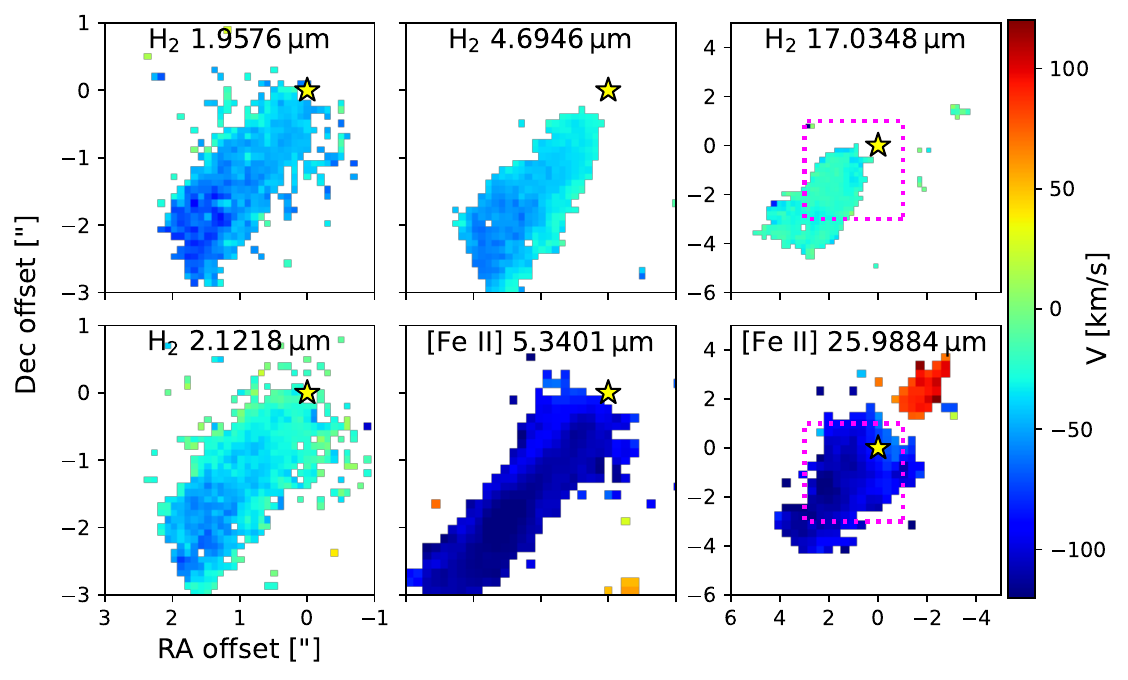}%
    \includegraphics[width=0.33\linewidth]{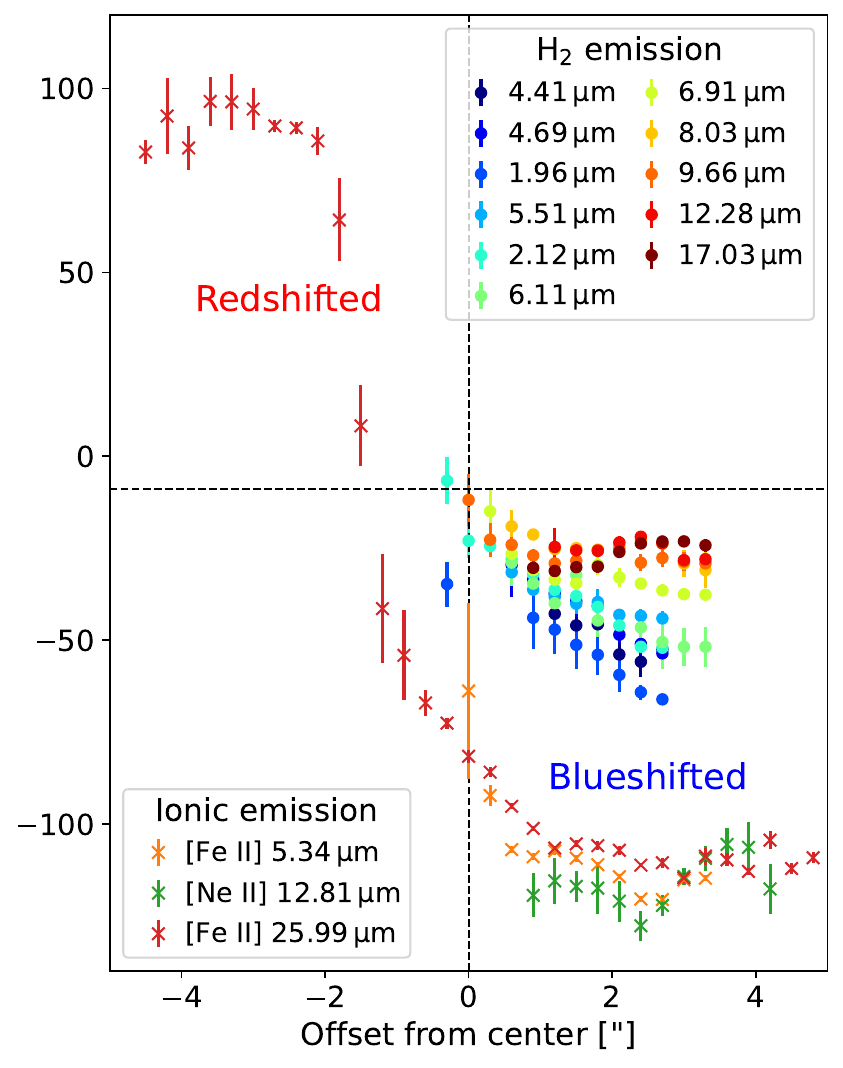}
    \caption{Velocity maps of selected lines (left). Note that the image size of the 17.03\,$\mu$m H$_2$ map and 25.99\,$\mu$m [Fe II] map is larger than the other maps of short wavelength transitions. The extent of the smaller maps is indicated by pink-dotted squares. Mean radial velocity curves of various emission lines along the jet (right). Positive offset is in the southeast direction, where the blueshifted outflow is located. Colors of the H$_2$ emission markers denote the upper energy levels of the transitions, with a gradient from red to blue representing increasing upper energy levels.}
    \label{fig:velmaps}
\end{figure*}

\subsection{Outflow Velocity Distribution}
We made velocity maps of strong emission lines to investigate the velocity distribution in the outflow. Using the continuum-subtracted data cubes, we fitted a Gaussian profile to each spaxel to derive the line-of-sight velocity at each position. The left panels of Figure \ref{fig:velmaps} show the velocity maps for selected strong lines. In the H$_2$ lines, we observe the blueshifted outflow, with velocity increasing along the outflow axis away from the central source. The [Fe II] and [Ne II] lines trace the jet. The blueshifted jet is detected in all transitions, but only in the [Fe II] line at 26\,$\mu$m, we detect the bipolar jet from both sides of the central source. The northwest lobe is red-shifted, while the southeast lobe is blue-shifted. \par
The right panel of Figure \ref{fig:velmaps} shows the change in velocity along the jet axis. Following a similar method to our jet width analysis, we placed rectangular apertures 0.3\arcsec\ long perpendicular to the jet axis and calculated the average velocity within each aperture. 
The H$_2$ transitions can be categorized into two groups. Lower excitation lines ($E_u < 5000$\,K), with wavelengths longer than 6.5\,$\mu$m maintain a relatively constant velocity of approximately 25\,km\,s$^{-1}$ along the jet axis. On the other hand, higher excitation lines exhibit significantly higher velocities, ranging between 40 and 60\,km\,s$^{-1}$. Within this group, lines with higher excitation energy have higher velocities. Furthermore, these lines show a linear increase in velocity with distance from the source. Assuming the inclination of the H$_2$ outflow is the same as that of the jet (32$^\circ$), the mean total velocity of each group is $\sim$30\,km\,s$^{-1}$ and $\sim$60\,km\,s$^{-1}$, respectively. \par
 
In contrast, the jet emission extends farther, and its velocity plateaus at $\sim$110\,km\,s$^{-1}$ beyond 2\arcsec\ from the center for both lobes. There is some wiggling, which can be either from observation uncertainty or precession of the jet \citep{cfLee2020}. The velocities of various jet lines are consistent with each other within 25\,km\,s$^{-1}$. After correcting for the inclination, the terminal velocity of the jet is estimated to be $\sim$130\,km\,s$^{-1}$. Notably, the zero-velocity position of the [Fe II] 25.99\,$\mu$m line is offset by $\sim$1.5\arcsec\ into the redshifted lobe, which is a large discrepancy that cannot be explained solely by projection effects. The origin of this discrepancy is discussed further in Section \ref{sec:jet_velocity}.

\begin{figure*}
    \centering
    \includegraphics[width=\linewidth]{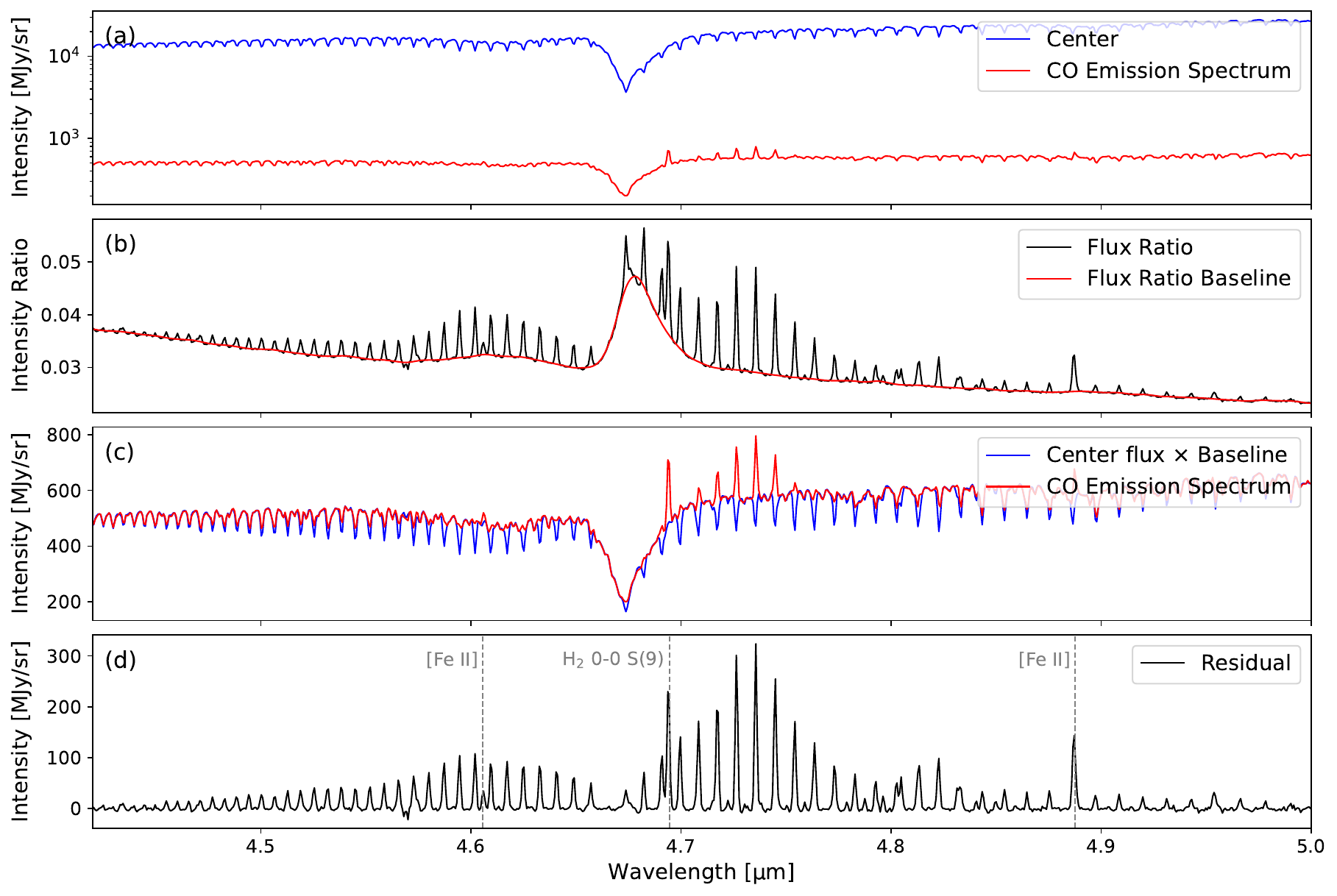}
    \caption{Derivation process of the hot CO emission. (a) Spectra extracted from the ALMA continuum center (blue) and a position with strong CO emission (red). (b) The ratio of the two spectra (black), with the fitted baseline (red), accounting for scattering and differences in foreground attenuation. (c) The central spectrum multiplied by the baseline (blue), compared with the observed CO emission spectrum (red). (d) The residual spectrum representing the hot CO component. Gray dashed lines mark the emission lines from other than CO.}
    \label{fig:co_process}
\end{figure*}

\begin{figure}
    \centering
    \includegraphics[width=\linewidth]{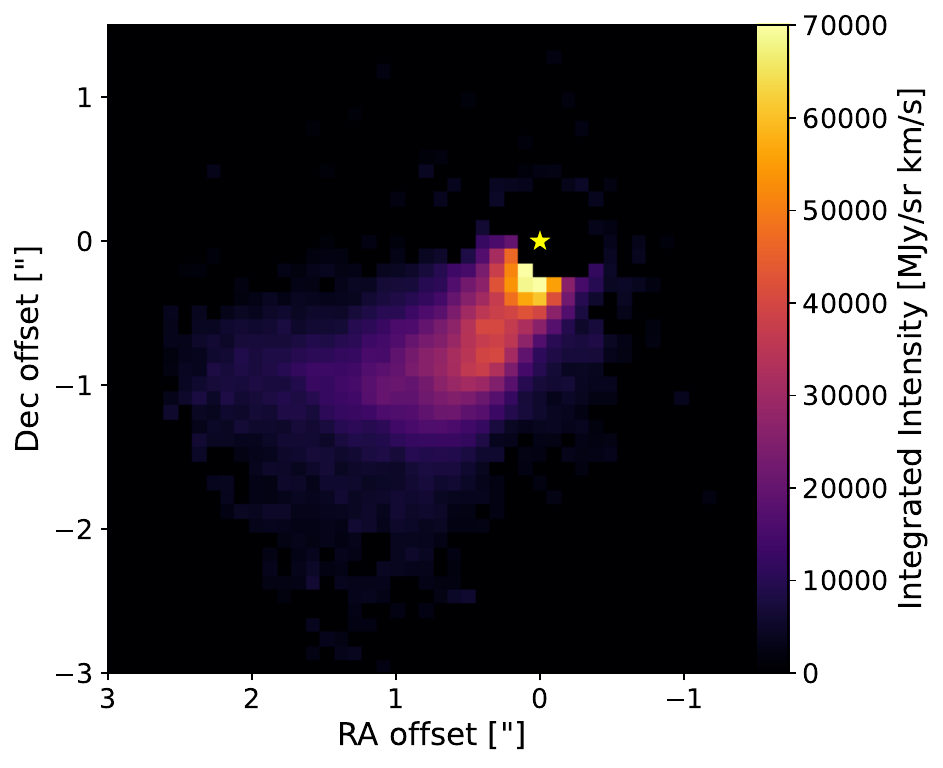}
    \caption{The residual integrated intensity map of CO v=1--0 P(8), after subtracting the scattered light, which includes the CO absorption spectra.}
    \label{fig:co_emission_residual}
\end{figure}

\subsection{Analysis of the mid-IR CO Emission} \label{sec:CO_origin}

The CO spectra, which are mixed with emission and absorption features detected inside the outflow cavity, are likely produced by two distinct steps. First, CO molecules near the central protostar produce absorption lines against the continuum, primarily originating from the heated disk midplane (Seokho Lee et al. submitted). This is observed in the spectra extracted toward the central protostar (blue spectrum of Figure \ref{fig:CO_spec}). Photons generated at the central source, including those with absorption features, are scattered along the line of sight through the outflow cavity, where hot CO gas is present. This CO gas is shock-heated by the outflow, producing emission lines. These emission lines are superimposed on the scattered continuum from the central source along our line of sight. If the emission lines are sufficiently strong, they can partially or fully fill in the absorption features seen in the scattered light. As a result, some CO lines appear in emission while others remain in absorption, as shown in the red spectrum of Figure~\ref{fig:CO_spec}. Finally, all radiation is further attenuated by dust and ice in the surrounding protostellar envelope. Multiple components of CO with different temperatures and origins have been reported in various young stellar objects \citep{Herczeg2011}, supporting this scenario. \par
Therefore, to study the emission distribution of the hot CO in the outflow, we extract the pure CO emission spectra following the process presented in Figure \ref{fig:co_process}.
The intensities that we observe from the center ($I_0'$) and the CO emission region ($I_1'$) can be expressed as follows:
\begin{gather}
    I_0' = I_0 e^{-\tau_0},\\
    I_1' = I_1 e^{-\tau_1} = (CI_0 + I_{CO})e^{-\tau_1}.
\end{gather}

$I_0'$ and $I_1'$ are the observed intensities obtained directly from the data (first panel of Figure \ref{fig:co_process}). $I_0$ represents the intrinsic intensity at the central position, including the CO absorption lines. $I_0$ is attenuated by the envelope optical depth $\tau_0$ before getting observed as $I_0'$. 
In regions affected by the outflow, $I_0$ is reduced by a factor of $C$ due to scattering. On the other hand, the hot CO gas shocked by the outflow adds the emission, $I_{CO}$, to the scattered emission, $CI_O$, making the total intensity, $I_1$. 
This total intensity is attenuated by the envelope optical depth ($\tau_1$) along the line of sight. 
However, the $I_1' / I_0'$ baseline, as presented with the red line in the second panel of Figure \ref{fig:co_process}, is independent of line emission and only contains information about scattering effects and foreground attenuation ($Ce^{\tau_0 - \tau_1}$). The baseline was obtained using the `mixture model' method from the \lstinline{pybaseline} Python package \citep{pybaselines}. The mixture model treats each data point as arising from a probabilistic combination of a baseline and a signal distribution, allowing robust separation of a smooth baseline from sharp spectral features by reducing the influence of outliers \citep{Ghojogh2019}. The third panel compares the light from the outflow and the central flux multiplied by the baseline. \par
The final panel shows the residual after subtracting the two spectra ($I_{CO}e^{-\tau_1}$). The final residual shows all CO lines in emission, with the P-branch stronger than the R-branch. Although the foreground attenuation ($e^{-\tau_1}$) is not fully corrected, the attenuation is fairly constant in the wavelength range except for the CO ice absorption, and would not change the existence of the observed asymmetry between the P- and R-branch. \par

\begin{figure*}
    \centering
    \includegraphics[width=0.9\linewidth]{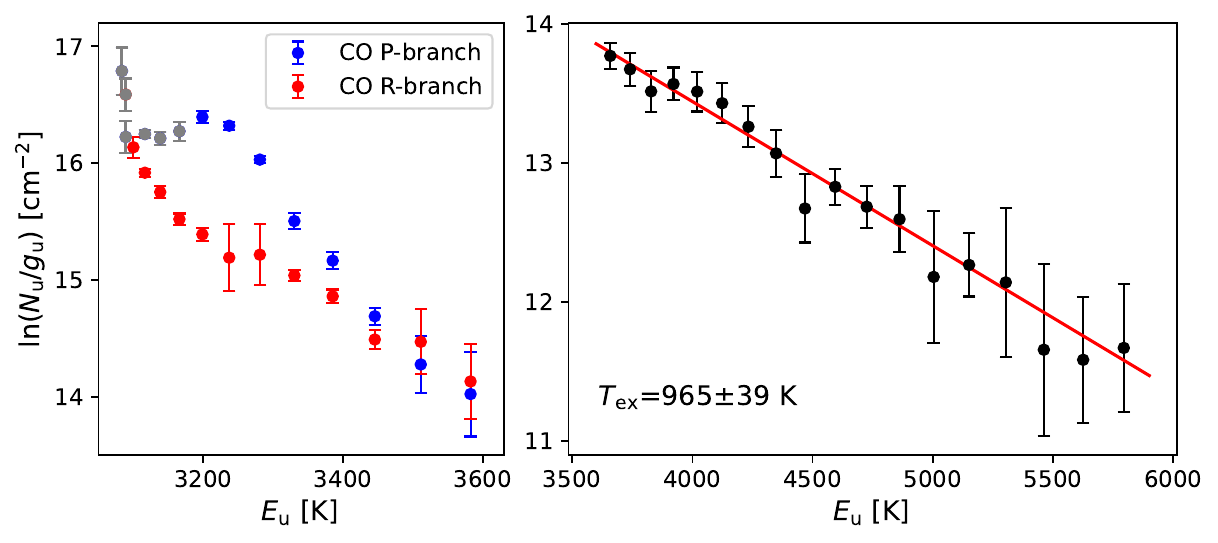}
    \caption{(Left) Excitation diagram of CO with low-J transitions. P-branch and R-branch transitions are marked in blue and red, respectively. Transitions contaminated by CO ice absorption are marked in gray. (Right) Excitation diagram of CO with high-J R-branch transitions.}
    \label{fig:CO_rot}
\end{figure*}

The procedure above can be applied to all pixels, including regions with no evident CO emission. Figure \ref{fig:co_emission_residual} shows the peak flux map of CO v=1--0 P(8) after subtracting the scattered continuum component, which includes the CO absorption features produced in the protostellar disk. Even in areas that originally showed CO absorption, the residual image shows emission if the absorption depth was less than that of the reference spectrum. The pixels near the center are masked because the absorption lines are stronger than those in the reference spectrum used for modeling the disk CO component. The spatial distribution of the residual emission resembles the H$_2$ emission, which traces hot gas in the outflow cavity. This suggests that the hot CO and H$_2$ may share a common origin. \par

\subsection{Excitation diagram of mid-IR CO}

We utilized an excitation diagram to estimate the physical properties of the CO gas. The analysis was done using the residual emission shown in the last panel of Figure \ref{fig:co_process}. Although the spectrum shows strong peaks of the $^{12}$CO v=1--0 lines, some transitions are heavily contaminated by CO ice absorption, H$_2$ emission, and $^{13}$CO emission.

Figure \ref{fig:CO_rot} shows the results. The low-J lines (J<11) clearly exhibit the P-R asymmetry. P-R asymmetry has been observed in many astronomical environments, from protostars \citep{Alfonso1998, Alfonso2002, Rubenstein2024} to AGN jets \citep{Buiten2024, Santaella2024}. Vibrationally excited CO molecules have very high critical densities ($\sim 10^{14}$\,cm$^{-3}$ at $T\sim1000$\,K), which are much higher than typical gas densities in outflow cavities ($10^5\sim10^6$\,cm$^{-3}$). In such low-density conditions, CO can deviate from LTE and become radiatively excited. In these cases, the P-branch emission can be stronger than the R-branch \citep{Alfonso2002, Lacy2013}. \par
The right panel shows the excitation diagram using the high-J R-branch transitions. Only the R-branch transitions were used, as the P-branch transitions were more contaminated. The observed intensities could be explained by a single-component model with an excitation temperature of 965\,K. The obtained excitation temperature of CO is similar to the excitation temperature of purely rotational H$_2$ lines. However, this temperature may not reflect the kinetic temperature of CO, as the LTE assumption is invalid. The $v=1$ levels of CO are primarily radiatively excited, and the excitation temperature is heavily influenced by the Planck temperature of the local radiation field \citep{Buiten2024}. Reliable non-LTE modeling requires collision coefficients for vibrationally excited CO with H$_2$. While some studies have modeled the emission using coefficients estimated from scaling relations \citep[e.g.,][]{Thi2013, Bosman2019}, accurate coefficients derived from quantum calculations are currently unavailable. Such modeling is beyond the scope of this paper and is left for future work.

\begin{figure*}
    \centering
    \includegraphics[width=0.39\linewidth]{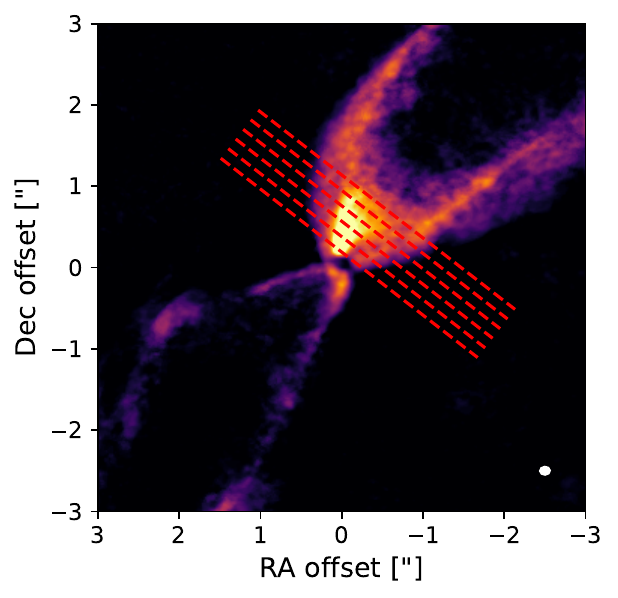}
    \includegraphics[width=0.58\linewidth]{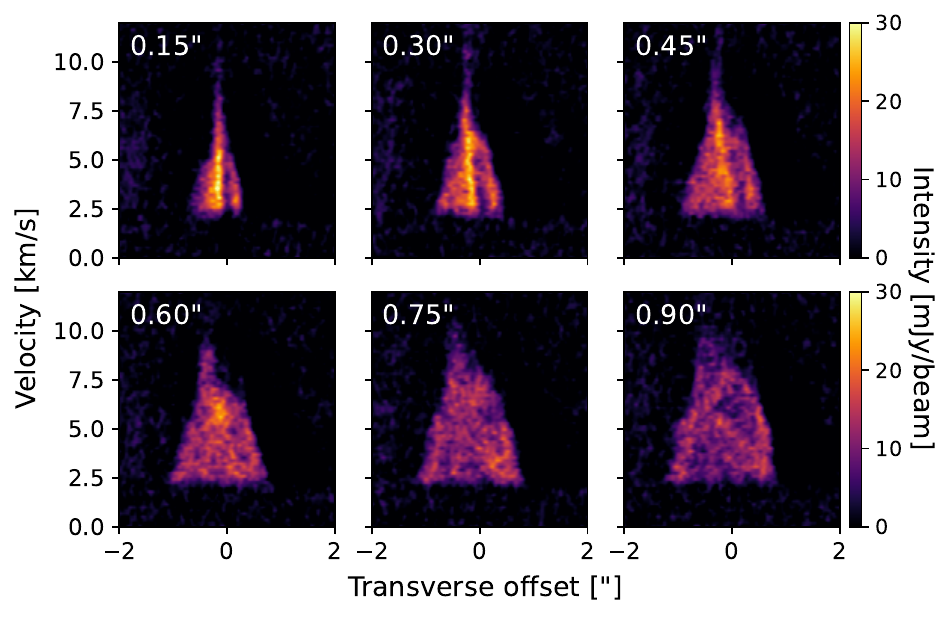}
    \caption{(Left) Integrated intensity map of the high-resolution ALMA CO observation. Red dashed lines denote the transverse PV cuts at intervals of 0.15\arcsec. (Right) Transverse PV diagrams for each PV cut.}
    \label{fig:CO_PV}
\end{figure*}

\subsection{Transverse Position-Velocity Diagram of ALMA CO Emission}

As shown in Figure \ref{fig:CO_PV}, the submillimeter CO emission traces both blueshifted and redshifted outflows, although large-scale emission suffers from the resolve-out effect due to the small maximum recoverable size (MRS\,$\sim1.63$\arcsec) of our ALMA observation. Nevertheless, the redshifted CO emission is well detected close to the central protostar, allowing the small-scale redshifted outflow emission to be used for kinematic analysis. For this, we made transverse position-velocity (PV) diagrams across the outflow axis with a 0.15\arcsec\ interval, as presented in the right panels of Figure \ref{fig:CO_PV}. The transverse PV diagrams consist of 2 components: a narrow high-velocity feature and a wide low-velocity feature. The low-velocity feature has a trapezoidal shape, with higher velocities near the jet axis and lower velocities on the edge. 

\section{Discussion} \label{sec:discussion}

\begin{figure*}
    \centering
    \includegraphics[width=\linewidth]{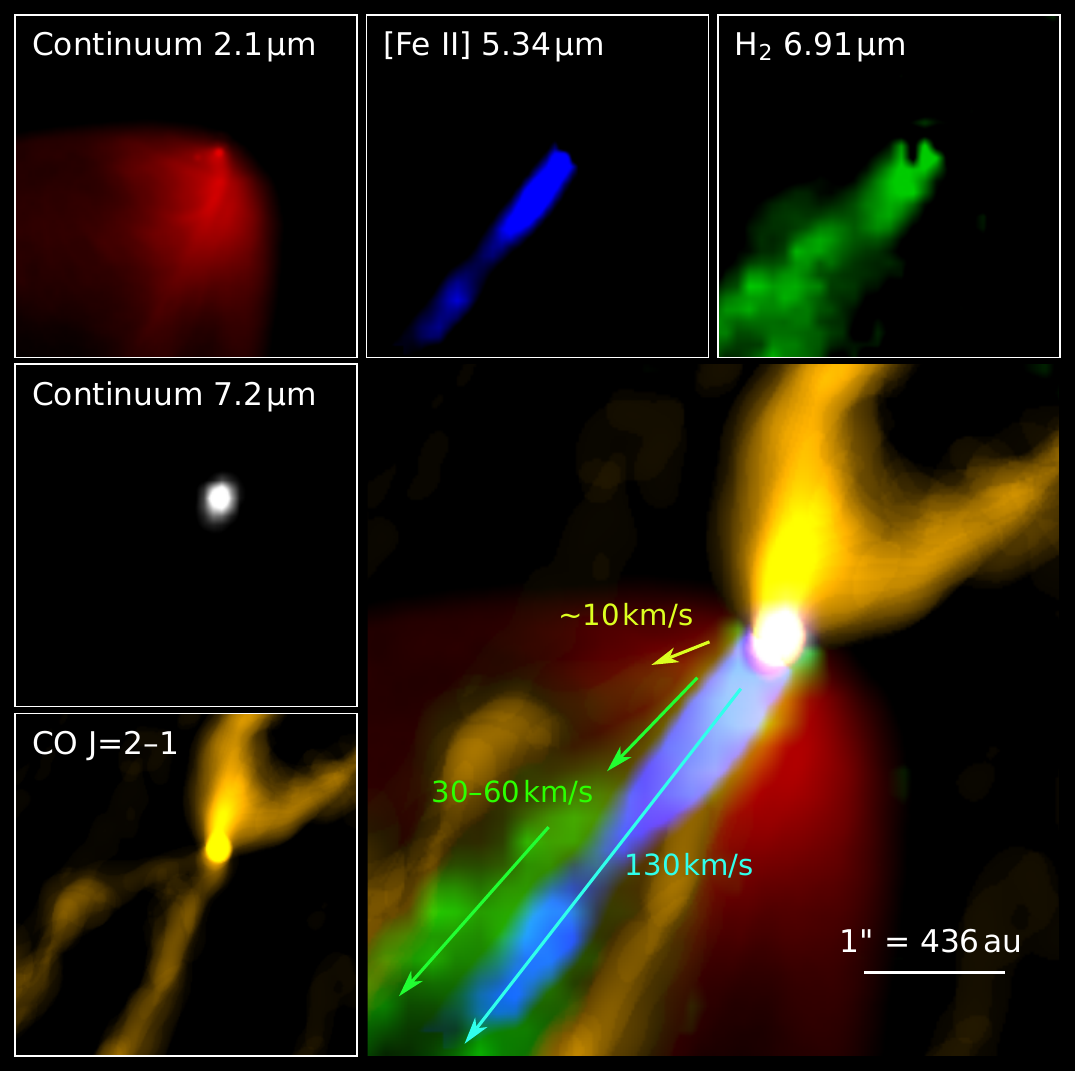}
    \caption{Images of different components of EC 53 (top and left) and their composite image (lower right). The color scale for each image is adjusted to best present its structure. The velocities of each component are marked with the same colors.}
    \label{fig:composite}
\end{figure*}

\subsection{Morphology and Possible Launching Mechanisms of the Outflow and Jet}

Figure \ref{fig:composite} presents the morphology of various outflow components of EC 53, observed with JWST and ALMA. The structure reveals an onion-like stratification within the wide outflow cavity traced by the short-wavelength NIRSpec continuum. Nested within this cavity are the cold CO outflow traced by ALMA, the hot H$_2$ emission, and the fast atomic jet. These components span a range of temperatures and velocities: the atomic jet traced with [Fe II] is well collimated and fast ($\sim$130\,km\,s$^{-1}$), the hot H$_2$ component shows intermediate velocities and temperatures, while the cold CO outflow is slower (<10\,km\,s$^{-1}$). \par

This layered structure with hotter, faster components enclosed within the cooler, slower ones is commonly seen in protostellar systems at various evolutionary stages \citep{CarattioGaratti2024, Delabrosse2024, Federman2024, Tychoniec2024, Pascucci2025}. The onion-like layered structure is often interpreted as a defining characteristic of magneto-centrifugal disk winds (MHD winds), which naturally produce stratified outflows \citep{Blandford1982, Ferreira2006}. In these models, hotter, faster material is launched from smaller radii in the disk, resulting in narrower opening angles and higher outflow velocities because of the higher Keplerian speeds at the base. The observed hierarchy in temperature and velocity among [Fe II], H$_2$, and cold CO in EC 53 supports this scenario. \par

Additionally, we find that this stratification persists within the H$_2$ gas itself, which shows a clear gradient in spatial extent, temperature, and velocity. The lower excitation transitions exhibit wider profiles ($\sim 1.7$\arcsec) that systematically narrow to a more collimated width ($\sim 1.1$\arcsec) at higher excitation. A similar trend is seen in temperature: transitions observed with MIRI yield excitation temperatures of $\sim900$\,K, while higher-energy lines observed with NIRSpec trace significantly hotter gas at $\sim 2500$\,K. The lower-energy transitions show slower, relatively constant velocities of $\sim 30$\,km\,s$^{-1}$ along the jet axis, whereas higher-excitation transitions reach average velocities of $\sim 60$\,km\,s$^{-1}$. \par
Even within the cold CO emission, the PV diagram shows an inner, faster component nested within a slower one, consistent with predictions of MHD wind models \citep[see Figure \ref{fig:CO_PV};][]{deValon2020}. \par

One way to further test the MHD disk wind scenario is to measure the angular momentum and compute the magnetic lever arm parameter, $\lambda$. The magnetic lever arm parameter is the ratio of the outflow's specific angular momentum to the Keplerian value at the launching radius. In MHD winds, $\lambda > 1.5$, indicating efficient angular momentum removal from the disk. However, the spectral resolution of JWST spectra is insufficient to detect the small velocity gradients that would be expected from rotation in H$_2$ or [Fe II]. \par

Another notable feature in EC 53 is the increase in H$_2$ velocity along the outflow axis, observed for higher energy transitions. This apparent acceleration could also arise from entrained envelope material rather than direct wind launching. Swept-up gas acceleration has been observed in several sources, including HH 211 \citep{cfLee2022}, Cep E \citep{Schutzer2022}, HOPS 373SW \citep{shlee2024}, and HOPS 315 \citep{Vleugels2025}. The wide-angle wind-driven shell model \citep{Li1996, cfLee2000} offers a plausible explanation: when a radial wind interacts with a flattened envelope, it drives a forward shock that sweeps up envelope material. This produces a radially expanding shell with velocity increasing linearly with distance from the source---a ``Hubble-law"–like acceleration profile. 
However, the fact that this acceleration is primarily seen in the narrower, higher energy lines suggests it may represent the acceleration of the inner wind entrained by the jet \citep{Smith1997, Rabenanahary2022}. The constant velocity of the lower-excitation H$_2$ may represent a more stable, outer component of the disk wind that has experienced less interaction with the high-speed central jet.

In summary, while the observed stratification and acceleration patterns in EC 53 are broadly consistent with MHD disk wind models, additional mechanisms, such as wide-angle winds or jet-driven entrainment, may also contribute. A full understanding should also explain the apparent separation between the dense, cold CO shell observed with ALMA and the broader outflow cavity traced by scattered continuum and C$_2$H. Higher spectral resolution and complementary observations will be necessary to disentangle these contributions and confirm the launching, confining, and entrainment mechanisms.

\begin{figure}
    \centering
    \includegraphics[width=\linewidth]{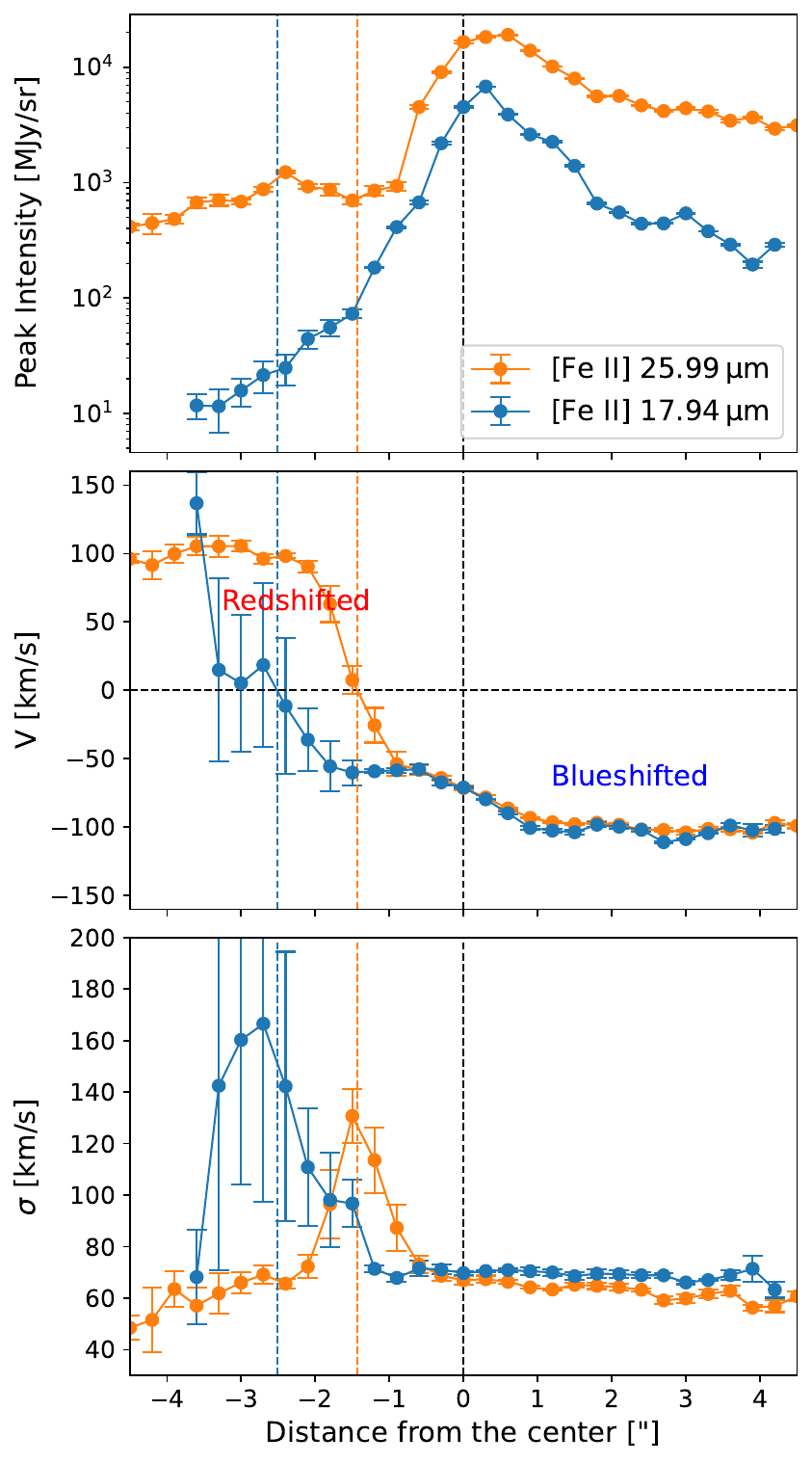}
    \caption{Peak intensity (top), Mean radial velocity (middle), and line width (bottom) of [Fe II] 25.99\,$\mu$m (orange) and 17.94\,$\mu$m (blue) line along the jet of EC 53. Positive offset is in the southeast direction, where the blueshifted jet is located. The velocities are corrected with the barycentric velocity of -8.9\,km\,s$^{-1}$, but not with the inclination. Vertical dashed lines indicate the zero-velocity positions for each line.}
    \label{fig:Fe_velocity}
\end{figure}

\subsection{Interpretation of the Jet Velocity Offset} \label{sec:jet_velocity}
Jets launched from a protostar at an inclined angle are expected to have shifted velocity in their lobes separated by the central protostar. However, the velocity maps in Figure \ref{fig:velmaps} reveal offsets between the zero-velocity position and the protostellar location. For instance, the transition from blueshift to redshift observed in the [Fe II] jet occurs approximately 1.5\arcsec\ northwest of the protostar, coinciding with the location of the redshifted submillimeter CO outflow.  \par
To investigate this phenomenon, we compare two [Fe II] lines, as shown in Figure \ref{fig:Fe_velocity}, which presents the peak intensities, radial velocities, and line widths of the 25.99\,$\mu$m and 17.94\,$\mu$m [Fe II] transitions. Both emission lines are detected on either side of the jet axis across the protostar. The blueshifted jet is brighter than the redshifted counterpart, with a stronger flux difference for the 17.94\,$\mu$m line. In both cases, the zero-velocity position is offset to the northwest of the protostellar position. Notably, the offset of the 17.94\,$\mu$m line ($\sim$2.5\arcsec) is larger than that of the 25.99\,$\mu$m line ($\sim$1.5\arcsec). The line width also varies with position, reaching a maximum near the zero-velocity region. \par
This shift in the zero-velocity position of the [Fe II] lines may result from the effects of PSF and extinction. Near the central region, the redshifted and blueshifted jets overlap along the line of sight due to the relatively large PSF (FWHM of 0.70\arcsec\ for 17.94\,$\mu$m and 0.96\arcsec\ for 25.99\,$\mu$m), leading to increased line broadening near the source. Moreover, compared to the blue-shifted emission, the red-shifted jet must traverse a thicker portion of the envelope, leading to stronger extinction. Consequently, the blended emission near the protostar appears blueshifted, shifting the apparent zero-velocity position toward the northwest. The larger offset observed in the 17.94\,$\mu$m line can be explained by the greater extinction at shorter wavelengths. We therefore assert that the observed jet velocity offset is due primarily to optical depth and angular resolution issues and expect that the true velocity along the jet shifts quickly from around $-$130\,km\,s$^{-1}$ to 130\,km\,s$^{-1}$ across the source.

\subsection{Effect of Variability on the Outflow} \label{variable_outflow}

EC 53 is a protostar showing a quasi-periodic variability with a period of approximately 1.5 years, observed across wavelengths from near-IR to sub-mm \citep[e.g.,][]{Hodapp1999, Hodapp2012, YLee2020, Francis2022}. The periodic brightening and fading are attributed to variable accretion \citep{Baek2020}, potentially triggered by interactions between the disk and a very close companion at a few AU separation \citep[e.g.,][]{Bonnell1992, Nayakshin2012}. If each accretion burst led to an increased ejection activity, they could appear as knots in the jet. In our observation, we do not see any significant knot-like features. If there was a newly formed knot ejected every 1.5 years, using intrinsic jet velocity of 130\,km\,s$^{-1}$ and inclination of 32$^\circ$, the apparent angular distance between the knots would be 0.033\arcsec. This is 3 times smaller than the NIRSpec pixel scale (0.1\arcsec). Even if such a series of knots by periodic accretion existed, it would not be resolved with the spatial resolution of JWST.

\section{Summary} \label{sec:summary}

We observed the Class I protostar EC 53 with JWST NIRSpec/IFU and MIRI/MRS, in addition to ALMA, and compared the results with archival NIRCam and lower-resolution ALMA data to obtain a comprehensive view of its outflow and jet structure. The main findings are summarized below:

\begin{enumerate}
    \item Scattered light from the outflow cavity is observed in NIRCam images (F140M, F210M, F360M, and F480M) and in the short-wavelength NIRSpec continuum. Only the southeast cavity, which faces the observer, is visible. At longer wavelengths ($>$5\,$\mu$m), emission from the central source dominates. ALMA continuum data resolve a compact disk with a radius of $\sim$0.1\arcsec.
    \item A total of 27 H$_2$ emission lines are detected with NIRSpec and MIRI. The H$_2$ emission traces a narrow, cone-like structure within the outflow cavity. Using an excitation diagram, we derived the gas temperature, column density, and extinction. Lower energy transitions exhibit a warm temperature ($\sim$900\,K), while the higher energy transitions trace a hotter component ($\sim$2500\,K). Lower-excitation H$_2$ maintains a constant, slow velocity, while the highly excited gas accelerates linearly along the outflow axis and exhibits a higher mean velocity.
    \item A high-velocity jet is detected in multiple metal forbidden lines ([Fe II], [Ni II], [Ne II]). These lines are predominantly blueshifted and dissect the H$_2$ emission region. Two lines (17.94\,$\mu$m and 25.99\,$\mu$m) also show redshifted emission. We derived a jet position angle of 142.2$^\circ$, an opening angle of 1.3$^\circ$, and an upper limit on the launching radius of 45.9\,au. The intrinsic velocity of the jet is $\sim$130\,km\,s$^{-1}$. The apparent zero-velocity position is offset to the redshifted side, likely due to extinction and PSF effects.
    \item MIR CO ro-vibrational lines are primarily seen in absorption, but some P-branch lines appear in emission. This is attributed to hot CO gas in the outflow cavity, which emits above the central source's absorption spectrum. The isolated CO emission shows stronger P-branch emission than the R-branch, consistent with radiative excitation in low-density regions such as the outflow cavity.
    \item Cold CO is detected with ALMA on both outflow lobes. A position-velocity (PV) diagram from high-resolution ALMA data shows that emission closer to the jet has a higher velocity. C$_2$H emission closely traces the outflow cavity walls.
    \item The atomic jet, hot gas (H$_2$, CO), and cold gas (CO) are spatially nested, consistent with predictions from MHD disk wind models. The acceleration of highly excited H$_2$ can also be explained by jet entrainment. The outflow-launching mechanism cannot be attributed to a single theory; multiple processes may act simultaneously.
\end{enumerate}

 Further detailed modeling and higher-spectral-resolution NIR spectroscopy will be essential to test and distinguish between these scenarios. Nevertheless, this study demonstrates the powerful synergy of ALMA and JWST in advancing our understanding of protostellar outflow and jet systems.\\

\begin{acknowledgments}
This work was supported by the National Research Foundation of Korea (NRF) grant funded by the Korea government (MSIT) (grant numbers 2021R1A2C1011718 and RS-2024-00416859). D.J. is supported by NRC Canada and by an NSERC Discovery Grant. G.J.H. is supported by grant IS23020 from the Beijing Natural Science Foundation. J.D.G. acknowledges support from the associated 3477 NASA observer grant. 
This work is based on observations made with the NASA/ESA/CSA James Webb Space Telescope. The data were obtained from the Mikulski Archive for Space Telescopes at the Space Telescope Science Institute, which is operated by the Association of Universities for Research in Astronomy, Inc., under NASA contract NAS 5-03127 for JWST. These observations are associated with JWST program \#3477 and \#1611. 
This paper makes use of the following ALMA data: ADS/JAO.ALMA\#2016.1.01304.T, ADS/JAO.ALMA\#2019.1.01792.S, and ADS/JAO.ALMA\#2022.1.00800.S. ALMA is a partnership of ESO (representing its member states), NSF (USA), and NINS (Japan), together with NRC (Canada), MOST, and ASIAA (Taiwan), and KASI (Republic of Korea), in cooperation with the Republic of Chile. The Joint ALMA Observatory is operated by ESO, AUI/NRAO, and NAOJ. Some of the data presented in this article were obtained from the Mikulski Archive for Space Telescopes (MAST) at the Space Telescope Science Institute. The specific observations analyzed can be accessed via \dataset[doi:10.17909/h56s-xr27]{https://doi.org/10.17909/h56s-xr27} and \dataset[doi:10.17909/pv1h-ta47]{https://doi.org/10.17909/pv1h-ta47}. This research also uses data from the ALMA archive.
We acknowledge the use of ChatGPT to check English grammar and improve the clarity of expressions. \par
\textit{Facilities} : JWST, ALMA \par
\textit{Software} : Numpy \citep{numpy}, Scipy \citep{scipy}, Astropy \citep{astropy}, Matplotlib \citep{matplotlib}, photutils \citep{photutils}, emcee \citep{emcee}, corner \citep{corner}, pybaselines \citep{pybaselines}.
\end{acknowledgments}

\bibliography{ref}{}

@ARTICLE{Aikawa2012,
       author = {{Aikawa}, Y. and {Kamuro}, D. and {Sakon}, I. and {Itoh}, Y. and {Terada}, H. and {Noble}, J.~A. and {Pontoppidan}, K.~M. and {Fraser}, H.~J. and {Tamura}, M. and {Kandori}, R. and {Kawamura}, A. and {Ueno}, M.},
        title = "{AKARI observations of ice absorption bands towards edge-on young stellar objects}",
      journal = {\aap},
         year = 2012,
        month = feb,
       volume = {538},
          eid = {A57},
        pages = {A57},
          doi = {10.1051/0004-6361/201015999},
       adsurl = {https://ui.adsabs.harvard.edu/abs/2012A&A...538A..57A}
}

@ARTICLE{Thi2013,
       author = {{Thi}, W.~F. and {Kamp}, I. and {Woitke}, P. and {van der Plas}, G. and {Bertelsen}, R. and {Wiesenfeld}, L.},
        title = "{Radiation thermo-chemical models of protoplanetary discs. IV. Modelling CO ro-vibrational emission from Herbig Ae discs}",
      journal = {\aap},
         year = 2013,
        month = mar,
       volume = {551},
          eid = {A49},
        pages = {A49},
          doi = {10.1051/0004-6361/201219210},
archivePrefix = {arXiv},
       eprint = {1210.7654},
 primaryClass = {astro-ph.GA},
       adsurl = {https://ui.adsabs.harvard.edu/abs/2013A&A...551A..49T}
}

@ARTICLE{shlee2020,
       author = {{Lee}, Seokho and {Lee}, Jeong-Eun and {Aikawa}, Yuri and {Herczeg}, Gregory and {Johnstone}, Doug},
        title = "{The Circumstellar Environment around the Embedded Protostar EC 53}",
      journal = {\apj},
         year = 2020,
        month = jan,
       volume = {889},
       number = {1},
          eid = {20},
        pages = {20},
          doi = {10.3847/1538-4357/ab5a7e},
archivePrefix = {arXiv},
       eprint = {1911.10318},
 primaryClass = {astro-ph.SR},
       adsurl = {https://ui.adsabs.harvard.edu/abs/2020ApJ...889...20L}
}

@INPROCEEDINGS{Arce2007,
       author = {{Arce}, H.~G. and {Shepherd}, D. and {Gueth}, F. and {Lee}, C. -F. and {Bachiller}, R. and {Rosen}, A. and {Beuther}, H.},
        title = "{Molecular Outflows in Low- and High-Mass Star-forming Regions}",
    booktitle = {Protostars and Planets V},
         year = 2007,
       editor = {{Reipurth}, Bo and {Jewitt}, David and {Keil}, Klaus},
        month = jan,
        pages = {245},
          doi = {10.48550/arXiv.astro-ph/0603071},
archivePrefix = {arXiv},
       eprint = {astro-ph/0603071},
 primaryClass = {astro-ph},
       adsurl = {https://ui.adsabs.harvard.edu/abs/2007prpl.conf..245A}
}

@ARTICLE{Bodenheimer1995,
       author = {{Bodenheimer}, Peter},
        title = "{Angular Momentum Evolution of Young Stars and Disks}",
      journal = {\araa},
         year = 1995,
        month = jan,
       volume = {33},
        pages = {199-238},
          doi = {10.1146/annurev.aa.33.090195.001215},
       adsurl = {https://ui.adsabs.harvard.edu/abs/1995ARA&A..33..199B}
}

@ARTICLE{Bachiller1996,
       author = {{Bachiller}, Rafael},
        title = "{Bipolar Molecular Outflows from Young Stars and Protostars}",
      journal = {\araa},
         year = 1996,
        month = jan,
       volume = {34},
        pages = {111-154},
          doi = {10.1146/annurev.astro.34.1.111},
       adsurl = {https://ui.adsabs.harvard.edu/abs/1996ARA&A..34..111B}
}

@ARTICLE{Baek2020,
       author = {{Baek}, Giseon and {MacFarlane}, Benjamin A. and {Lee}, Jeong-Eun and {Stamatellos}, Dimitris and {Herczeg}, Gregory and {Johnstone}, Doug and {Pe{\~n}a}, Carlos Contreras and {Varricatt}, Watson and {Hodapp}, Klaus W. and {Chen}, Huei-Ru Vivien and {Kang}, Sung-Ju},
        title = "{Radiative Transfer Modeling of EC 53: An Episodically Accreting Class I Young Stellar Object}",
      journal = {\apj},
         year = 2020,
        month = may,
       volume = {895},
       number = {1},
          eid = {27},
        pages = {27},
          doi = {10.3847/1538-4357/ab8ad4},
archivePrefix = {arXiv},
       eprint = {2004.05600},
 primaryClass = {astro-ph.SR},
       adsurl = {https://ui.adsabs.harvard.edu/abs/2020ApJ...895...27B}
}

@ARTICLE{Bally2016,
       author = {{Bally}, John},
        title = "{Protostellar Outflows}",
      journal = {\araa},
         year = 2016,
        month = sep,
       volume = {54},
        pages = {491-528},
          doi = {10.1146/annurev-astro-081915-023341},
       adsurl = {https://ui.adsabs.harvard.edu/abs/2016ARA&A..54..491B}
}

@ARTICLE{Dunham2015,
       author = {{Dunham}, Michael M. and {Allen}, Lori E. and {Evans}, Neal J., II and {Broekhoven-Fiene}, Hannah and {Cieza}, Lucas A. and {Di Francesco}, James and {Gutermuth}, Robert A. and {Harvey}, Paul M. and {Hatchell}, Jennifer and {Heiderman}, Amanda and {Huard}, Tracy L. and {Johnstone}, Doug and {Kirk}, Jason M. and {Matthews}, Brenda C. and {Miller}, Jennifer F. and {Peterson}, Dawn E. and {Young}, Kaisa E.},
        title = "{Young Stellar Objects in the Gould Belt}",
      journal = {\apjs},
         year = 2015,
        month = sep,
       volume = {220},
       number = {1},
          eid = {11},
        pages = {11},
          doi = {10.1088/0067-0049/220/1/11},
archivePrefix = {arXiv},
       eprint = {1508.03199},
 primaryClass = {astro-ph.GA},
       adsurl = {https://ui.adsabs.harvard.edu/abs/2015ApJS..220...11D}
}

@INPROCEEDINGS{Frank2014,
       author = {{Frank}, A. and {Ray}, T.~P. and {Cabrit}, S. and {Hartigan}, P. and {Arce}, H.~G. and {Bacciotti}, F. and {Bally}, J. and {Benisty}, M. and {Eisl{\"o}ffel}, J. and {G{\"u}del}, M. and {Lebedev}, S. and {Nisini}, B. and {Raga}, A.},
        title = "{Jets and Outflows from Star to Cloud: Observations Confront Theory}",
    booktitle = {Protostars and Planets VI},
         year = 2014,
       editor = {{Beuther}, Henrik and {Klessen}, Ralf S. and {Dullemond}, Cornelis P. and {Henning}, Thomas},
        month = jan,
        pages = {451},
          doi = {10.2458/azu\_uapress\_9780816531240-ch020},
archivePrefix = {arXiv},
       eprint = {1402.3553},
 primaryClass = {astro-ph.SR},
       adsurl = {https://ui.adsabs.harvard.edu/abs/2014prpl.conf..451F}
}

@ARTICLE{Jorgensen2007,
       author = {{J{\o}rgensen}, Jes K. and {Bourke}, Tyler L. and {Myers}, Philip C. and {Di Francesco}, James and {van Dishoeck}, Ewine F. and {Lee}, Chin-Fei and {Ohashi}, Nagayoshi and {Sch{\"o}ier}, Fredrik L. and {Takakuwa}, Shigehisa and {Wilner}, David J. and {Zhang}, Qizhou},
        title = "{PROSAC: A Submillimeter Array Survey of Low-Mass Protostars. I. Overview of Program: Envelopes, Disks, Outflows, and Hot Cores}",
      journal = {\apj},
         year = 2007,
        month = apr,
       volume = {659},
       number = {1},
        pages = {479-498},
          doi = {10.1086/512230},
archivePrefix = {arXiv},
       eprint = {astro-ph/0701115},
 primaryClass = {astro-ph},
       adsurl = {https://ui.adsabs.harvard.edu/abs/2007ApJ...659..479J}
}

@ARTICLE{cfLee2020,
       author = {{Lee}, Chin-Fei},
        title = "{Molecular jets from low-mass young protostellar objects}",
      journal = {\aapr},
         year = 2020,
        month = mar,
       volume = {28},
       number = {1},
          eid = {1},
        pages = {1},
          doi = {10.1007/s00159-020-0123-7},
archivePrefix = {arXiv},
       eprint = {2002.05823},
 primaryClass = {astro-ph.GA},
       adsurl = {https://ui.adsabs.harvard.edu/abs/2020A&ARv..28....1L}
}

@INPROCEEDINGS{McMullin2007,
       author = {{McMullin}, J.~P. and {Waters}, B. and {Schiebel}, D. and {Young}, W. and {Golap}, K.},
        title = "{CASA Architecture and Applications}",
    booktitle = {Astronomical Data Analysis Software and Systems XVI},
         year = 2007,
       editor = {{Shaw}, R.~A. and {Hill}, F. and {Bell}, D.~J.},
       series = {Astronomical Society of the Pacific Conference Series},
       volume = {376},
        month = oct,
        pages = {127},
       adsurl = {https://ui.adsabs.harvard.edu/abs/2007ASPC..376..127M}
}

@ARTICLE{Shu1987,
       author = {{Shu}, Frank H. and {Adams}, Fred C. and {Lizano}, Susana},
        title = "{Star formation in molecular clouds: observation and theory.}",
      journal = {\araa},
         year = 1987,
        month = jan,
       volume = {25},
        pages = {23-81},
          doi = {10.1146/annurev.aa.25.090187.000323},
       adsurl = {https://ui.adsabs.harvard.edu/abs/1987ARA&A..25...23S}
}

@ARTICLE{shu1994,
       author = {{Shu}, Frank and {Najita}, Joan and {Ostriker}, Eve and {Wilkin}, Frank and {Ruden}, Steven and {Lizano}, Susana},
        title = "{Magnetocentrifugally Driven Flows from Young Stars and Disks. I. A Generalized Model}",
      journal = {\apj},
         year = 1994,
        month = jul,
       volume = {429},
        pages = {781},
          doi = {10.1086/174363},
       adsurl = {https://ui.adsabs.harvard.edu/abs/1994ApJ...429..781S}
}

@ARTICLE{Tychoniec2021,
       author = {{Tychoniec}, {\L}ukasz and {van Dishoeck}, Ewine F. and {van't Hoff}, Merel L.~R. and {van Gelder}, Martijn L. and {Tabone}, Beno{\^\i}t and {Chen}, Yuan and {Harsono}, Daniel and {Hull}, Charles L.~H. and {Hogerheijde}, Michiel R. and {Murillo}, Nadia M. and {Tobin}, John J.},
        title = "{Which molecule traces what: Chemical diagnostics of protostellar sources}",
      journal = {\aap},
         year = 2021,
        month = nov,
       volume = {655},
          eid = {A65},
        pages = {A65},
          doi = {10.1051/0004-6361/202140692},
archivePrefix = {arXiv},
       eprint = {2107.03696},
 primaryClass = {astro-ph.SR},
       adsurl = {https://ui.adsabs.harvard.edu/abs/2021A&A...655A..65T}
}

@ARTICLE{Francis2022,
       author = {{Francis}, Logan and {Johnstone}, Doug and {Lee}, Jeong-Eun and {Herczeg}, Gregory J. and {Long}, Feng and {Mairs}, Steve and {Contreras Pe{\~n}a}, Carlos and {Moriarty-Schieven}, Gerald and {JCMT Transient Team}},
        title = "{Accretion Burst Echoes as Probes of Protostellar Environments and Episodic Mass Assembly}",
      journal = {\apj},
         year = 2022,
        month = sep,
       volume = {937},
       number = {1},
          eid = {29},
        pages = {29},
          doi = {10.3847/1538-4357/ac8a9e},
archivePrefix = {arXiv},
       eprint = {2208.13568},
 primaryClass = {astro-ph.SR},
       adsurl = {https://ui.adsabs.harvard.edu/abs/2022ApJ...937...29F}
}

@ARTICLE{Yang2022,
       author = {{Yang}, Yao-Lun and {Green}, Joel D. and {Pontoppidan}, Klaus M. and {Bergner}, Jennifer B. and {Cleeves}, L. Ilsedore and {Evans}, Neal J., II and {Garrod}, Robin T. and {Jin}, Mihwa and {Kim}, Chul Hwan and {Kim}, Jaeyeong and {Lee}, Jeong-Eun and {Sakai}, Nami and {Shingledecker}, Christopher N. and {Shope}, Brielle and {Tobin}, John J. and {van Dishoeck}, Ewine F.},
        title = "{CORINOS. I. JWST/MIRI Spectroscopy and Imaging of a Class 0 Protostar IRAS 15398-3359}",
      journal = {\apjl},
         year = 2022,
        month = dec,
       volume = {941},
       number = {1},
          eid = {L13},
        pages = {L13},
          doi = {10.3847/2041-8213/aca289},
archivePrefix = {arXiv},
       eprint = {2208.10673},
 primaryClass = {astro-ph.SR},
       adsurl = {https://ui.adsabs.harvard.edu/abs/2022ApJ...941L..13Y}
}

@ARTICLE{Bonnell1992,
       author = {{Bonnell}, Ian and {Bastien}, Pierre},
        title = "{A Binary Origin for FU Orionis Stars}",
      journal = {\apjl},
         year = 1992,
        month = dec,
       volume = {401},
        pages = {L31},
          doi = {10.1086/186663},
       adsurl = {https://ui.adsabs.harvard.edu/abs/1992ApJ...401L..31B}
}

@ARTICLE{Hodapp1999,
       author = {{Hodapp}, Klaus W.},
        title = "{Proper Motions of H\_2 Jets and Variability of Young Stars in the Serpens NW Region}",
      journal = {\aj},
         year = 1999,
        month = sep,
       volume = {118},
       number = {3},
        pages = {1338-1346},
          doi = {10.1086/301003},
       adsurl = {https://ui.adsabs.harvard.edu/abs/1999AJ....118.1338H}
}

@ARTICLE{Hodapp2012,
       author = {{Hodapp}, Klaus W. and {Chini}, Rolf and {Watermann}, Ramon and {Lemke}, Roland},
        title = "{Eruptive Variable Stars and Outflows in Serpens NW}",
      journal = {\apj},
         year = 2012,
        month = jan,
       volume = {744},
       number = {1},
          eid = {56},
        pages = {56},
          doi = {10.1088/0004-637X/744/1/56},
archivePrefix = {arXiv},
       eprint = {1109.3728},
 primaryClass = {astro-ph.SR},
       adsurl = {https://ui.adsabs.harvard.edu/abs/2012ApJ...744...56H}
}

@ARTICLE{YLee2020,
       author = {{Lee}, Yong-Hee and {Johnstone}, Doug and {Lee}, Jeong-Eun and {Herczeg}, Gregory and {Mairs}, Steve and {Varricatt}, Watson and {Hodapp}, Klaus W. and {Naylor}, Tim and {Pe{\~n}a}, Carlos Contreras and {Baek}, Giseon and {Haas}, Martin and {Chini}, Rolf and {JCMT Transient Team}},
        title = "{Young Faithful: The Eruptions of EC 53 as It Cycles through Filling and Draining the Inner Disk}",
      journal = {\apj},
         year = 2020,
        month = nov,
       volume = {903},
       number = {1},
          eid = {5},
        pages = {5},
          doi = {10.3847/1538-4357/abb6fe},
archivePrefix = {arXiv},
       eprint = {2009.05268},
 primaryClass = {astro-ph.SR},
       adsurl = {https://ui.adsabs.harvard.edu/abs/2020ApJ...903....5L}
}

@ARTICLE{Nayakshin2012,
       author = {{Nayakshin}, Sergei and {Lodato}, Giuseppe},
        title = "{Fu Ori outbursts and the planet-disc mass exchange}",
      journal = {\mnras},
         year = 2012,
        month = oct,
       volume = {426},
       number = {1},
        pages = {70-90},
          doi = {10.1111/j.1365-2966.2012.21612.x},
archivePrefix = {arXiv},
       eprint = {1110.6316},
 primaryClass = {astro-ph.EP},
       adsurl = {https://ui.adsabs.harvard.edu/abs/2012MNRAS.426...70N}
}

@ARTICLE{Narang2024,
       author = {{Narang}, Mayank and {Manoj}, P. and {Tyagi}, Himanshu and {Watson}, Dan M. and {Megeath}, S. Thomas and {Federman}, Samuel and {Rubinstein}, Adam E. and {Gutermuth}, Robert and {Caratti o Garatti}, Alessio and {Beuther}, Henrik and {Bourke}, Tyler L. and {Van Dishoeck}, Ewine F. and {Evans}, Neal J. and {Anglada}, Guillem and {Osorio}, Mayra and {Stanke}, Thomas and {Muzerolle}, James and {Looney}, Leslie W. and {Yang}, Yao-Lun and {Klaassen}, Pamela and {Karnath}, Nicole and {Atnagulov}, Prabhani and {Brunken}, Nashanty and {Fischer}, William J. and {Furlan}, Elise and {Green}, Joel and {Habel}, Nolan and {Hartmann}, Lee and {Linz}, Hendrik and {Nazari}, Pooneh and {Pokhrel}, Riwaj and {Rahatgaonkar}, Rohan and {Rocha}, Will R.~M. and {Sheehan}, Patrick and {Slavicinska}, Katerina and {Stutz}, Amelia M. and {Tobin}, John J. and {Tychoniec}, Lukasz and {Wolk}, Scott},
        title = "{Discovery of a Collimated Jet from the Low-luminosity Protostar IRAS 16253‑2429 in a Quiescent Accretion Phase with the JWST}",
      journal = {\apjl},
         year = 2024,
        month = feb,
       volume = {962},
       number = {1},
          eid = {L16},
        pages = {L16},
          doi = {10.3847/2041-8213/ad1de3},
archivePrefix = {arXiv},
       eprint = {2310.14061},
 primaryClass = {astro-ph.SR},
       adsurl = {https://ui.adsabs.harvard.edu/abs/2024ApJ...962L..16N}
}

@ARTICLE{Assani2024,
       author = {{Assani}, K.~D. and {Harsono}, D. and {Ramsey}, J.~P. and {Li}, Z. -Y. and {Bjerkeli}, P. and {Pontoppidan}, K.~M. and {Tychoniec}, {\L}. and {Calcutt}, H. and {Kristensen}, L.~E. and {J{\o}rgensen}, J.~K. and {Plunkett}, A. and {van Gelder}, M.~L. and {Francis}, L.},
        title = "{The asymmetric bipolar [Fe II] jet and H$_{2}$ outflow of TMC1A resolved with the JWST NIRSpec Integral Field Unit}",
      journal = {\aap},
         year = 2024,
        month = aug,
       volume = {688},
          eid = {A26},
        pages = {A26},
          doi = {10.1051/0004-6361/202449745},
archivePrefix = {arXiv},
       eprint = {2404.18334},
 primaryClass = {astro-ph.SR},
       adsurl = {https://ui.adsabs.harvard.edu/abs/2024A&A...688A..26A}
}

@misc{NIST2023,
    author = {{Kramida}, A. and {Ralchenko}, Yu. and {Reader}, J. and NIST ASD Team},
    title = "NIST Atomic Spectra Database (version 5.11)",
    year = 2023,
    publisher = {https://physics.nist.gov/asd}
}

@ARTICLE{vanHoof2018,
       author = {{van Hoof}, Peter A.~M.},
        title = "{Recent Development of the Atomic Line List}",
      journal = {Galaxies},
         year = 2018,
        month = jun,
       volume = {6},
       number = {2},
          eid = {63},
        pages = {63},
          doi = {10.3390/galaxies6020063},
       adsurl = {https://ui.adsabs.harvard.edu/abs/2018Galax...6...63V}
}

@ARTICLE{Law2023,
       author = {{Law}, David R. and {E. Morrison}, Jane and {Argyriou}, Ioannis and {Patapis}, Polychronis and {{\'A}lvarez-M{\'a}rquez}, J. and {Labiano}, Alvaro and {Vandenbussche}, Bart},
        title = "{A 3D Drizzle Algorithm for JWST and Practical Application to the MIRI Medium Resolution Spectrometer}",
      journal = {\aj},
         year = 2023,
        month = aug,
       volume = {166},
       number = {2},
          eid = {45},
        pages = {45},
          doi = {10.3847/1538-3881/acdddc},
archivePrefix = {arXiv},
       eprint = {2306.05520},
 primaryClass = {astro-ph.IM},
       adsurl = {https://ui.adsabs.harvard.edu/abs/2023AJ....166...45L}
}

@ARTICLE{Ortiz-Leon2017,
       author = {{Ortiz-Le{\'o}n}, Gisela N. and {Dzib}, Sergio A. and {Kounkel}, Marina A. and {Loinard}, Laurent and {Mioduszewski}, Amy J. and {Rodr{\'\i}guez}, Luis F. and {Torres}, Rosa M. and {Pech}, Gerardo and {Rivera}, Juana L. and {Hartmann}, Lee and {Boden}, Andrew F. and {Evans}, Neal J., II and {Brice{\~n}o}, Cesar and {Tobin}, John J. and {Galli}, Phillip A.~B.},
        title = "{The Gould{\textquoteright}s Belt Distances Survey (GOBELINS). III. The Distance to the Serpens/Aquila Molecular Complex}",
      journal = {\apj},
         year = 2017,
        month = jan,
       volume = {834},
       number = {2},
          eid = {143},
        pages = {143},
          doi = {10.3847/1538-4357/834/2/143},
archivePrefix = {arXiv},
       eprint = {1610.03128},
 primaryClass = {astro-ph.SR},
       adsurl = {https://ui.adsabs.harvard.edu/abs/2017ApJ...834..143O}
}

@ARTICLE{Federman2024,
       author = {{Federman}, Samuel A. and {Megeath}, S. Thomas and {Rubinstein}, Adam E. and {Gutermuth}, Robert and {Narang}, Mayank and {Tyagi}, Himanshu and {Manoj}, P. and {Anglada}, Guillem and {Atnagulov}, Prabhani and {Beuther}, Henrik and {Bourke}, Tyler L. and {Brunken}, Nashanty and {Caratti o Garatti}, Alessio and {Evans}, Neal J. and {Fischer}, William J. and {Furlan}, Elise and {Green}, Joel D. and {Habel}, Nolan and {Hartmann}, Lee and {Karnath}, Nicole and {Klaassen}, Pamela and {Linz}, Hendrik and {Looney}, Leslie W. and {Osorio}, Mayra and {Muzerolle Page}, James and {Nazari}, Pooneh and {Pokhrel}, Riwaj and {Rahatgaonkar}, Rohan and {Rocha}, Will R.~M. and {Sheehan}, Patrick and {Slavicinska}, Katerina and {Stanke}, Thomas and {Stutz}, Amelia M. and {Tobin}, John J. and {Tychoniec}, Lukasz and {Van Dishoeck}, Ewine F. and {Watson}, Dan M. and {Wolk}, Scott and {Yang}, Yao-Lun},
        title = "{Investigating Protostellar Accretion-driven Outflows across the Mass Spectrum: JWST NIRSpec Integral Field Unit 3{\textendash}5 {\ensuremath{\mu}}m Spectral Mapping of Five Young Protostars}",
      journal = {\apj},
         year = 2024,
        month = may,
       volume = {966},
       number = {1},
          eid = {41},
        pages = {41},
          doi = {10.3847/1538-4357/ad2fa0},
archivePrefix = {arXiv},
       eprint = {2310.03803},
 primaryClass = {astro-ph.SR},
       adsurl = {https://ui.adsabs.harvard.edu/abs/2024ApJ...966...41F}
}

@ARTICLE{Nisini2024,
       author = {{Nisini}, Brunella and {Navarro}, Maria Gabriela and {Giannini}, Teresa and {Antoniucci}, Simone and {Kavanagh}, Patrick, J. and {Hartigan}, Patrick and {Bacciotti}, Francesca and {Caratti o Garatti}, Alessio and {Noriega-Crespo}, Alberto and {van Dishoeck}, Ewine F. and {Whelan}, Emma T. and {Arce}, Hector G. and {Cabrit}, Sylvie and {Coffey}, Deirdre and {Fedele}, Davide and {Eisl{\"o}ffel}, Jochen and {Palumbo}, Maria Elisabetta and {Podio}, Linda and {Ray}, Tom P. and {Schultze}, Megan and {Urso}, Riccardo G. and {Alcal{\'a}}, Juan M. and {Bautista}, Manuel A. and {Codella}, Claudio and {Greene}, Thomas P. and {Manara}, Carlo F.},
        title = "{PROJECT-J: JWST Observations of HH46 IRS and Its Outflow. Overview and First Results}",
      journal = {\apj},
         year = 2024,
        month = jun,
       volume = {967},
       number = {2},
          eid = {168},
        pages = {168},
          doi = {10.3847/1538-4357/ad3d5a},
archivePrefix = {arXiv},
       eprint = {2404.06878},
 primaryClass = {astro-ph.GA},
       adsurl = {https://ui.adsabs.harvard.edu/abs/2024ApJ...967..168N}
}

@ARTICLE{Gelder2024,
       author = {{van Gelder}, M.~L. and {Ressler}, M.~E. and {van Dishoeck}, E.~F. and {Nazari}, P. and {Tabone}, B. and {Black}, J.~H. and {Tychoniec}, {\L}. and {Francis}, L. and {Barsony}, M. and {Beuther}, H. and {Caratti o Garatti}, A. and {Chen}, Y. and {Gieser}, C. and {le Gouellec}, V.~J.~M. and {Kavanagh}, P.~J. and {Klaassen}, P.~D. and {Lew}, B.~W.~P. and {Linnartz}, H. and {Majumdar}, L. and {Perotti}, G. and {Rocha}, W.~R.~M.},
        title = "{JOYS+: Mid-infrared detection of gas-phase SO$_{2}$ emission in a low-mass protostar. The case of NGC 1333 IRAS 2A: Hot core or accretion shock?}",
      journal = {\aap},
         year = 2024,
        month = feb,
       volume = {682},
          eid = {A78},
        pages = {A78},
          doi = {10.1051/0004-6361/202348118},
archivePrefix = {arXiv},
       eprint = {2311.17161},
 primaryClass = {astro-ph.SR},
       adsurl = {https://ui.adsabs.harvard.edu/abs/2024A&A...682A..78V}
}

@ARTICLE{Pontoppidan2024,
       author = {{Pontoppidan}, Klaus M. and {Evans}, Neal and {Bergner}, Jennifer and {Yang}, Yao-Lun},
        title = "{A Constrained Dust Opacity for Models of Dense Clouds and Protostellar Envelopes}",
      journal = {Research Notes of the American Astronomical Society},
         year = 2024,
        month = mar,
       volume = {8},
       number = {3},
          eid = {68},
        pages = {68},
          doi = {10.3847/2515-5172/ad303f},
       adsurl = {https://ui.adsabs.harvard.edu/abs/2024RNAAS...8...68P}
}

@software{pybaselines,
       author = {{Erb}, Donald},
        title = "{pybaselines: A Python library of algorithms for the baseline correction of experimental data}",
         year = 2024,
        month = feb,
          eid = {10.5281/zenodo.10676584},
          doi = {10.5281/zenodo.10676584},
      version = {v1.1.0},
    publisher = {Zenodo},
       adsurl = {https://ui.adsabs.harvard.edu/abs/2024zndo..10676584E}
}

@ARTICLE{Alfonso1998,
       author = {{Gonz{\'a}lez-Alfonso}, Eduardo and {Cernicharo}, Jos{\'e} and {van Dishoeck}, Ewine F. and {Wright}, Christopher M. and {Heras}, Ana},
        title = "{Radiative Transfer Models of Emission and Absorption in the H$_{2}$O 6 Micron Vibration-Rotation Band toward Orion-BN-KL}",
      journal = {\apjl},
         year = 1998,
        month = aug,
       volume = {502},
       number = {2},
        pages = {L169-L172},
          doi = {10.1086/311503},
       adsurl = {https://ui.adsabs.harvard.edu/abs/1998ApJ...502L.169G}
}

@ARTICLE{Alfonso2002,
       author = {{Gonz{\'a}lez-Alfonso}, E. and {Wright}, C.~M. and {Cernicharo}, J. and {Rosenthal}, D. and {Boonman}, A.~M.~S. and {van Dishoeck}, E.~F.},
        title = "{CO and H$_{2}$O vibrational emission toward Orion Peak 1 and Peak 2}",
      journal = {\aap},
         year = 2002,
        month = may,
       volume = {386},
        pages = {1074-1102},
          doi = {10.1051/0004-6361:20020362},
       adsurl = {https://ui.adsabs.harvard.edu/abs/2002A&A...386.1074G}
}

@ARTICLE{Lacy2013,
       author = {{Lacy}, John H.},
        title = "{Interpretation of Infrared Vibration-rotation Spectra of Interstellar and Circumstellar Molecules}",
      journal = {\apj},
         year = 2013,
        month = mar,
       volume = {765},
       number = {2},
          eid = {130},
        pages = {130},
          doi = {10.1088/0004-637X/765/2/130},
archivePrefix = {arXiv},
       eprint = {1302.5003},
 primaryClass = {astro-ph.GA},
       adsurl = {https://ui.adsabs.harvard.edu/abs/2013ApJ...765..130L}
}

@ARTICLE{Rubenstein2024,
       author = {{Rubinstein}, Adam E. and {Evans}, Neal J. and {Tyagi}, Himanshu and {Narang}, Mayank and {Nazari}, Pooneh and {Gutermuth}, Robert and {Federman}, Samuel and {Manoj}, P. and {Green}, Joel D. and {Watson}, Dan M. and {Megeath}, S. Thomas and {Rocha}, Will R.~M. and {Brunken}, Nashanty G.~C. and {Slavicinska}, Katerina and {van Dishoeck}, Ewine F. and {Beuther}, Henrik and {Bourke}, Tyler L. and {Caratti o Garatti}, Alessio and {Hartmann}, Lee and {Klaassen}, Pamela and {Linz}, Hendrik and {Looney}, Leslie W. and {Muzerolle}, James and {Stanke}, Thomas and {Tobin}, John J. and {Wolk}, Scott J. and {Yang}, Yao-Lun},
        title = "{IPA: Class 0 Protostars Viewed in CO Emission Using JWST}",
      journal = {\apj},
         year = 2024,
        month = oct,
       volume = {974},
       number = {1},
          eid = {112},
        pages = {112},
          doi = {10.3847/1538-4357/ad6b92},
archivePrefix = {arXiv},
       eprint = {2312.07807},
 primaryClass = {astro-ph.SR},
       adsurl = {https://ui.adsabs.harvard.edu/abs/2024ApJ...974..112R}
}

@ARTICLE{Santaella2024,
       author = {{Pereira-Santaella}, M. and {Gonz{\'a}lez-Alfonso}, E. and {Garc{\'\i}a-Bernete}, I. and {Garc{\'\i}a-Burillo}, S. and {Rigopoulou}, D.},
        title = "{The CO-to-H$_{2}$ conversion factor of molecular outflows. Rovibrational CO emission in NGC 3256-S resolved by JWST/NIRSpec}",
      journal = {\aap},
         year = 2024,
        month = jan,
       volume = {681},
          eid = {A117},
        pages = {A117},
          doi = {10.1051/0004-6361/202347942},
archivePrefix = {arXiv},
       eprint = {2309.06486},
 primaryClass = {astro-ph.GA},
       adsurl = {https://ui.adsabs.harvard.edu/abs/2024A&A...681A.117P}
}

@ARTICLE{Buiten2024,
       author = {{Buiten}, Victorine A. and {van der Werf}, Paul P. and {Viti}, Serena and {Armus}, Lee and {Barr}, Andrew G. and {Barcos-Mu{\~n}oz}, Loreto and {Evans}, Aaron S. and {Inami}, Hanae and {Linden}, Sean T. and {Privon}, George C. and {Song}, Yiqing and {Rich}, Jeffrey A. and {Aalto}, Susanne and {Appleton}, Philip N. and {B{\"o}ker}, Torsten and {Charmandaris}, Vassilis and {Diaz-Santos}, Tanio and {Hayward}, Christopher C. and {Lai}, Thomas S. -Y. and {Medling}, Anne M. and {Ricci}, Claudio and {U}, Vivian},
        title = "{GOALS-JWST: Mid-infrared Molecular Gas Excitation Probes the Local Conditions of Nuclear Star Clusters and the Active Galactic Nucleus in the LIRG VV 114}",
      journal = {\apj},
         year = 2024,
        month = may,
       volume = {966},
       number = {2},
          eid = {166},
        pages = {166},
          doi = {10.3847/1538-4357/ad344b},
archivePrefix = {arXiv},
       eprint = {2312.01945},
 primaryClass = {astro-ph.GA},
       adsurl = {https://ui.adsabs.harvard.edu/abs/2024ApJ...966..166B}
}

@ARTICLE{Matt2005,
       author = {{Matt}, Sean and {Pudritz}, Ralph E.},
        title = "{Accretion-powered Stellar Winds as a Solution to the Stellar Angular Momentum Problem}",
      journal = {\apjl},
         year = 2005,
        month = oct,
       volume = {632},
       number = {2},
        pages = {L135-L138},
          doi = {10.1086/498066},
archivePrefix = {arXiv},
       eprint = {astro-ph/0510060},
 primaryClass = {astro-ph},
       adsurl = {https://ui.adsabs.harvard.edu/abs/2005ApJ...632L.135M}
}

@ARTICLE{Blandford1982,
       author = {{Blandford}, R.~D. and {Payne}, D.~G.},
        title = "{Hydromagnetic flows from accretion disks and the production of radio jets.}",
      journal = {\mnras},
         year = 1982,
        month = jun,
       volume = {199},
        pages = {883-903},
          doi = {10.1093/mnras/199.4.883},
       adsurl = {https://ui.adsabs.harvard.edu/abs/1982MNRAS.199..883B}
}

@ARTICLE{Ray2021,
       author = {{Ray}, T.~P. and {Ferreira}, J.},
        title = "{Jets from young stars}",
      journal = {\nar},
         year = 2021,
        month = dec,
       volume = {93},
          eid = {101615},
        pages = {101615},
          doi = {10.1016/j.newar.2021.101615},
archivePrefix = {arXiv},
       eprint = {2009.00547},
 primaryClass = {astro-ph.SR},
       adsurl = {https://ui.adsabs.harvard.edu/abs/2021NewAR..9301615R}
}

@ARTICLE{Pudritz2019,
       author = {{Pudritz}, Ralph E. and {Ray}, Tom P.},
        title = "{The Role of Magnetic Fields in Protostellar Outflows and Star Formation}",
      journal = {Frontiers in Astronomy and Space Sciences},
         year = 2019,
        month = jul,
       volume = {6},
          eid = {54},
        pages = {54},
          doi = {10.3389/fspas.2019.00054},
archivePrefix = {arXiv},
       eprint = {1912.05605},
 primaryClass = {astro-ph.SR},
       adsurl = {https://ui.adsabs.harvard.edu/abs/2019FrASS...6...54P}
}

@ARTICLE{Lopez-Vazquez2024,
       author = {{L{\'o}pez-V{\'a}zquez}, J.~A. and {Lee}, Chin-Fei and {Shang}, Hsien and {Cabrit}, Sylvie and {Krasnopolsky}, Ruben and {Codella}, Claudio and {Liu}, Chun-Fan and {Podio}, Linda and {Dutta}, Somnath and {Murphy}, A. and {Wiseman}, Jennifer},
        title = "{Multiple Components of the Outflow in the Protostellar System HH 212: Outer Outflow Shell, Rotating Wind, Shocked Wind, and Jet}",
      journal = {\apj},
         year = 2024,
        month = dec,
       volume = {977},
       number = {1},
          eid = {126},
        pages = {126},
          doi = {10.3847/1538-4357/ad8eb4},
archivePrefix = {arXiv},
       eprint = {2411.01728},
 primaryClass = {astro-ph.SR},
       adsurl = {https://ui.adsabs.harvard.edu/abs/2024ApJ...977..126L}
}

@ARTICLE{cfLee2018,
       author = {{Lee}, Chin-Fei and {Li}, Zhi-Yun and {Codella}, Claudio and {Ho}, Paul T.~P. and {Podio}, Linda and {Hirano}, Naomi and {Shang}, Hsien and {Turner}, Neal J. and {Zhang}, Qizhou},
        title = "{A 100 au Wide Bipolar Rotating Shell Emanating from the HH 212 Protostellar Disk: A Disk Wind?}",
      journal = {\apj},
         year = 2018,
        month = mar,
       volume = {856},
       number = {1},
          eid = {14},
        pages = {14},
          doi = {10.3847/1538-4357/aaae6d},
archivePrefix = {arXiv},
       eprint = {1802.03668},
 primaryClass = {astro-ph.GA},
       adsurl = {https://ui.adsabs.harvard.edu/abs/2018ApJ...856...14L}
}

@ARTICLE{cfLee2017,
       author = {{Lee}, Chin-Fei and {Ho}, Paul. T.~P. and {Li}, Zhi-Yun and {Hirano}, Naomi and {Zhang}, Qizhou and {Shang}, Hsien},
        title = "{A rotating protostellar jet launched from the innermost disk of HH 212}",
      journal = {Nature Astronomy},
         year = 2017,
        month = jul,
       volume = {1},
          eid = {0152},
        pages = {0152},
          doi = {10.1038/s41550-017-0152},
archivePrefix = {arXiv},
       eprint = {1706.06343},
 primaryClass = {astro-ph.GA},
       adsurl = {https://ui.adsabs.harvard.edu/abs/2017NatAs...1E.152L}
}

@ARTICLE{Ray2023,
       author = {{Ray}, T.~P. and {McCaughrean}, M.~J. and {Caratti o Garatti}, A. and {Kavanagh}, P.~J. and {Justtanont}, K. and {van Dishoeck}, E.~F. and {Reitsma}, M. and {Beuther}, H. and {Francis}, L. and {Gieser}, C. and {Klaassen}, P. and {Perotti}, G. and {Tychoniec}, L. and {van Gelder}, M. and {Colina}, L. and {Greve}, Th. R. and {G{\"u}del}, M. and {Henning}, Th. and {Lagage}, P.~O. and {{\"O}stlin}, G. and {Vandenbussche}, B. and {Waelkens}, C. and {Wright}, G.},
        title = "{Outflows from the youngest stars are mostly molecular}",
      journal = {\nat},
         year = 2023,
        month = oct,
       volume = {622},
       number = {7981},
        pages = {48-52},
          doi = {10.1038/s41586-023-06551-1},
       adsurl = {https://ui.adsabs.harvard.edu/abs/2023Natur.622...48R}
}

@ARTICLE{CarattioGaratti2024,
       author = {{Caratti o Garatti}, A. and {Ray}, T.~P. and {Kavanagh}, P.~J. and {McCaughrean}, M.~J. and {Gieser}, C. and {Giannini}, T. and {van Dishoeck}, E.~F. and {Justtanont}, K. and {van Gelder}, M.~L. and {Francis}, L. and {Beuther}, H. and {Tychoniec}, {\L}. and {Nisini}, B. and {Navarro}, M.~G. and {Devaraj}, R. and {Reyes}, S. and {Nazari}, P. and {Klaassen}, P. and {G{\"u}del}, M. and {Henning}, Th. and {Lagage}, P.~O. and {{\"O}stlin}, G. and {Vandenbussche}, B. and {Waelkens}, C. and {Wright}, G.},
        title = "{JWST Observations of Young protoStars (JOYS): HH211: Textbook case of a protostellar jet and outflow}",
      journal = {\aap},
         year = 2024,
        month = nov,
       volume = {691},
          eid = {A134},
        pages = {A134},
          doi = {10.1051/0004-6361/202451350},
archivePrefix = {arXiv},
       eprint = {2409.16061},
 primaryClass = {astro-ph.SR},
       adsurl = {https://ui.adsabs.harvard.edu/abs/2024A&A...691A.134C}
}

@ARTICLE{Narang2025,
       author = {{Narang}, Mayank and {Ohashi}, Nagayoshi and {Tobin}, John J. and {McClure}, M.~K. and {J{\o}rgensen}, Jes K. and {Sai (Insa Choi)}, Jinshi and {eDisk + IceAge Team}},
        title = "{An Embedded Disk (eDisk) in the IceAge: Investigating the Jet and Outflow from Ced 110 IRS4}",
      journal = {\aj},
         year = 2025,
        month = apr,
       volume = {169},
       number = {4},
          eid = {192},
        pages = {192},
          doi = {10.3847/1538-3881/adb1ba},
archivePrefix = {arXiv},
       eprint = {2502.00394},
 primaryClass = {astro-ph.SR},
       adsurl = {https://ui.adsabs.harvard.edu/abs/2025AJ....169..192N}
}

@ARTICLE{Delabrosse2024,
       author = {{Delabrosse}, V. and {Dougados}, C. and {Cabrit}, S. and {Tabone}, B. and {Tychoniec}, L. and {Ray}, T. and {Podio}, L. and {McClure}, M.},
        title = "{JWST study of the DG Tau B disk-wind candidate. I. Overview and nested H$_{2}$-CO outflows}",
      journal = {\aap},
         year = 2024,
        month = aug,
       volume = {688},
          eid = {A173},
        pages = {A173},
          doi = {10.1051/0004-6361/202449176},
archivePrefix = {arXiv},
       eprint = {2403.19400},
 primaryClass = {astro-ph.SR},
       adsurl = {https://ui.adsabs.harvard.edu/abs/2024A&A...688A.173D}
}

@ARTICLE{Harsono2023,
       author = {{Harsono}, D. and {Bjerkeli}, P. and {Ramsey}, J.~P. and {Pontoppidan}, K.~M. and {Kristensen}, L.~E. and {J{\o}rgensen}, J.~K. and {Calcutt}, H. and {Li}, Z. -Y. and {Plunkett}, A.},
        title = "{JWST Peers into the Class I Protostar TMC1A: Atomic Jet and Spatially Resolved Dissociative Shock Region}",
      journal = {\apjl},
         year = 2023,
        month = jul,
       volume = {951},
       number = {2},
          eid = {L32},
        pages = {L32},
          doi = {10.3847/2041-8213/acdfca},
archivePrefix = {arXiv},
       eprint = {2306.08380},
 primaryClass = {astro-ph.SR},
       adsurl = {https://ui.adsabs.harvard.edu/abs/2023ApJ...951L..32H}
}

@ARTICLE{deValon2020,
       author = {{de Valon}, A. and {Dougados}, C. and {Cabrit}, S. and {Louvet}, F. and {Zapata}, L.~A. and {Mardones}, D.},
        title = "{ALMA reveals a large structured disk and nested rotating outflows in DG Tauri B}",
      journal = {\aap},
         year = 2020,
        month = feb,
       volume = {634},
          eid = {L12},
        pages = {L12},
          doi = {10.1051/0004-6361/201936950},
archivePrefix = {arXiv},
       eprint = {2001.09776},
 primaryClass = {astro-ph.SR},
       adsurl = {https://ui.adsabs.harvard.edu/abs/2020A&A...634L..12D}
}

@ARTICLE{Tychoniec2024,
       author = {{Tychoniec}, {\L}ukasz and {van Gelder}, Martijn L. and {van Dishoeck}, Ewine F. and {Francis}, Logan and {Rocha}, Will R.~M. and {Caratti o Garatti}, Alessio and {Beuther}, Henrik and {Gieser}, Caroline and {Justtanont}, Kay and {Linnartz}, Harold and {Le Gouellec}, Valentin J.~M. and {Perotti}, Giulia and {Devaraj}, Rangaswamy and {Tabone}, Beno{\^\i}t and {Ray}, Thomas P. and {Brunken}, Nashanty G.~C. and {Chen}, Yuan and {Kavanagh}, Patrick J. and {Klaassen}, Pamela and {Slavicinska}, Katerina and {G{\"u}del}, Manuel and {{\"O}stlin}, Goran},
        title = "{JWST Observations of Young protoStars (JOYS). Linked accretion and ejection in a Class I protobinary system}",
      journal = {\aap},
         year = 2024,
        month = jul,
       volume = {687},
          eid = {A36},
        pages = {A36},
          doi = {10.1051/0004-6361/202348889},
archivePrefix = {arXiv},
       eprint = {2402.04343},
 primaryClass = {astro-ph.SR},
       adsurl = {https://ui.adsabs.harvard.edu/abs/2024A&A...687A..36T}
}

@ARTICLE{Okoda2025,
       author = {{Okoda}, Yuki and {Yang}, Yao-Lun and {Evans}, II, Neal J. and {Kim}, Jaeyeong and {Jin}, Mihwa and {Garrod}, Robin T. and {Francis}, Logan and {Johnstone}, Doug and {Ceccarelli}, Cecilia and {Codella}, Claudio and {Chandler}, Claire J. and {Yamamoto}, Satoshi and {Sakai}, Nami},
        title = "{CORINOS. III. Outflow Shocked Regions of the Low-mass Protostellar Source IRAS 15398{\textendash}3359 with JWST and ALMA}",
      journal = {\apj},
         year = 2025,
        month = apr,
       volume = {982},
       number = {2},
          eid = {149},
        pages = {149},
          doi = {10.3847/1538-4357/adb83f},
archivePrefix = {arXiv},
       eprint = {2503.03050},
 primaryClass = {astro-ph.SR},
       adsurl = {https://ui.adsabs.harvard.edu/abs/2025ApJ...982..149O}
}

@ARTICLE{Matt2008,
       author = {{Matt}, Sean and {Pudritz}, Ralph E.},
        title = "{Accretion-powered Stellar Winds. II. Numerical Solutions for Stellar Wind Torques}",
      journal = {\apj},
         year = 2008,
        month = may,
       volume = {678},
       number = {2},
        pages = {1109-1118},
          doi = {10.1086/533428},
archivePrefix = {arXiv},
       eprint = {0801.0436},
 primaryClass = {astro-ph},
       adsurl = {https://ui.adsabs.harvard.edu/abs/2008ApJ...678.1109M}
}

@ARTICLE{Balbus1991,
       author = {{Balbus}, Steven A. and {Hawley}, John F.},
        title = "{A Powerful Local Shear Instability in Weakly Magnetized Disks. I. Linear Analysis}",
      journal = {\apj},
         year = 1991,
        month = jul,
       volume = {376},
        pages = {214},
          doi = {10.1086/170270},
       adsurl = {https://ui.adsabs.harvard.edu/abs/1991ApJ...376..214B}
}

@ARTICLE{Ercolano2017,
       author = {{Ercolano}, Barbara and {Pascucci}, Ilaria},
        title = "{The dispersal of planet-forming discs: theory confronts observations}",
      journal = {Royal Society Open Science},
         year = 2017,
        month = apr,
       volume = {4},
       number = {4},
          eid = {170114},
        pages = {170114},
          doi = {10.1098/rsos.170114},
archivePrefix = {arXiv},
       eprint = {1704.00214},
 primaryClass = {astro-ph.EP},
       adsurl = {https://ui.adsabs.harvard.edu/abs/2017RSOS....470114E}
}

@ARTICLE{Lesur2021,
       author = {{Lesur}, Geoffroy R.~J.},
        title = "{Systematic description of wind-driven protoplanetary discs}",
      journal = {\aap},
         year = 2021,
        month = jun,
       volume = {650},
          eid = {A35},
        pages = {A35},
          doi = {10.1051/0004-6361/202040109},
archivePrefix = {arXiv},
       eprint = {2101.10349},
 primaryClass = {astro-ph.SR},
       adsurl = {https://ui.adsabs.harvard.edu/abs/2021A&A...650A..35L}
}

@INPROCEEDINGS{Pascucci2023,
       author = {{Pascucci}, I. and {Cabrit}, S. and {Edwards}, S. and {Gorti}, U. and {Gressel}, O. and {Suzuki}, T.~K.},
        title = "{The Role of Disk Winds in the Evolution and Dispersal of Protoplanetary Disks}",
    booktitle = {Protostars and Planets VII},
         year = 2023,
       editor = {{Inutsuka}, S. and {Aikawa}, Y. and {Muto}, T. and {Tomida}, K. and {Tamura}, M.},
       series = {Astronomical Society of the Pacific Conference Series},
       volume = {534},
        month = jul,
        pages = {567},
          doi = {10.48550/arXiv.2203.10068},
archivePrefix = {arXiv},
       eprint = {2203.10068},
 primaryClass = {astro-ph.EP},
       adsurl = {https://ui.adsabs.harvard.edu/abs/2023ASPC..534..567P}
}

@ARTICLE{Rabenanahary2022,
       author = {{Rabenanahary}, M. and {Cabrit}, S. and {Meliani}, Z. and {Pineau des For{\^e}ts}, G.},
        title = "{Wide-angle protostellar outflows driven by narrow jets in stratified cores}",
      journal = {\aap},
         year = 2022,
        month = aug,
       volume = {664},
          eid = {A118},
        pages = {A118},
          doi = {10.1051/0004-6361/202243139},
archivePrefix = {arXiv},
       eprint = {2204.05850},
 primaryClass = {astro-ph.SR},
       adsurl = {https://ui.adsabs.harvard.edu/abs/2022A&A...664A.118R}
}

@ARTICLE{Green2024,
       author = {{Green}, Joel D. and {Pontoppidan}, Klaus M. and {Reiter}, Megan and {Watson}, Dan M. and {Shenoy}, Sachindev S. and {Manoj}, P. and {Narang}, Mayank},
        title = "{Why Are (Almost) All the Protostellar Outflows Aligned in Serpens Main?}",
      journal = {\apj},
         year = 2024,
        month = sep,
       volume = {972},
       number = {1},
          eid = {5},
        pages = {5},
          doi = {10.3847/1538-4357/ad5a02},
archivePrefix = {arXiv},
       eprint = {2406.13084},
 primaryClass = {astro-ph.SR},
       adsurl = {https://ui.adsabs.harvard.edu/abs/2024ApJ...972....5G}
}

@ARTICLE{Hensley2021,
       author = {{Hensley}, Brandon S. and {Draine}, B.~T.},
        title = "{Observational Constraints on the Physical Properties of Interstellar Dust in the Post-Planck Era}",
      journal = {\apj},
         year = 2021,
        month = jan,
       volume = {906},
       number = {2},
          eid = {73},
        pages = {73},
          doi = {10.3847/1538-4357/abc8f1},
archivePrefix = {arXiv},
       eprint = {2009.00018},
 primaryClass = {astro-ph.GA},
       adsurl = {https://ui.adsabs.harvard.edu/abs/2021ApJ...906...73H}
}

@ARTICLE{Chapman2009,
       author = {{Chapman}, Nicholas L. and {Mundy}, Lee G. and {Lai}, Shih-Ping and {Evans}, II, Neal J.},
        title = "{The Mid-Infrared Extinction Law in the Ophiuchus, Perseus, and Serpens Molecular Clouds}",
      journal = {\apj},
         year = 2009,
        month = jan,
       volume = {690},
       number = {1},
        pages = {496-511},
          doi = {10.1088/0004-637X/690/1/496},
archivePrefix = {arXiv},
       eprint = {0809.1106},
 primaryClass = {astro-ph},
       adsurl = {https://ui.adsabs.harvard.edu/abs/2009ApJ...690..496C}
}

@ARTICLE{McKee2007,
       author = {{McKee}, Christopher F. and {Ostriker}, Eve C.},
        title = "{Theory of Star Formation}",
      journal = {\araa},
         year = 2007,
        month = sep,
       volume = {45},
       number = {1},
        pages = {565-687},
          doi = {10.1146/annurev.astro.45.051806.110602},
archivePrefix = {arXiv},
       eprint = {0707.3514},
 primaryClass = {astro-ph},
       adsurl = {https://ui.adsabs.harvard.edu/abs/2007ARA&A..45..565M}
}

@ARTICLE{Ferreira2006,
       author = {{Ferreira}, J. and {Dougados}, C. and {Cabrit}, S.},
        title = "{Which jet launching mechanism(s) in T Tauri stars?}",
      journal = {\aap},
         year = 2006,
        month = jul,
       volume = {453},
       number = {3},
        pages = {785-796},
          doi = {10.1051/0004-6361:20054231},
archivePrefix = {arXiv},
       eprint = {astro-ph/0604053},
 primaryClass = {astro-ph},
       adsurl = {https://ui.adsabs.harvard.edu/abs/2006A&A...453..785F}
}

@ARTICLE{Bai2013,
       author = {{Bai}, Xue-Ning and {Stone}, James M.},
        title = "{Wind-driven Accretion in Protoplanetary Disks. I. Suppression of the Magnetorotational Instability and Launching of the Magnetocentrifugal Wind}",
      journal = {\apj},
         year = 2013,
        month = may,
       volume = {769},
       number = {1},
          eid = {76},
        pages = {76},
          doi = {10.1088/0004-637X/769/1/76},
archivePrefix = {arXiv},
       eprint = {1301.0318},
 primaryClass = {astro-ph.EP},
       adsurl = {https://ui.adsabs.harvard.edu/abs/2013ApJ...769...76B}
}

@INPROCEEDINGS{Pudritz2007,
       author = {{Pudritz}, R.~E. and {Ouyed}, R. and {Fendt}, Ch. and {Brandenburg}, A.},
        title = "{Disk Winds, Jets, and Outflows: Theoretical and Computational Foundations}",
    booktitle = {Protostars and Planets V},
         year = 2007,
       editor = {{Reipurth}, Bo and {Jewitt}, David and {Keil}, Klaus},
        month = jan,
        pages = {277},
          doi = {10.48550/arXiv.astro-ph/0603592},
archivePrefix = {arXiv},
       eprint = {astro-ph/0603592},
 primaryClass = {astro-ph},
       adsurl = {https://ui.adsabs.harvard.edu/abs/2007prpl.conf..277P}
}

@ARTICLE{Tafalla2000,
       author = {{Tafalla}, M. and {Myers}, P.~C. and {Mardones}, D. and {Bachiller}, R.},
        title = "{L483: a protostar in transition from Class 0 to Class I}",
      journal = {\aap},
         year = 2000,
        month = jul,
       volume = {359},
        pages = {967-976},
          doi = {10.48550/arXiv.astro-ph/0005525},
archivePrefix = {arXiv},
       eprint = {astro-ph/0005525},
 primaryClass = {astro-ph},
       adsurl = {https://ui.adsabs.harvard.edu/abs/2000A&A...359..967T}
}

@ARTICLE{Hirano2010,
       author = {{Hirano}, Naomi and {Ho}, Paul P.~T. and {Liu}, Sheng-Yuan and {Shang}, Hsien and {Lee}, Chin-Fei and {Bourke}, Tyler L.},
        title = "{Extreme Active Molecular Jets in L1448C}",
      journal = {\apj},
         year = 2010,
        month = jul,
       volume = {717},
       number = {1},
        pages = {58-73},
          doi = {10.1088/0004-637X/717/1/58},
archivePrefix = {arXiv},
       eprint = {1005.0703},
 primaryClass = {astro-ph.SR},
       adsurl = {https://ui.adsabs.harvard.edu/abs/2010ApJ...717...58H}
}

@ARTICLE{Fuente1993,
       author = {{Fuente}, A. and {Martin-Pintado}, J. and {Cernicharo}, J. and {Bachiller}, R.},
        title = "{A chemical study of the photodissociation region NGC 7023.}",
      journal = {\aap},
         year = 1993,
        month = sep,
       volume = {276},
        pages = {473-488},
       adsurl = {https://ui.adsabs.harvard.edu/abs/1993A&A...276..473F}
}

@ARTICLE{Pascucci2025,
       author = {{Pascucci}, Ilaria and {Beck}, Tracy L. and {Cabrit}, Sylvie and {Bajaj}, Naman S. and {Edwards}, Suzan and {Louvet}, Fabien and {Najita}, Joan R. and {Skinner}, Bennett N. and {Gorti}, Uma and {Salyk}, Colette and {Brittain}, Sean D. and {Krijt}, Sebastiaan and {Muzerolle Page}, James and {Ruaud}, Maxime and {Schwarz}, Kamber and {Semenov}, Dmitry and {Duch{\^e}ne}, Gaspard and {Villenave}, Marion},
        title = "{The nested morphology of disk winds from young stars revealed by JWST/NIRSpec observations}",
      journal = {Nature Astronomy},
         year = 2025,
        month = jan,
       volume = {9},
        pages = {81-89},
          doi = {10.1038/s41550-024-02385-7},
archivePrefix = {arXiv},
       eprint = {2410.18033},
 primaryClass = {astro-ph.EP},
       adsurl = {https://ui.adsabs.harvard.edu/abs/2025NatAs...9...81P}
}

@ARTICLE{shlee2024,
       author = {{Lee}, Seokho and {Lee}, Jeong-Eun and {Johnstone}, Doug and {Herczeg}, Gregory J. and {Aikawa}, Yuri},
        title = "{Multiple Jets in the Bursting Protostar HOPS 373SW}",
      journal = {\apj},
         year = 2024,
        month = mar,
       volume = {964},
       number = {1},
          eid = {34},
        pages = {34},
          doi = {10.3847/1538-4357/ad21e3},
archivePrefix = {arXiv},
       eprint = {2401.12449},
 primaryClass = {astro-ph.GA},
       adsurl = {https://ui.adsabs.harvard.edu/abs/2024ApJ...964...34L}
}

@ARTICLE{cfLee2022,
       author = {{Lee}, Chin-Fei and {Li}, Zhi-Yun and {Shang}, Hsien and {Hirano}, Naomi},
        title = "{Magnetocentrifugal Origin for Protostellar Jets Validated through Detection of Radial Flow at the Jet Base}",
      journal = {\apjl},
         year = 2022,
        month = mar,
       volume = {927},
       number = {2},
          eid = {L27},
        pages = {L27},
          doi = {10.3847/2041-8213/ac59c0},
archivePrefix = {arXiv},
       eprint = {2203.00961},
 primaryClass = {astro-ph.GA},
       adsurl = {https://ui.adsabs.harvard.edu/abs/2022ApJ...927L..27L}
}

@ARTICLE{Schutzer2022,
       author = {{de A. Schutzer}, A. and {Rivera-Ortiz}, P.~R. and {Lefloch}, B. and {Gusdorf}, A. and {Favre}, C. and {Segura-Cox}, D. and {L{\'o}pez-Sepulcre}, A. and {Neri}, R. and {Ospina-Zamudio}, J. and {De Simone}, M. and {Codella}, C. and {Viti}, S. and {Podio}, L. and {Pineda}, J. and {O'Donoghue}, R. and {Ceccarelli}, C. and {Caselli}, P. and {Alves}, F. and {Bachiller}, R. and {Balucani}, N. and {Bianchi}, E. and {Bizzocchi}, L. and {Bottinelli}, S. and {Caux}, E. and {Chac{\'o}n-Tanarro}, A. and {Dulieu}, F. and {Enrique-Romero}, J. and {Fontani}, F. and {Feng}, S. and {Holdship}, J. and {Jim{\'e}nez-Serra}, I. and {Jaber Al-Edhari}, A. and {Kahane}, C. and {Lattanzi}, V. and {Oya}, Y. and {Punanova}, A. and {Rimola}, A. and {Sakai}, N. and {Spezzano}, S. and {Sims}, I.~R. and {Taquet}, V. and {Testi}, L. and {Theul{\'e}}, P. and {Ugliengo}, P. and {Vastel}, C. and {Vasyunin}, A.~I. and {Vazart}, F. and {Yamamoto}, S. and {Witzel}, A.},
        title = "{SOLIS. XVI. Mass ejection and time variability in protostellar outflows: Cep E}",
      journal = {\aap},
         year = 2022,
        month = jun,
       volume = {662},
          eid = {A104},
        pages = {A104},
          doi = {10.1051/0004-6361/202142931},
archivePrefix = {arXiv},
       eprint = {2203.09383},
 primaryClass = {astro-ph.SR},
       adsurl = {https://ui.adsabs.harvard.edu/abs/2022A&A...662A.104D}
}

@ARTICLE{Vleugels2025,
       author = {{Vleugels}, C. and {McClure}, M. and {Sturm}, A. and {Vlasblom}, M.},
        title = "{The H$_{2}$ jet and disk wind of the Class I protostar HOPS 315}",
      journal = {\aap},
         year = 2025,
        month = mar,
       volume = {695},
          eid = {A145},
        pages = {A145},
          doi = {10.1051/0004-6361/202452475},
       adsurl = {https://ui.adsabs.harvard.edu/abs/2025A&A...695A.145V}
}

@ARTICLE{cfLee2000,
       author = {{Lee}, Chin-Fei and {Mundy}, Lee G. and {Reipurth}, Bo and {Ostriker}, Eve C. and {Stone}, James M.},
        title = "{CO Outflows from Young Stars: Confronting the Jet and Wind Models}",
      journal = {\apj},
         year = 2000,
        month = oct,
       volume = {542},
       number = {2},
        pages = {925-945},
          doi = {10.1086/317056},
       adsurl = {https://ui.adsabs.harvard.edu/abs/2000ApJ...542..925L}
}

@ARTICLE{Li1996,
       author = {{Li}, Zhi-Yun and {Shu}, Frank H.},
        title = "{Magnetized Singular Isothermal Toroids}",
      journal = {\apj},
         year = 1996,
        month = nov,
       volume = {472},
        pages = {211},
          doi = {10.1086/178056},
       adsurl = {https://ui.adsabs.harvard.edu/abs/1996ApJ...472..211L}
}

@ARTICLE{Smith1997,
       author = {{Smith}, M.~D. and {Suttner}, G. and {Yorke}, H.~W.},
        title = "{Numerical hydrodynamic simulations of jet-driven bipolar outflows.}",
      journal = {\aap},
         year = 1997,
        month = jul,
       volume = {323},
        pages = {223-230},
       adsurl = {https://ui.adsabs.harvard.edu/abs/1997A&A...323..223S}
}

@ARTICLE{Assani2025,
       author = {{Assani}, K.~D. and {Li}, Z.-Y. and {Ramsey}, J.~P. and {Tychoniec}, {\L}. and {Francis}, L. and {Le Gouellec}, V.~J.~M. and {Caratti o Garatti}, A. and {Giannini}, T. and {McClure}, M. and {Bjerkeli}, P. and {Calcutt}, H. and {Beuther}, H. and {Devaraj}, R. and {Liu}, X. and {Plunkett}, A. and {Navarro}, M.~G. and {van Dishoeck}, E.~F. and {Harsono}, D.},
        title = "{Mid-infrared extinction curve for protostellar envelopes from JWST-detected embedded jet emission: The case of TMC1A}",
      journal = {\aap},
         year = 2025,
        month = sep,
       volume = {701},
          eid = {A175},
        pages = {A175},
          doi = {10.1051/0004-6361/202555016},
archivePrefix = {arXiv},
       eprint = {2504.02136},
 primaryClass = {astro-ph.SR},
       adsurl = {https://ui.adsabs.harvard.edu/abs/2025A&A...701A.175A}
}

@ARTICLE{corner,
       author = {{Foreman-Mackey}, Daniel},
        title = "{corner.py: Scatterplot matrices in Python}",
      journal = {The Journal of Open Source Software},
         year = 2016,
        month = jun,
       volume = {1},
        pages = {24},
          doi = {10.21105/joss.00024},
       adsurl = {https://ui.adsabs.harvard.edu/abs/2016JOSS....1...24F}
}

@ARTICLE{Hsieh2024,
       author = {{Hsieh}, Cheng-Han and {Arce}, H{\'e}ctor G. and {Maureira}, Mar{\'\i}a Jos{\'e} and {Pineda}, Jaime E. and {Segura-Cox}, Dominique and {Mardones}, Diego and {Dunham}, Michael M. and {Arun}, Aiswarya},
        title = "{The ALMA Legacy Survey of Class 0/I Disks in Corona australis, Aquila, chaMaeleon, oPhiuchus north, Ophiuchus, Serpens (CAMPOS). I. Evolution of Protostellar Disk Radii}",
      journal = {\apj},
         year = 2024,
        month = oct,
       volume = {973},
       number = {2},
          eid = {138},
        pages = {138},
          doi = {10.3847/1538-4357/ad6152},
archivePrefix = {arXiv},
       eprint = {2404.02809},
 primaryClass = {astro-ph.SR},
       adsurl = {https://ui.adsabs.harvard.edu/abs/2024ApJ...973..138H}
}

@ARTICLE{Ghojogh2019,
       author = {{Ghojogh}, Benyamin and {Ghojogh}, Aydin and {Crowley}, Mark and {Karray}, Fakhri},
        title = "{Fitting A Mixture Distribution to Data: Tutorial}",
      journal = {arXiv e-prints},
         year = 2019,
        month = jan,
          eid = {arXiv:1901.06708},
        pages = {arXiv:1901.06708},
          doi = {10.48550/arXiv.1901.06708},
archivePrefix = {arXiv},
       eprint = {1901.06708},
 primaryClass = {stat.OT},
       adsurl = {https://ui.adsabs.harvard.edu/abs/2019arXiv190106708G}
}

@ARTICLE{Herczeg2011,
       author = {{Herczeg}, G.~J. and {Brown}, J.~M. and {van Dishoeck}, E.~F. and {Pontoppidan}, K.~M.},
        title = "{Disks and outflows in CO rovibrational emission from embedded, low-mass young stellar objects}",
      journal = {\aap},
         year = 2011,
        month = sep,
       volume = {533},
          eid = {A112},
        pages = {A112},
          doi = {10.1051/0004-6361/201016246},
archivePrefix = {arXiv},
       eprint = {1106.5391},
 primaryClass = {astro-ph.SR},
       adsurl = {https://ui.adsabs.harvard.edu/abs/2011A&A...533A.112H}
}

@ARTICLE{Terebey1984,
       author = {{Terebey}, S. and {Shu}, F.~H. and {Cassen}, P.},
        title = "{The collapse of the cores of slowly rotating isothermal clouds}",
      journal = {\apj},
         year = 1984,
        month = nov,
       volume = {286},
        pages = {529-551},
          doi = {10.1086/162628},
       adsurl = {https://ui.adsabs.harvard.edu/abs/1984ApJ...286..529T}
}

@ARTICLE{Liang2020,
       author = {{Liang}, Lichen and {Johnstone}, Doug and {Cabrit}, Sylvie and {Kristensen}, Lars E.},
        title = "{Steady Wind-blown Cavities within Infalling Rotating Envelopes: Application to the Broad Velocity Component in Young Protostars}",
      journal = {\apj},
         year = 2020,
        month = sep,
       volume = {900},
       number = {1},
          eid = {15},
        pages = {15},
          doi = {10.3847/1538-4357/aba830},
archivePrefix = {arXiv},
       eprint = {2007.13744},
 primaryClass = {astro-ph.SR},
       adsurl = {https://ui.adsabs.harvard.edu/abs/2020ApJ...900...15L}
}

@ARTICLE{matplotlib,
       author = {{Hunter}, John D.},
        title = "{Matplotlib: A 2D Graphics Environment}",
      journal = {Computing in Science and Engineering},
         year = 2007,
        month = may,
       volume = {9},
       number = {3},
        pages = {90-95},
          doi = {10.1109/MCSE.2007.55},
       adsurl = {https://ui.adsabs.harvard.edu/abs/2007CSE.....9...90H}
}

@ARTICLE{astropy,
       author = {{Astropy Collaboration} and {Price-Whelan}, Adrian M. and {Lim}, Pey Lian and {Earl}, Nicholas and {Starkman}, Nathaniel and {Bradley}, Larry and {Shupe}, David L. and {Patil}, Aarya A. and {Corrales}, Lia and {Brasseur}, C.~E. and {N{\"o}the}, Maximilian and {Donath}, Axel and {Tollerud}, Erik and {Morris}, Brett M. and {Ginsburg}, Adam and {Vaher}, Eero and {Weaver}, Benjamin A. and {Tocknell}, James and {Jamieson}, William and {van Kerkwijk}, Marten H. and {Robitaille}, Thomas P. and {Merry}, Bruce and {Bachetti}, Matteo and {G{\"u}nther}, H. Moritz and {Aldcroft}, Thomas L. and {Alvarado-Montes}, Jaime A. and {Archibald}, Anne M. and {B{\'o}di}, Attila and {Bapat}, Shreyas and {Barentsen}, Geert and {Baz{\'a}n}, Juanjo and {Biswas}, Manish and {Boquien}, M{\'e}d{\'e}ric and {Burke}, D.~J. and {Cara}, Daria and {Cara}, Mihai and {Conroy}, Kyle E. and {Conseil}, Simon and {Craig}, Matthew W. and {Cross}, Robert M. and {Cruz}, Kelle L. and {D'Eugenio}, Francesco and {Dencheva}, Nadia and {Devillepoix}, Hadrien A.~R. and {Dietrich}, J{\"o}rg P. and {Eigenbrot}, Arthur Davis and {Erben}, Thomas and {Ferreira}, Leonardo and {Foreman-Mackey}, Daniel and {Fox}, Ryan and {Freij}, Nabil and {Garg}, Suyog and {Geda}, Robel and {Glattly}, Lauren and {Gondhalekar}, Yash and {Gordon}, Karl D. and {Grant}, David and {Greenfield}, Perry and {Groener}, Austen M. and {Guest}, Steve and {Gurovich}, Sebastian and {Handberg}, Rasmus and {Hart}, Akeem and {Hatfield-Dodds}, Zac and {Homeier}, Derek and {Hosseinzadeh}, Griffin and {Jenness}, Tim and {Jones}, Craig K. and {Joseph}, Prajwel and {Kalmbach}, J. Bryce and {Karamehmetoglu}, Emir and {Ka{\l}uszy{\'n}ski}, Miko{\l}aj and {Kelley}, Michael S.~P. and {Kern}, Nicholas and {Kerzendorf}, Wolfgang E. and {Koch}, Eric W. and {Kulumani}, Shankar and {Lee}, Antony and {Ly}, Chun and {Ma}, Zhiyuan and {MacBride}, Conor and {Maljaars}, Jakob M. and {Muna}, Demitri and {Murphy}, N.~A. and {Norman}, Henrik and {O'Steen}, Richard and {Oman}, Kyle A. and {Pacifici}, Camilla and {Pascual}, Sergio and {Pascual-Granado}, J. and {Patil}, Rohit R. and {Perren}, Gabriel I. and {Pickering}, Timothy E. and {Rastogi}, Tanuj and {Roulston}, Benjamin R. and {Ryan}, Daniel F. and {Rykoff}, Eli S. and {Sabater}, Jose and {Sakurikar}, Parikshit and {Salgado}, Jes{\'u}s and {Sanghi}, Aniket and {Saunders}, Nicholas and {Savchenko}, Volodymyr and {Schwardt}, Ludwig and {Seifert-Eckert}, Michael and {Shih}, Albert Y. and {Jain}, Anany Shrey and {Shukla}, Gyanendra and {Sick}, Jonathan and {Simpson}, Chris and {Singanamalla}, Sudheesh and {Singer}, Leo P. and {Singhal}, Jaladh and {Sinha}, Manodeep and {Sip{\H{o}}cz}, Brigitta M. and {Spitler}, Lee R. and {Stansby}, David and {Streicher}, Ole and {{\v{S}}umak}, Jani and {Swinbank}, John D. and {Taranu}, Dan S. and {Tewary}, Nikita and {Tremblay}, Grant R. and {de Val-Borro}, Miguel and {Van Kooten}, Samuel J. and {Vasovi{\'c}}, Zlatan and {Verma}, Shresth and {de Miranda Cardoso}, Jos{\'e} Vin{\'\i}cius and {Williams}, Peter K.~G. and {Wilson}, Tom J. and {Winkel}, Benjamin and {Wood-Vasey}, W.~M. and {Xue}, Rui and {Yoachim}, Peter and {Zhang}, Chen and {Zonca}, Andrea and {Astropy Project Contributors}},
        title = "{The Astropy Project: Sustaining and Growing a Community-oriented Open-source Project and the Latest Major Release (v5.0) of the Core Package}",
      journal = {\apj},
         year = 2022,
        month = aug,
       volume = {935},
       number = {2},
          eid = {167},
        pages = {167},
          doi = {10.3847/1538-4357/ac7c74},
archivePrefix = {arXiv},
       eprint = {2206.14220},
 primaryClass = {astro-ph.IM},
       adsurl = {https://ui.adsabs.harvard.edu/abs/2022ApJ...935..167A}
}

@ARTICLE{numpy,
       author = {{Harris}, Charles R. and {Millman}, K. Jarrod and {van der Walt}, St{\'e}fan J. and {Gommers}, Ralf and {Virtanen}, Pauli and {Cournapeau}, David and {Wieser}, Eric and {Taylor}, Julian and {Berg}, Sebastian and {Smith}, Nathaniel J. and {Kern}, Robert and {Picus}, Matti and {Hoyer}, Stephan and {van Kerkwijk}, Marten H. and {Brett}, Matthew and {Haldane}, Allan and {del R{\'\i}o}, Jaime Fern{\'a}ndez and {Wiebe}, Mark and {Peterson}, Pearu and {G{\'e}rard-Marchant}, Pierre and {Sheppard}, Kevin and {Reddy}, Tyler and {Weckesser}, Warren and {Abbasi}, Hameer and {Gohlke}, Christoph and {Oliphant}, Travis E.},
        title = "{Array programming with NumPy}",
      journal = {\nat},
         year = 2020,
        month = sep,
       volume = {585},
       number = {7825},
        pages = {357-362},
          doi = {10.1038/s41586-020-2649-2},
archivePrefix = {arXiv},
       eprint = {2006.10256},
 primaryClass = {cs.MS},
       adsurl = {https://ui.adsabs.harvard.edu/abs/2020Natur.585..357H}
}

@ARTICLE{scipy,
       author = {{Virtanen}, Pauli and {Gommers}, Ralf and {Oliphant}, Travis E. and {Haberland}, Matt and {Reddy}, Tyler and {Cournapeau}, David and {Burovski}, Evgeni and {Peterson}, Pearu and {Weckesser}, Warren and {Bright}, Jonathan and {van der Walt}, St{\'e}fan J. and {Brett}, Matthew and {Wilson}, Joshua and {Millman}, K. Jarrod and {Mayorov}, Nikolay and {Nelson}, Andrew R.~J. and {Jones}, Eric and {Kern}, Robert and {Larson}, Eric and {Carey}, C.~J. and {Polat}, {\.I}lhan and {Feng}, Yu and {Moore}, Eric W. and {VanderPlas}, Jake and {Laxalde}, Denis and {Perktold}, Josef and {Cimrman}, Robert and {Henriksen}, Ian and {Quintero}, E.~A. and {Harris}, Charles R. and {Archibald}, Anne M. and {Ribeiro}, Ant{\^o}nio H. and {Pedregosa}, Fabian and {van Mulbregt}, Paul and {SciPy 1. 0 Contributors}},
        title = "{SciPy 1.0: fundamental algorithms for scientific computing in Python}",
      journal = {Nature Methods},
         year = 2020,
        month = feb,
       volume = {17},
        pages = {261-272},
          doi = {10.1038/s41592-019-0686-2},
archivePrefix = {arXiv},
       eprint = {1907.10121},
 primaryClass = {cs.MS},
       adsurl = {https://ui.adsabs.harvard.edu/abs/2020NatMe..17..261V}
}

@software{photutils,
       author = {{Bradley}, Larry and {Sipocz}, Brigitta and {Robitaille}, Thomas and {Tollerud}, Erik and {Deil}, Christoph and {Vin{\'\i}cius}, Z{\`e} and {Barbary}, Kyle and {G{\"u}nther}, Hans Moritz and {Bostroem}, Azalee and {Droettboom}, Michael and {Bray}, Erik and {Bratholm}, Lars Andersen and {Pickering}, T.~E. and {Craig}, Matt and {Pascual}, Sergio and {Greco}, Johnny and {Donath}, Axel and {Kerzendorf}, Wolfgang and {Littlefair}, Stuart and {Barentsen}, Geert and {D'Eugenio}, Francesco and {Weaver}, Benjamin Alan},
        title = "{Photutils: Photometry tools}",
 howpublished = {Astrophysics Source Code Library, record ascl:1609.011},
         year = 2016,
        month = sep,
          eid = {ascl:1609.011},
       adsurl = {https://ui.adsabs.harvard.edu/abs/2016ascl.soft09011B}
}

@ARTICLE{emcee,
       author = {{Foreman-Mackey}, Daniel and {Hogg}, David W. and {Lang}, Dustin and {Goodman}, Jonathan},
        title = "{emcee: The MCMC Hammer}",
      journal = {\pasp},
         year = 2013,
        month = mar,
       volume = {125},
       number = {925},
        pages = {306},
          doi = {10.1086/670067},
archivePrefix = {arXiv},
       eprint = {1202.3665},
 primaryClass = {astro-ph.IM},
       adsurl = {https://ui.adsabs.harvard.edu/abs/2013PASP..125..306F}
}

@ARTICLE{dEugenio2024,
       author = {{D'Eugenio}, Francesco and {P{\'e}rez-Gonz{\'a}lez}, Pablo G. and {Maiolino}, Roberto and {Scholtz}, Jan and {Perna}, Michele and {Circosta}, Chiara and {{\"U}bler}, Hannah and {Arribas}, Santiago and {B{\"o}ker}, Torsten and {Bunker}, Andrew J. and {Carniani}, Stefano and {Charlot}, Stephane and {Chevallard}, Jacopo and {Cresci}, Giovanni and {Curtis-Lake}, Emma and {Jones}, Gareth C. and {Kumari}, Nimisha and {Lamperti}, Isabella and {Looser}, Tobias J. and {Parlanti}, Eleonora and {Rix}, Hans-Walter and {Robertson}, Brant and {Rodr{\'\i}guez Del Pino}, Bruno and {Tacchella}, Sandro and {Venturi}, Giacomo and {Willott}, Chris J.},
        title = "{A fast-rotator post-starburst galaxy quenched by supermassive black-hole feedback at z = 3}",
      journal = {Nature Astronomy},
         year = 2024,
        month = nov,
       volume = {8},
        pages = {1443-1456},
          doi = {10.1038/s41550-024-02345-1},
archivePrefix = {arXiv},
       eprint = {2308.06317},
 primaryClass = {astro-ph.GA},
       adsurl = {https://ui.adsabs.harvard.edu/abs/2024NatAs...8.1443D}
}

@ARTICLE{Kristensen2023,
       author = {{Kristensen}, L.~E. and {Godard}, B. and {Guillard}, P. and {Gusdorf}, A. and {Pineau des For{\^e}ts}, G.},
        title = "{Shock excitation of H$_{2}$ in the James Webb Space Telescope era}",
      journal = {\aap},
         year = 2023,
        month = jul,
       volume = {675},
          eid = {A86},
        pages = {A86},
          doi = {10.1051/0004-6361/202346254},
archivePrefix = {arXiv},
       eprint = {2307.04178},
 primaryClass = {astro-ph.GA},
       adsurl = {https://ui.adsabs.harvard.edu/abs/2023A&A...675A..86K}
}

@ARTICLE{vanDishoeck2025,
       author = {{van Dishoeck}, E.~F. and {Tychoniec}, {\L}. and {Rocha}, W.~R.~M. and {Slavicinska}, K. and {Francis}, L. and {van Gelder}, M.~L. and {Ray}, T.~P. and {Beuther}, H. and {Caratti o Garatti}, A. and {Brunken}, N.~G.~C. and {Chen}, Y. and {Devaraj}, R. and {Geers}, V.~C. and {Gieser}, C. and {Greene}, T.~P. and {Justtanont}, K. and {Le Gouellec}, V.~J.~M. and {Kavanagh}, P.~J. and {Klaassen}, P.~D. and {Janssen}, A.~G.~M. and {Navarro}, M.~G. and {Nazari}, P. and {Notsu}, S. and {Perotti}, G. and {Ressler}, M.~E. and {Reyes}, S.~D. and {Sellek}, A.~D. and {Tabone}, B. and {Tap}, C. and {Theijssen}, N.~C.~M.~A. and {Colina}, L. and {G{\"u}del}, M. and {Henning}, Th. and {Lagage}, P.-O. and {{\"O}stlin}, G. and {Vandenbussche}, B. and {Wright}, G.~S.},
        title = "{JWST Observations of Young protoStars (JOYS): Overview of program and early results}",
      journal = {\aap},
         year = 2025,
        month = jul,
       volume = {699},
          eid = {A361},
        pages = {A361},
          doi = {10.1051/0004-6361/202554444},
archivePrefix = {arXiv},
       eprint = {2505.08002},
 primaryClass = {astro-ph.GA},
       adsurl = {https://ui.adsabs.harvard.edu/abs/2025A&A...699A.361V}
}

@ARTICLE{Kaplan2021,
       author = {{Kaplan}, Kyle F. and {Dinerstein}, Harriet L. and {Kim}, Hwihyun and {Jaffe}, Daniel T.},
        title = "{A Near-infrared Survey of UV-excited Molecular Hydrogen in Photodissociation Regions}",
      journal = {\apj},
         year = 2021,
        month = sep,
       volume = {919},
       number = {1},
          eid = {27},
        pages = {27},
          doi = {10.3847/1538-4357/ac0899},
archivePrefix = {arXiv},
       eprint = {2108.08484},
 primaryClass = {astro-ph.GA},
       adsurl = {https://ui.adsabs.harvard.edu/abs/2021ApJ...919...27K}
}

@ARTICLE{Navarro2025,
       author = {{Navarro}, Maria Gabriela and {Nisini}, Brunella and {Giannini}, Teresa and {Kavanagh}, Patrick J. and {Caratti o Garatti}, Alessio and {Antoniucci}, Simone and {Arce}, Hector G. and {Bacciotti}, Francesca and {Cabrit}, Sylvie and {Coffey}, Deirdre and {Dougados}, Catherine and {Eisl{\"o}ffel}, Jochen and {Hartigan}, Patrick and {Crespo}, Alberto Noriega- and {Podio}, Linda and {van Dishoeck}, Ewine F. and {Whelan}, Emma T.},
        title = "{PROJECT-J: The Shocking H$_{2}$ Outflow from HH 46}",
      journal = {\apj},
         year = 2025,
        month = dec,
       volume = {995},
       number = {2},
          eid = {199},
        pages = {199},
          doi = {10.3847/1538-4357/ae1f8f},
archivePrefix = {arXiv},
       eprint = {2511.17712},
 primaryClass = {astro-ph.GA},
       adsurl = {https://ui.adsabs.harvard.edu/abs/2025ApJ...995..199N}
}

@ARTICLE{Lee2026,
        author = {{Lee}, Jeong-Eun and {Kim}, Chul-Hwan and {Kim}, Jaeyoung and {Lee}, Seokho and {Kim}, Young-Jun and {Lee}, Seonjae and {Baek}, Giseon and {Green}, Joel D. and {Herczeg}, Gregory D. and {Johnstone}, Doug and {Pontoppidan}, Klaus M. and {Aikawa}, Yuri and {Yang}, Yao-Lun and {Francis}, Logan and {Jin}, Mihwa and {Jang}, Hyerin},
        title = "Accretion bursts crystallize silicates in a planet-forming disk",
journal = {Nature},
year = 2026,
month = jan,
volume = {649},
pages = {853}, 
doi = {10.1038/s41586-025-09939-3}
}

@ARTICLE{Bosman2019,
       author = {{Bosman}, Arthur D. and {Banzatti}, Andrea and {Bruderer}, Simon and {Tielens}, Alexander G.~G.~M. and {Blake}, Geoffrey A. and {van Dishoeck}, Ewine F.},
        title = "{Probing planet formation and disk substructures in the inner disk of Herbig Ae stars with CO rovibrational emission}",
      journal = {\aap},
         year = 2019,
        month = nov,
       volume = {631},
          eid = {A133},
        pages = {A133},
          doi = {10.1051/0004-6361/201935910},
archivePrefix = {arXiv},
       eprint = {1909.02031},
 primaryClass = {astro-ph.EP},
       adsurl = {https://ui.adsabs.harvard.edu/abs/2019A&A...631A.133B}
}

@article{Gordon2026,
title = {The HITRAN2024 molecular spectroscopic database},
journal = {Journal of Quantitative Spectroscopy and Radiative Transfer},
pages = {109807},
year = {2026},
issn = {0022-4073},
doi = {https://doi.org/10.1016/j.jqsrt.2026.109807},
url = {https://www.sciencedirect.com/science/article/pii/S0022407326000014},
author = {I.E. Gordon and L.S. Rothman and R.J. Hargreaves and F.M. Gomez and T. Bertin and C. Hill and R.V. Kochanov and Y. Tan and P. Wcisło and V. Yu. Makhnev and P.F. Bernath and M. Birk and V. Boudon and A. Campargue and A. Coustenis and B.J. Drouin and R.R. Gamache and J.T. Hodges and D. Jacquemart and E.J. Mlawer and A.V. Nikitin and V.I. Perevalov and M. Rotger and S. Robert and J. Tennyson and G.C. Toon and H. Tran and V.G. Tyuterev and E.M. Adkins and A. Barbe and D.M. Bailey and K. Bielska and L. Bizzocchi and T.A. Blake and C.A. Bowesman and P. Cacciani and P. Čermák and A.G. Császár and L. Denis and S.C. Egbert and O. Egorov and A. Yu. Ermilov and A.J. Fleisher and H. Fleurbaey and A. Foltynowicz and T. Furtenbacher and M. Germann and E.R. Guest and J.J. Harrison and J.-M. Hartmann and A. Hjältén and S.-M. Hu and X. Huang and T.J. Johnson and H. Jóźwiak and S. Kassi and M.V. Khan and F. Kwabia-Tchana and T.J. Lee and D. Lisak and A.-W. Liu and O.M. Lyulin and N.A. Malarich and L. Manceron and A.A. Marinina and S.T. Massie and J. Mascio and E.S. Medvedev and V.V. Meshkov and G. Ch. Mellau and M. Melosso and S.N. Mikhailenko and D. Mondelain and H.S.P. Müller and M. O’Donnell and A. Owens and A. Perrin and O.L. Polyansky and P.L. Raston and Z.D. Reed and M. Rey and C. Richard and G.B. Rieker and C. Röske and S.W. Sharpe and E. Starikova and N. Stolarczyk and A.V. Stolyarov and K. Sung and F. Tamassia and J. Terragni and V.G. Ushakov and S. Vasilchenko and B. Vispoel and K.L. Vodopyanov and G. Wagner and S. Wójtewicz and S.N. Yurchenko and N.F. Zobov},
}
\bibliographystyle{aasjournalv7}

\appendix
\restartappendixnumbering

\section{ALMA Continuum of EC 53} \label{sec:ALMA_continuum}

\begin{figure}
    \centering
    \includegraphics[width=\linewidth]{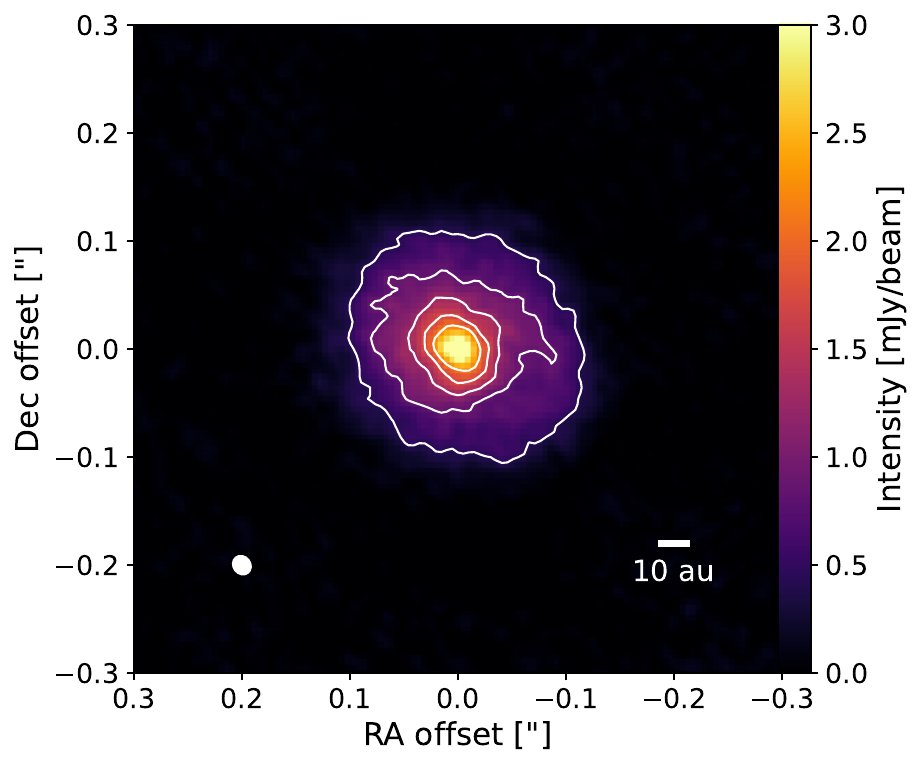}
    \caption{ALMA Continuum of EC 53. The contours are 10, 20, 30, 40, 50$\sigma$ ($\sigma = 0.043$\,mJy\,beam$^{-1}$). The beam size is marked as a white ellipse.}
    \label{fig:EC53_ALMA_continuum}
\end{figure}

In this section, the ALMA Band 7 continuum of EC 53 is presented in Figure \ref{fig:EC53_ALMA_continuum}. As explained in Section \ref{sec:continuum}, the emission is from the protostar and the surrounding protoplanetary disk.

\section{Hot/Cold pixel extraction} \label{sec:hotpixel}
\begin{figure*}
    \centering
    \includegraphics[width=\linewidth]{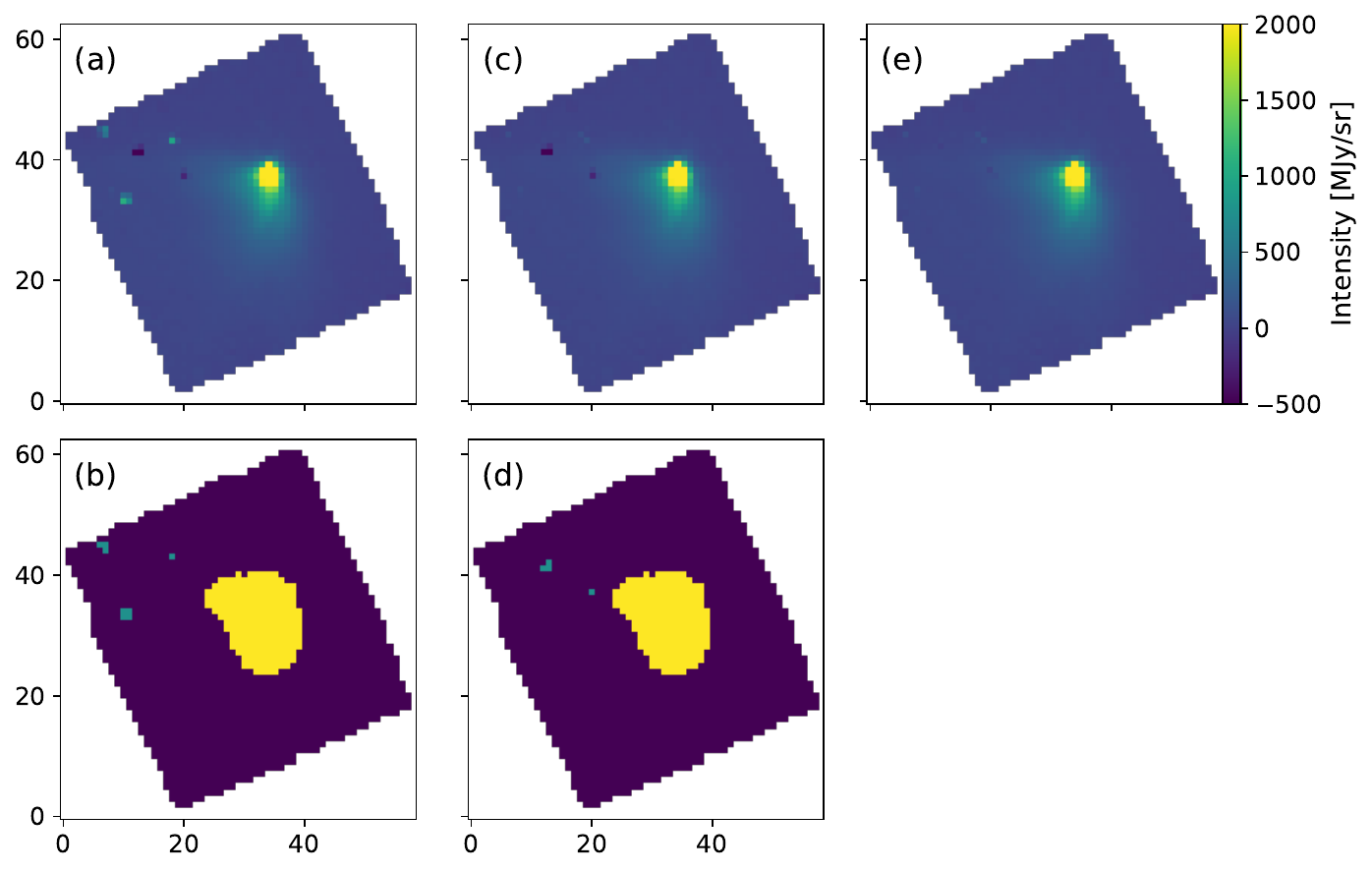}
    \caption{Hot and cold pixel subtraction process for the image slice at 2.7476\,$\mu$m. (a) Raw image slice. (b) `Sources' detected by the \lstinline{detect_sources} function. The yellow patch indicates the central source, while green pixels are detected as hot pixels. (c) Image after hot pixel flagging and median filtering. (d) Same with panel (b), but for cold pixels. (e) Final image slice, after correcting for cold pixels.}
    \label{fig:hotpixel}
\end{figure*}

The original JWST Calibration Pipeline was insufficient for fully removing hot and cold pixels, particularly in NIRSpec data. Although most artifacts appeared near the image edges and did not affect the scientific analysis, a small number of hot and cold pixels were present within the outflow region. To address this, we implemented a two-step method to remove artifacts. First, we used the \lstinline{detect_sources} function from the \lstinline{photutils} Python package to identify and remove hot pixels. For each image slice, we detected sources with an intensity threshold of 90\% of the image maximum. The largest detected source corresponds to the target itself, while the remaining detections were treated as artifacts. These artifacts were flagged and removed to produce the final science image. We used a median filter with size 5$\times$5 to smooth out the flagged pixels. Second, for the hot-pixel-removed images, we computed the background intensity level only from the unflagged pixels. The cold pixels with brightness more than 3$\sigma$ below the mean background were also flagged and median-filtered. Figure~\ref{fig:hotpixel} illustrates this procedure. \par
This two-step method effectively removed a significant portion of the hot and cold pixels; however, it is not able to remove low-intensity artifacts. Lowering the detection threshold could enable us to remove such features, but it would increase the area identified as the central source. Consequently, any hot/cold pixels located close to the central source may be incorporated into the central footprint and falsely preserved rather than flagged as artifacts. To avoid this misidentification, we used conservative threshold values.

\section{Comparison of flux between spectral bands} \label{sec:app_fluxcal}

\begin{figure*}
    \centering
    \includegraphics[width=\linewidth]{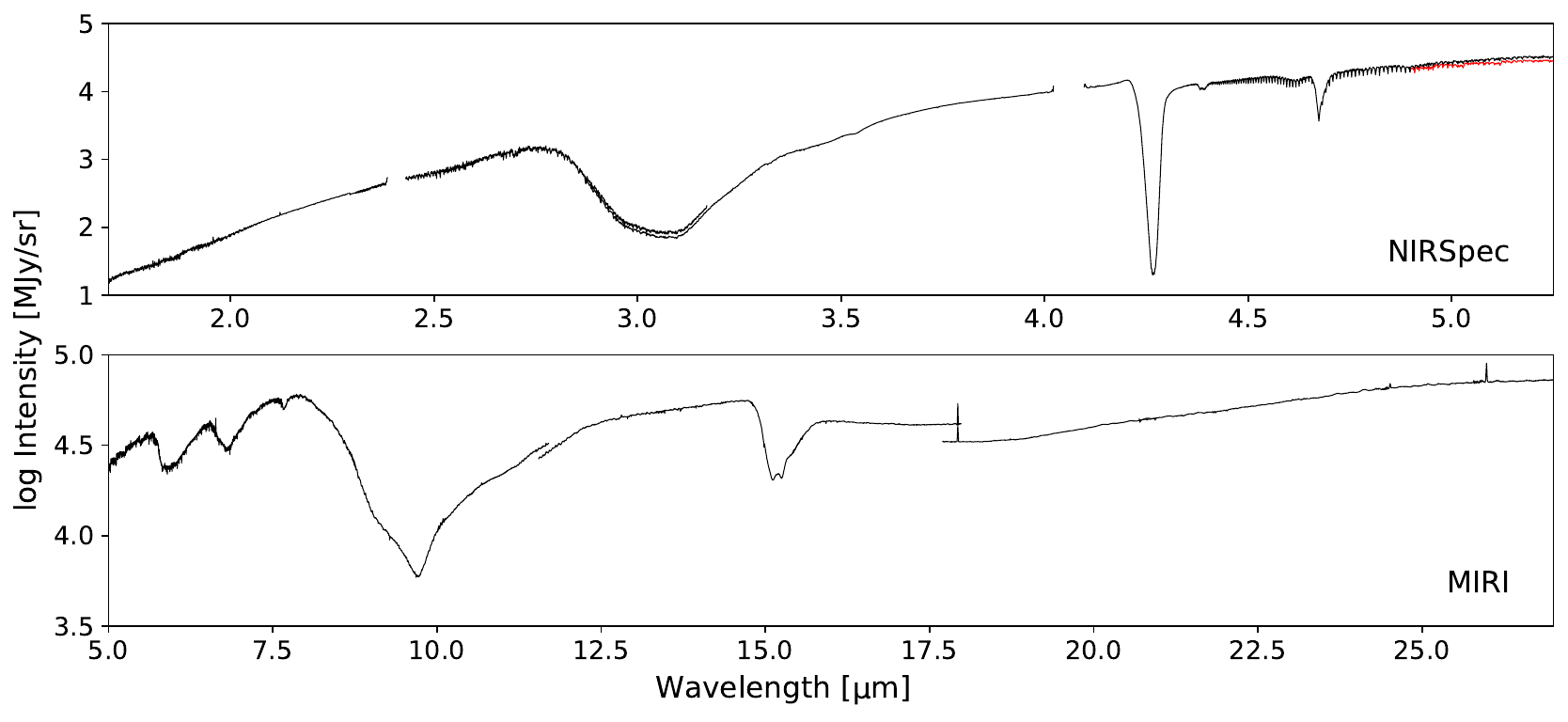}
    \vspace{1cm}
    \includegraphics[width=\linewidth]{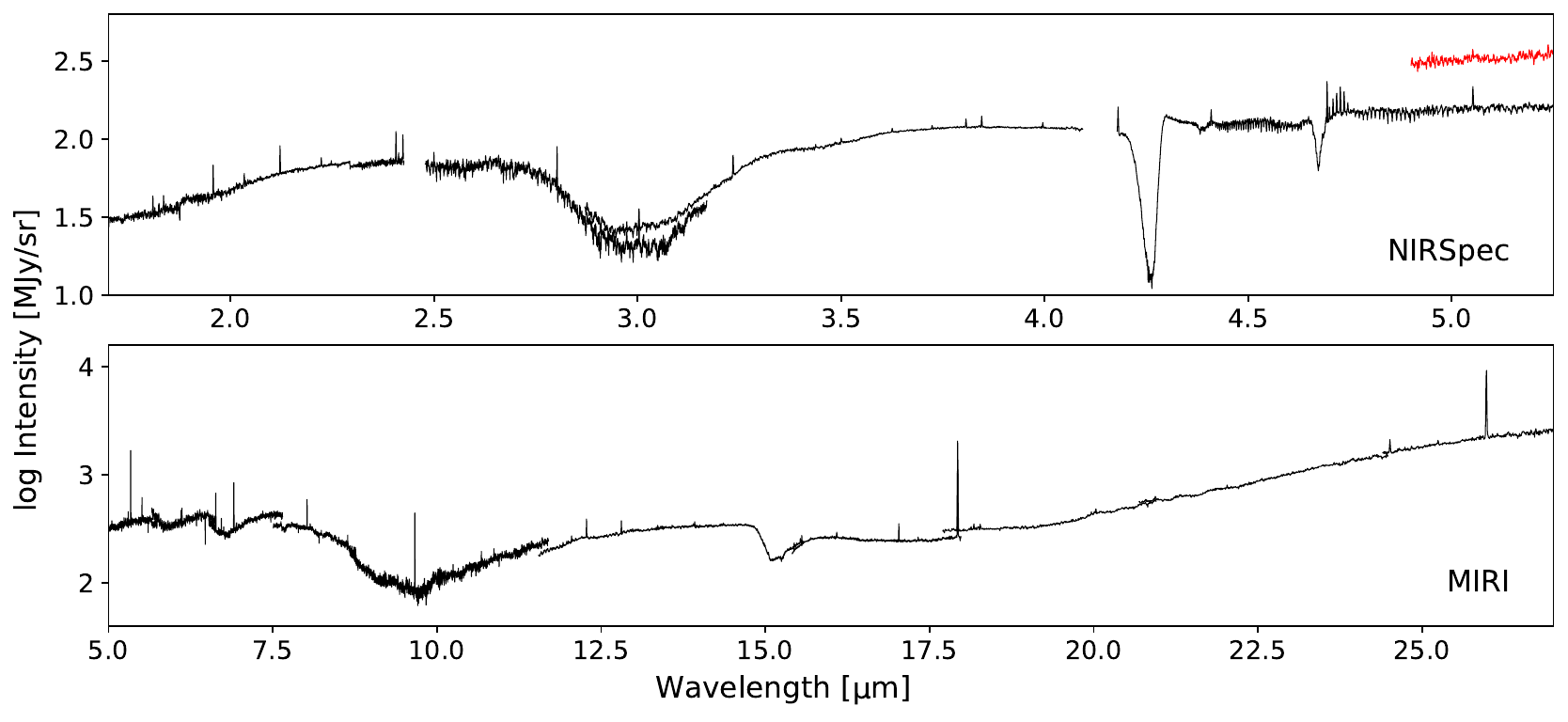}
    \caption{The upper two panels show spectra extracted from the ALMA continuum center using the aperture described in Section \ref{sec:fluxcal}. The red spectrum on the top panel is the MIRI spectrum over the wavelength range that overlaps with NIRSpec. The lower two panels show the same comparison but for spectra extracted from the outflow position of Figure \ref{fig:spectrum_center_outflow}.}
    \label{fig:spectrum_noalign}
\end{figure*}

Figure \ref{fig:spectrum_noalign} shows spectra extracted from two positions before flux stitching. At both points, the longest wavelength of NIRSpec and the shortest wavelength region of MIRI overlap with each other. While the ratio between median values is near unity at the center, it is much higher (>6) in the jet position. This discrepancy raises an issue in the JWST calibration pipeline that needs to be addressed. \par
\begin{figure*}
    \centering
    \includegraphics[width=\linewidth]{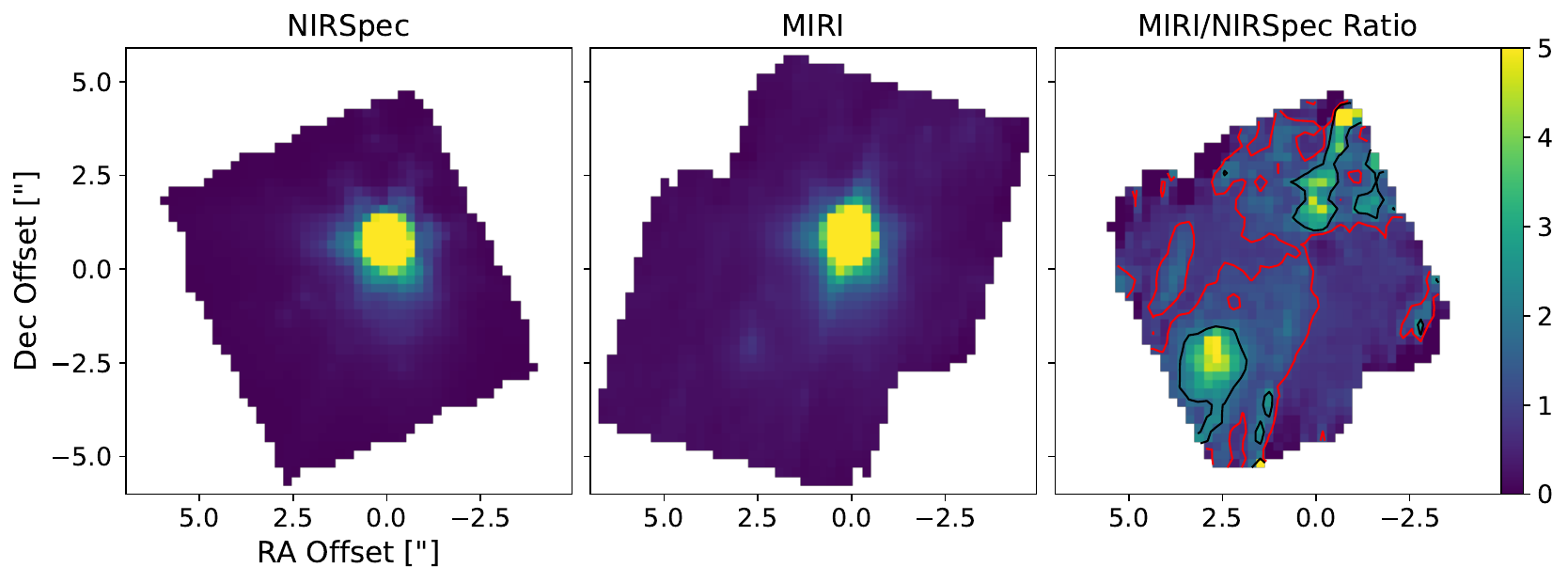}
    \caption{Median intensity maps between 5.0\,$\mu$m and 5.2\,$\mu$m for NIRSpec (left), MIRI (middle), and their ratio (right). In the ratio map, contours at values of 1 and 2 are drawn in red and black, respectively.}
    \label{fig:flux_ratio_2d_5um}
\end{figure*}

Figure \ref{fig:flux_ratio_2d_5um} shows the intensity ratio between the shortest wavelength range of MIRI and the longest wavelength range of NIRSpec. The NIRSpec image was resampled to match the pixel size of MIRI band 1. Although the ratio is below 2 in most images, it is significantly higher ($\sim$6) in some regions near the edges of the NIRSpec field. Notably, there is a 'blob' in the southeast corner where the ratio is elevated and overlaps with the blueshifted outflow jet.

\clearpage

\section{Emission Line list}
Table \ref{tb:h2lines} lists the parameters of all observed H$_2$ lines, sorted by vibrational level. All data are taken from the HITRAN database \citep{Gordon2026}.
Table \ref{tb:atomlines} lists the parameters of all atomic lines detected in our data, sorted by species and wavelength. All data are taken from the \href{https://physics.nist.gov/PhysRefData/ASD/lines_form.html}{NIST Atomic Spectra Database}. For some lines, the \href{https://linelist.pa.uky.edu/newpage}{Atomic Line List v3.00b5} \citep{vanHoof2018} reports slightly different wavelengths. Table A.1. of \cite{Assani2024} also presents a similar list of atomic lines and proposes that the two databases differ in their upper energy level (E$_{\rm up}$) values. However, these differences arise solely from the use of different units: the NIST database uses cm$^{-1}$ as its energy unit, while the Atomic Linelist uses K. After converting the units (using $hc/k_B = 1.439$), the values from both databases are consistent. \par

\begin{deluxetable*}{cccccc}
    \tablecaption{Spectral lines and parameters for H$_2$ emission lines detected in our observations.\label{tb:h2lines}}
    \tablehead{
    \colhead{Wavelength [$\mu$m]} & \colhead{Configuration} & \colhead{A$_{ul}$ [s$^{-1}$]} & \colhead{E$_{\rm up}$ [K]} & \colhead{g$_{\rm up}$} & \colhead{spw}}
    \startdata
    \multicolumn{6}{c}{ H$_2$ 1--0 }                           \\ \hline
    1.8358 & 1--0 S(5)  & 3.95 ($-7$) & 10341 & 45  & 235H        \\
    1.9576 & 1--0 S(3)  & 4.20 ($-7$) & 8365  & 33  & 235H        \\
    2.0338 & 1--0 S(2)  & 3.98 ($-7$) & 7584  & 9   & 235H        \\
    2.1218 & 1--0 S(1)  & 3.47 ($-7$) & 6951  & 21  & 235H        \\
    2.2233 & 1--0 S(0)  & 2.52 ($-7$) & 6471  & 5   & 235H        \\
    2.4066 & 1--0 Q(1)  & 4.29 ($-7$) & 6149  & 9   & 235H        \\
    2.5000 & 1--0 Q(7)  & 2.34 ($-7$) & 10341 & 45  & 235H        \\
    2.8025 & 1--0 O(3)  & 4.22 ($-7$) & 6149  & 9   & 235H        \\
    3.0039 & 1--0 O(4)  & 2.89 ($-7$) & 6471  & 5   & 235H, 395H \\
    3.2350 & 1--0 O(5)  & 2.08 ($-7$) & 6951  & 21  & 395H        \\
    3.5008 & 1--0 O(6)  & 1.50 ($-7$) & 7584  & 9   & 395H        \\
    3.8074 & 1--0 O(7)  & 1.06 ($-7$) & 8365  & 33  & 395H        \\ \hline \hline
    \multicolumn{6}{c}{ H$_2$ 0--0 }                           \\ \hline
    3.6262 & 0--0 S(15) & 2.40 ($-6$) & 21411 & 117 & 395H        \\
    3.7244 & 0--0 S(14) & 1.99 ($-6$) & 19403 & 33  & 395H        \\
    3.8461 & 0--0 S(13) & 1.61 ($-6$) & 17443 & 105 & 395H        \\
    3.9961 & 0--0 S(12) & 1.27 ($-6$) & 15540 & 29  & 395H        \\
    4.1811 & 0--0 S(11) & 9.60 ($-7$) & 13702 & 81  & 395H        \\
    4.4098 & 0--0 S(10) & 7.01 ($-7$) & 11940 & 25  & 395H        \\
    4.6946 & 0--0 S(9)  & 4.89 ($-7$) & 10261 & 69  & 395H        \\
    5.0531 & 0--0 S(8)  & 3.23 ($-7$) & 8677  & 21  & 395H        \\
    5.5112 & 0--0 S(7)  & 2.00 ($-7$) & 7197  & 57  & CH1-SHORT        \\
    6.1086 & 0--0 S(6)  & 1.14 ($-7$) & 5830  & 17  & CH1-MEDIUM        \\
    6.9095 & 0--0 S(5)  & 5.87 ($-8$) & 4586  & 45  & CH1-LONG        \\
    8.0250 & 0--0 S(4)  & 2.64 ($-8$) & 3474  & 13  & CH2-SHORT        \\
    9.6649 & 0--0 S(3)  & 9.83 ($-9$) & 2504  & 33  & CH2-MEDIUM        \\
    12.2786& 0--0 S(2)  & 2.75 ($-9$) & 1682  & 9   & CH3-SHORT        \\
    17.0348& 0--0 S(1)  & 4.76 ($-10$)& 1015  & 21  & CH3-LONG        \\
    \enddata
    \tablecomments{a($-$b) indicates a$\times 10^{-\rm b}$.}
\end{deluxetable*}

\begin{deluxetable*}{cccccc}
    \tablecaption{Spectral lines and parameters for atomic lines detected in our observations.\label{tb:atomlines}}
    \tablehead{
    \colhead{Wavelength [$\mu$m]} & \colhead{Configuration} & \colhead{A$_{ul}$ [s$^{-1}$]} & \colhead{E$_{\rm up}$ [K]} & \colhead{g$_{\rm up}$} & \colhead{spw}}
    \startdata
    \multicolumn{6}{c}{{[}Fe II{]}}                                                                      \\ \hline
    1.8094 \tablenotemark{a}    & a$^4$F 7/2 -- a$^4$D 7/2 & 9.99 ($-4$)   & 11446 & 8  & 235H           \\
    4.8891                      & a$^4$F 7/2 -- a$^6$D 7/2 & 9.3 ($-5$)    & 3496  & 8  & 395H           \\
    5.3401                      & a$^4$F 9/2 -- a$^6$D 9/2 & 1.3 ($-4$)    & 2694  & 10 & CH1-SHORT           \\
    6.7213                      & a$^4$F 9/2 -- a$^6$D 7/2 & 1.16 ($-5$)   & 2694  & 10 & CH1-LONG           \\
    17.936                      & a$^4$F 7/2 -- a$^4$F 9/2 & 5.843 ($-3$)  & 3496  & 8  & CH3-LONG, CH4-SHORT    \\
    24.5192                     & a$^4$F 5/2 -- a$^4$F 7/2 & 3.934 ($-3$)  & 4083  & 6  & CH4-LONG           \\
    25.9884                     & a$^6$D 7/2 -- a$^6$D 9/2 & 2.151 ($-3$)  & 554   & 8  & CH4-LONG           \\ \hline \hline
    \multicolumn{6}{c}{{[}Ne II{]}}                                                                      \\ \hline
    12.8135                     & 2Po 1/2 -- 2Po 3/2       & 8.59 ($-3$)   & 1123  & 2  & CH3-SHORT           \\ \hline \hline
    \multicolumn{6}{c}{{[}Ni II{]}}                                                                      \\ \hline
    6.6360                      & $^2$D 3/2 -- $^2$D 5/2   & 5.54 ($-2$)   & 2168  & 4  & CH1-LONG           \\
    10.6822                     & $^4$F 7/2 -- $^4$F 9/2   & 2.71 ($-2$)   & 13424 & 8  & CH2-LONG           \\
    \enddata
    \tablecomments{a($-$b) indicates a$\times 10^{-\rm b}$.}
    \tablenotetext{b}{The \href{https://linelist.pa.uky.edu/newpage}{Atomic Line List} reports the wavelength to be 1.8099\,$\mu$m.}
\end{deluxetable*}

\begin{figure*}
    \centering
    \includegraphics[width=\linewidth]{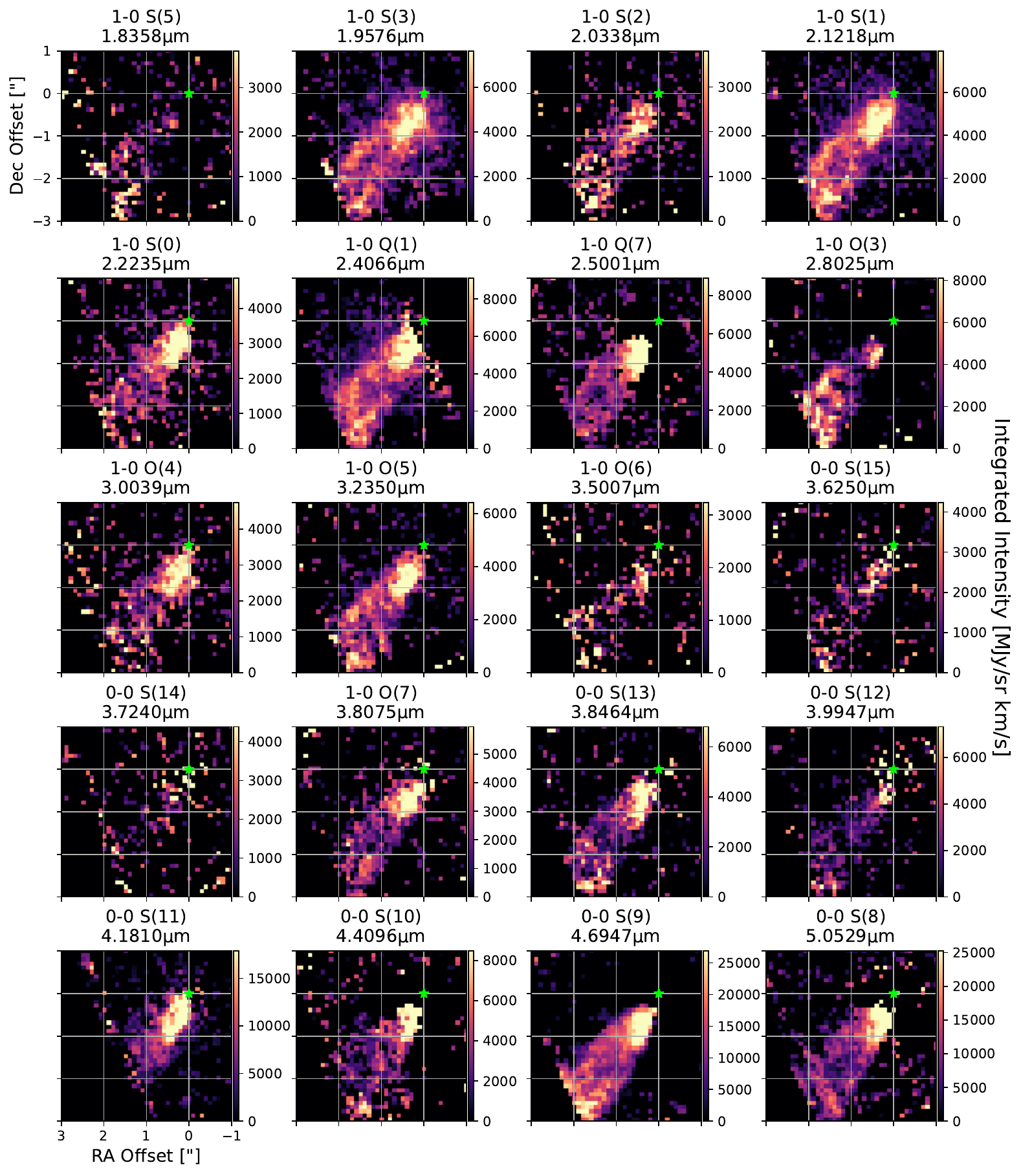}
    \caption{Integrated intensity maps of the H$_2$ lines detected with NIRSpec. The position of the central source is marked with a green star.}
    \label{fig:H2_nirspec_total_maps}
\end{figure*}

\begin{figure*}
    \centering
    \includegraphics[width=\linewidth]{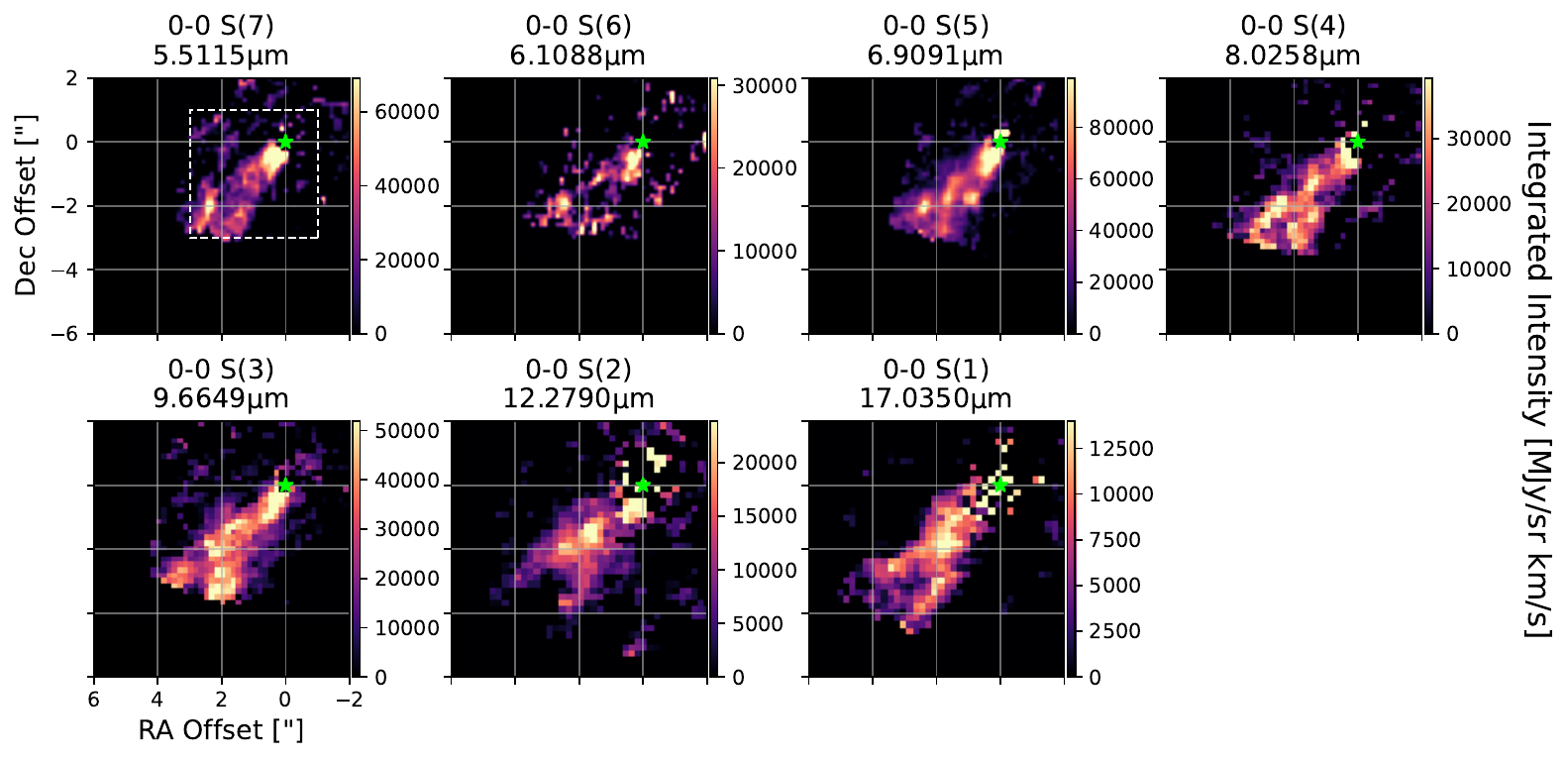}
    \caption{Integrated intensity maps of the H$_2$ lines detected with MIRI. The field of view of the NIRSpec data (Figure \ref{fig:H2_nirspec_total_maps}) is indicated by a white dashed square on the first panel.}
    \label{fig:H2_miri_total_maps}
\end{figure*}

\begin{figure*}
    \centering
    \includegraphics[width=\linewidth]{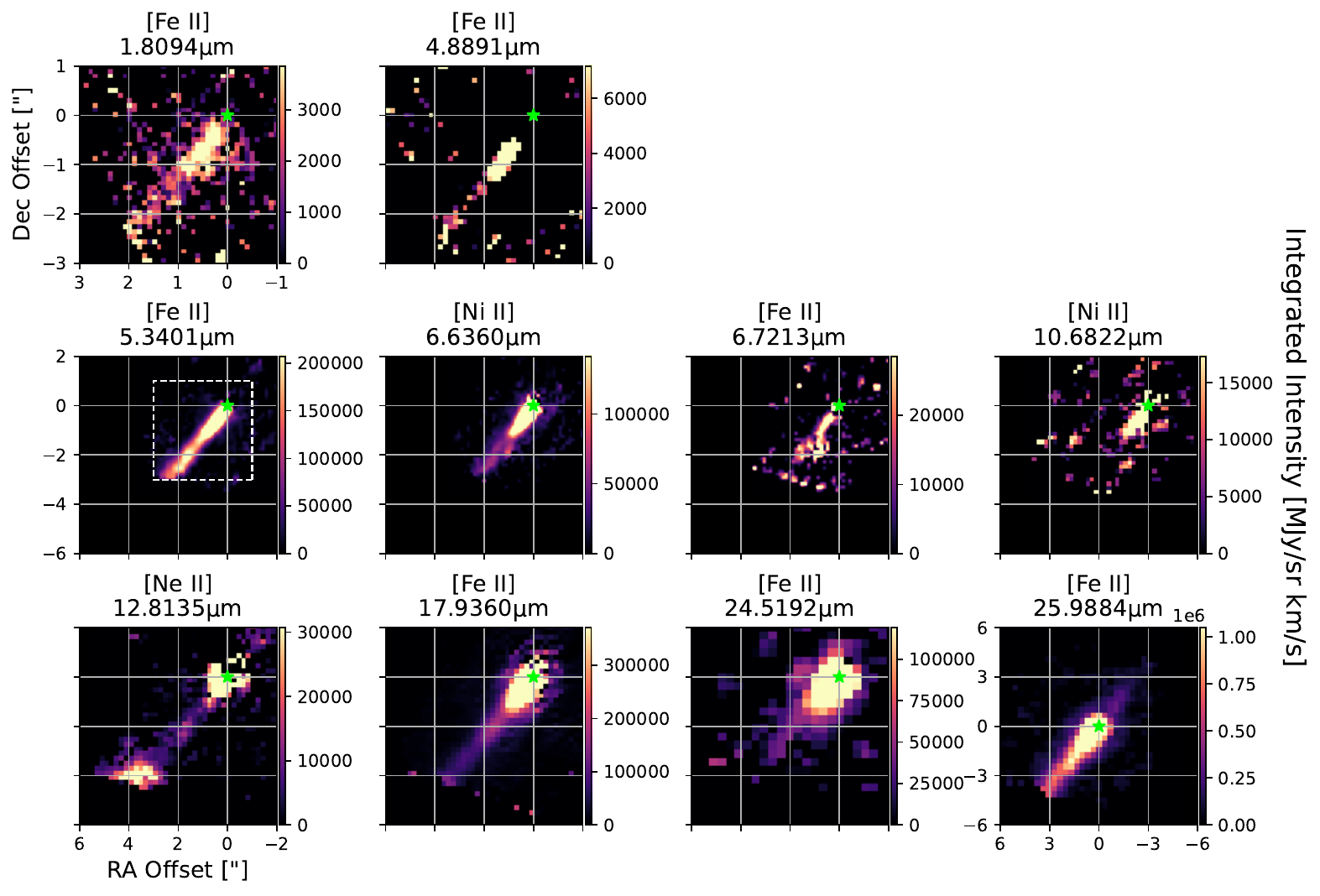}
    \caption{Integrated intensity maps of the detected atomic lines. The two transitions in the first row are from the NIRSpec wavelength range, while the remaining lines are from the MIRI range. A white dashed square on the first panel of the MIRI wavelength maps indicates the field of view of the NIRSpec data.}
    \label{fig:Atomic_total_maps}
\end{figure*}

\section{H$_2$ Excitation Diagram Description} \label{sec:H2_appendix}

\subsection{MCMC fitting} \label{sec:H2_MCMC}

\begin{figure*}[h]
    \centering
    \includegraphics[width=0.49\linewidth]{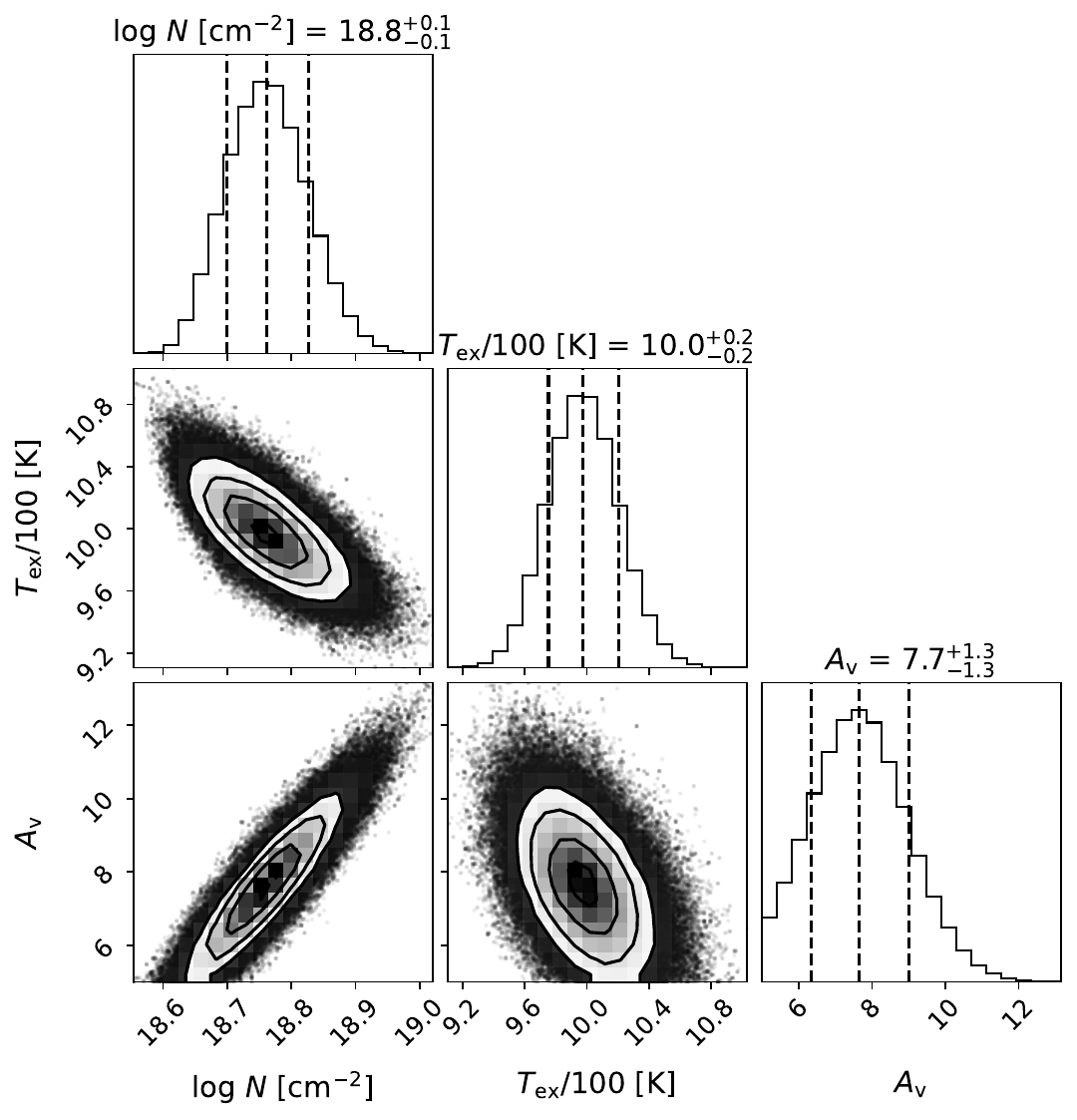}
    \includegraphics[width=0.49\linewidth]{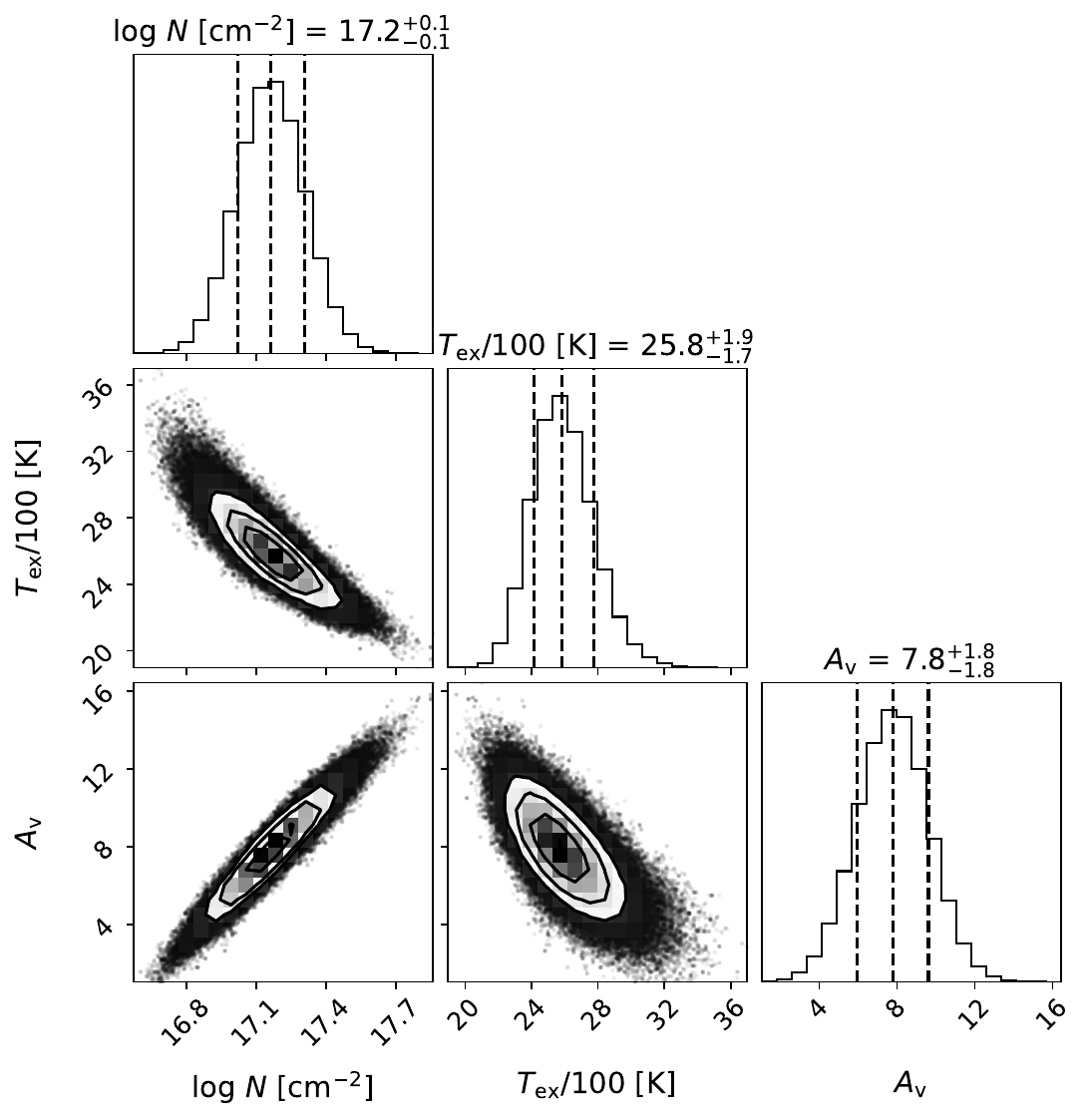}
    \caption{(left) Corner plot of the MCMC fit of Figure \ref{fig:H2_rotation}a. (right) Same with the left panel, but for Figure \ref{fig:H2_rotation}b.}
    \label{fig:H2_corner}
\end{figure*}

For the excitation diagram analysis of H$_2$, we performed a Bayesian analysis using Markov-Chain Monte Carlo (MCMC) sampling with the publicly available python package \lstinline{emcee} \citep{emcee}. We applied this to each pixel of the spatially convolved map. We assumed a Gaussian likelihood function, where the log-likelihood is proportional to $-\frac{1}{2}\chi^2$. The measurement uncertainties used to calculate $\chi^2$ were derived from the Gaussian fit error of each line. We adopted uniform priors for the parameters within reasonable ranges: $15 < \log N$\,[\rm cm$^{-2}$] $< 21$, $500 < T$\,[K] $< 2500$, and $5 < A_V < 30$. The same prior was used for fitting MIRI and NIRSpec data. We used 1000 walkers and 5000 steps, after the initial 1000 steps as burn-in. Figure \ref{fig:H2_corner} shows the marginalized posterior distributions in corner plots for the top panels of Figure \ref{fig:H2_rotation}, visualized with \lstinline{corner.py} \citep{corner}. The posterior sampling converged well to a single value. The uncertainties of individual components are derived from the 16th and 84th percentiles of the posterior distribution.

\subsection{Effect of Extinction Curves in H$_2$ Excitation Diagram} \label{sec:H2_extinction}

\begin{figure*}[h]
    \centering
    \includegraphics[width=\linewidth]{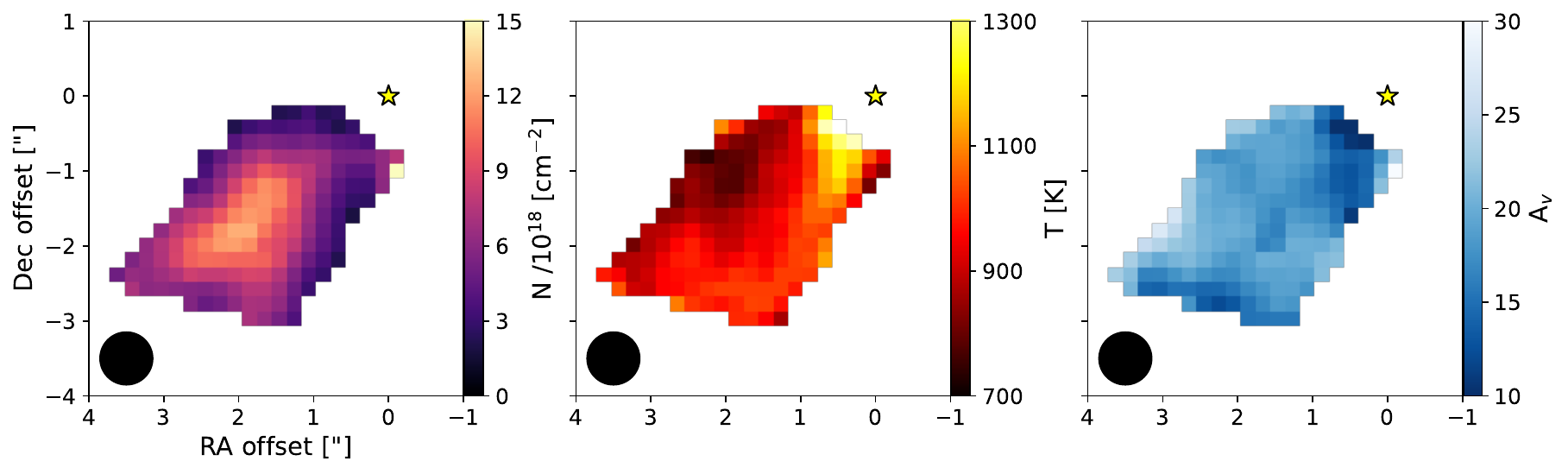}
    \includegraphics[width=\linewidth]{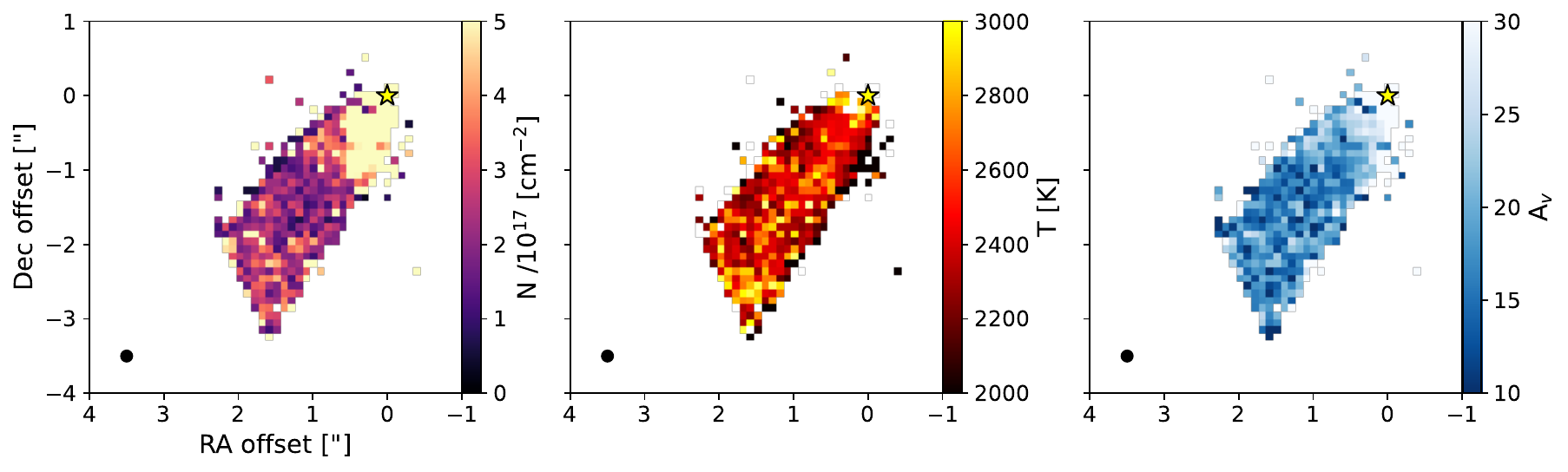}
    \caption{Same as the bottom 2 rows of Figure \ref{fig:H2_rotation}, but using the extinction curve of \cite{Hensley2021}.}
    \label{fig:H2_morph_Hensley}
\end{figure*}

In Section \ref{sec:H2_diagram}, we adopted the extinction curve of \cite{Pontoppidan2024} to correct for the interstellar extinction since it is designed explicitly for dense molecular clouds with ice mantles. However, as \cite{Chapman2009} pointed out, extinction laws vary across different molecular clouds and environments due to variations in dust properties. \cite{Assani2025} also showed that grain growth in protostellar envelopes can change the extinction curve, deviating from widely used relations. It is, therefore, important to check whether changing the assumed extinction curve significantly affects the results of the excitation diagram analysis. 

If the extinction from the ambient material dominates over that from the protostellar envelope, it may be more appropriate to use the extinction curve of \citet{Hensley2021}, which characterizes the diffuse ISM, rather than that of \citet{Pontoppidan2024}.
Therefore, we repeated the excitation diagram analysis using the extinction curve of \cite{Hensley2021}. Figure \ref{fig:H2_morph_Hensley} shows the results. For the long wavelengths, the distribution of column density and temperature is broadly consistent between the two models. However, the extinction values differ significantly, with those from \cite{Hensley2021} approximately twice as large. Meanwhile, for the NIRSpec data, there is a significant increase in column density near the central source when using the extinction curve of \cite{Hensley2021}. This is because many H$_2$ lines in shorter wavelengths get affected by water ice absorption, which only \cite{Pontoppidan2024} takes into account.



\end{document}